\documentclass[twocolumn]{aastex61}

\usepackage{amssymb}
\usepackage{amsmath}
\usepackage{latexsym}
\usepackage{fnpos}
\usepackage{scrextend}
\usepackage{scalerel}
\usepackage{hyperref}
\usepackage{tabularx}
\usepackage{xspace}
\usepackage{enumerate}
\usepackage{graphicx}
\usepackage{url}
\usepackage{longtable}
\usepackage{rotating}
\newcommand{\eal}[2]{\ifmmode{\mathrm{#1\,#2}}\else{#1\textsc{$\,$\lowercase{#2}}}\fi\xspace}
\newcommand{\feal}[2]{\ifmmode{\mathrm{#1\,#2}}\else{[#1\textsc{$\,$\lowercase{#2}}]}\fi\xspace}
\newcommand{\hfeal}[2]{\ifmmode{\mathrm{#1\,#2}}\else{#1\textsc{$\,$\lowercase{#2}}]}\fi\xspace}

\received{2021-08-18}
\accepted{2022-09-09}

\shorttitle{The peculiar outburst in V1047 Cen}
\shortauthors{A\MakeLowercase{ydi et al.}}

\begin{document}

\title{The 2019 outburst of the 2005 classical nova V1047~Cen:\\
a record breaking dwarf nova outburst or a new phenomenon?}

\correspondingauthor{Elias Aydi; NHFP Hubble Fellow}
\email{aydielia@msu.edu}
\author[0000-0001-8525-3442]{E.~Aydi}
\affiliation{Center for Data Intensive and Time Domain Astronomy, Department of Physics and Astronomy, Michigan State University, East Lansing, MI 48824, USA \\}

\author[0000-0001-5991-6863]{K.~V.~Sokolovsky}
\affiliation{Center for Data Intensive and Time Domain Astronomy, Department of Physics and Astronomy, Michigan State University, East Lansing, MI 48824, USA \\}
\affiliation{Sternberg Astronomical Institute, Moscow State University, Universitetskii~pr.~13, 119992~Moscow, Russia\\}

\author{J.~S.~Bright}
\affiliation{Astrophysics, Department of Physics, University of Oxford, Keble Road, Oxford, OX1 3RH, UK \\}

\author[0000-0002-4039-6703]{E.~Tremou}
\affiliation{National Radio Astronomy Observatory, P.O. Box O, Socorro, NM 87801, USA \\}

\author{M.~M.~Nyamai}
\affiliation{Department of Astronomy, University of Cape Town, Private Bag X3, Rondebosch 7701, South Africa \\}

\author[0000-0002-3142-8953]{A.~Evans}
\affiliation{Astrophysics Group, Keele University, Keele, Staffordshire, ST5 5BG, UK \\}

\author{J.~Strader}
\affiliation{Center for Data Intensive and Time Domain Astronomy, Department of Physics and Astronomy, Michigan State University, East Lansing, MI 48824, USA \\}

\author{L.~Chomiuk}
\affiliation{Center for Data Intensive and Time Domain Astronomy, Department of Physics and Astronomy, Michigan State University, East Lansing, MI 48824, USA \\}

\author{G.~Myers}
\affiliation{American Association of Variable Star Observers, 5 Inverness Way, Hillsborough, CA 94010\\}

\author{F-J.~Hambsch}
\affiliation{Vereniging Voor Sterrenkunde (VVS), Oostmeers 122 C, 8000 Brugge, Belgium \\}
\affiliation{Groupe Europ\'{e}en d'Observations Stellaires (GEOS), 23 Parc de Levesville, 28300 Bailleau l'Ev\^{e}que, France \\}
\affiliation{Bundesdeutsche Arbeitsgemeinschaft f\"{u}r Ver\"{a}nderliche Sterne (BAV), Munsterdamm 90, 12169 Berlin, Germany \\}

\author[0000-0001-5624-2613]{K.~L.~Page}
\affiliation{School of Physics \& Astronomy, University of Leicester, LE1 7RH, UK \\}

\author{D.~A.~H.~Buckley}
\affiliation{South African Astronomical Observatory, P.O.\ Box 9, 7935 Observatory, South Africa \\}
\affiliation{Department of Astronomy, University of Cape Town, Private Bag X3, Rondebosch 7701, South Africa \\}

\author{C.~E.~Woodward}
\affiliation{Minnesota Institute for Astrophysics, School of Physics \& Astronomy, 116 Church Street SE, University of Minnesota, Minneapolis, MN 55455, USA \\}

\author[0000-0001-7796-1756]{F.~M.~Walter}
\affiliation{Department of Physics \& Astronomy, Stony Brook University, Stony Brook, NY 11794-2100 USA}

\author[0000-0001-7016-1692]{P. Mr\'oz}
\affiliation{Division of Physics, Mathematics, and Astronomy, California Institute of Technology, Pasadena, CA 91125, USA}
\affiliation{Astronomical Observatory, University of Warsaw, Al. Ujazdowskie 4, 00-478 Warszawa, Poland \\}

\author{P.~J.~Vallely}
\affiliation{Department of Astronomy, The Ohio State University, 140 West 18th Avenue, Columbus, OH 43210, USA \\}

\author{T.~R.~Geballe}
\affiliation{Gemini Observatory/NSF's NOIRLab, 670 N. A\'ohoku Place Hilo, Hawaii, 96720, USA \\}

\author{D.~P.~K.~Banerjee}
\affiliation{Physical Research Laboratory, Navrangpura, Ahmedabad, Gujarat 380009, India \\}

\author{R.~D.~Gehrz}
\affiliation{Minnesota Institute for Astrophysics, School of Physics \& Astronomy, 116 Church Street SE, University of Minnesota, Minneapolis, MN 55455, USA \\}


\author{R.~P.~Fender}
\affiliation{Astrophysics, Department of Physics, University of Oxford, Keble Road, Oxford, OX1 3RH, UK \\}
\affiliation{Department of Astronomy, University of Cape Town, Private Bag X3, Rondebosch 7701, South Africa \\}

\author[0000-0002-1650-1518]{M.~Gromadzki}
\affiliation{Astronomical Observatory, University of Warsaw, Al. Ujazdowskie 4, 00-478 Warszawa, Poland \\}

\author{A.~Kawash}
\affiliation{Center for Data Intensive and Time Domain Astronomy, Department of Physics and Astronomy, Michigan State University, East Lansing, MI 48824, USA \\}

\author{C.~Knigge}
\affiliation{School of Physics and Astronomy, University of Southampton, Highfield, Southampton, SO17 1BJ, UK \\}

\author[0000-0002-8286-8094]{K.~Mukai}
\affiliation{CRESST II and X-ray Astrophysics Laboratory, NASA/GSFC, Greenbelt, MD 20771, USA \\}
\affiliation{Department of Physics, University of Maryland, Baltimore County, 1000 Hilltop Circle, Baltimore, MD 21250, USA \\}

\author{U.~Munari}
\affiliation{INAF Astronomical Observatory of Padova, 36012 Asiago (VI), Italy \\}

\author{M.~Orio}
\affiliation{INAF--Osservatorio di Padova, vicolo dell`Osservatorio 5, I-35122 Padova, Italy \\}
\affiliation{Department of Astronomy, University of Wisconsin, 475 N.\ Charter St., Madison, WI 53704, USA \\}

\author{V.~A.~R.~M.~Ribeiro}
\affiliation{Instituto de Telecomunica\c{c}\~{o}es, Campus Universit \'{a}rio de Santiago, 3810-193 Aveiro, Portugal \\}
\affiliation{Departamento de F\'{i}sica, Universidade de Aveiro, Campus Universit \'{a}rio de Santiago, 3810-193 Aveiro, Portugal \\}

\author{J.~L.\ Sokoloski}
\affiliation{Columbia Astrophysics Laboratory and Department of Physics, Columbia University, New York, NY 10027, USA\\}

\author{S.~Starrfield}
\affiliation{School of Earth and Space Exploration, Arizona State University Tempe, AZ 85287-1404, USA \\}

\author[0000-0001-5207-5619]{A.~Udalski}
\affiliation{Astronomical Observatory, University of Warsaw, Al. Ujazdowskie 4, 00-478 Warszawa, Poland \\}

\author[0000-0002-6896-1655]{P.~A.~Woudt}
\affiliation{Department of Astronomy, University of Cape Town, Private Bag X3, Rondebosch 7701, South Africa \\}



\begin{abstract}
We present a detailed study of the 2019 outburst of the cataclysmic variable V1047~Cen, which hosted a classical nova eruption in 2005.  The peculiar outburst occurred 14 years after the classical nova event and lasted for more than 400 days, reaching an amplitude of around 6 magnitudes in the optical. Early spectral follow-up revealed what could be a dwarf nova (accretion disk instability) outburst. However, the outburst duration, high velocity ($>$2000\,km\,s$^{-1}$) features in the optical line profiles, luminous optical emission, and presence of prominent long-lasting radio emission together suggest a phenomenon more exotic and energetic than a dwarf nova outburst. The outburst amplitude, radiated energy, and spectral evolution are also not consistent with a classical nova eruption. There are similarities between V1047~Cen's 2019 outburst and those of classical symbiotic stars, but pre-2005 images of the field of V1047~Cen indicate that the system likely hosts a dwarf companion, implying a typical cataclysmic variable system. Based on our multi-wavelength observations, we suggest that the outburst may have started with a brightening of the disk due to enhanced mass transfer or disk instability, possibly leading to enhanced nuclear shell burning on the white dwarf, which was already experiencing some level of quasi-steady shell burning. This eventually led to the generation of a wind and/or bipolar, collimated outflows. The 2019 outburst of V1047~Cen appears to be unique, and nothing similar has been observed in a typical cataclysmic variable system before, hinting at a potentially new astrophysical phenomenon.

\end{abstract}




\section{Introduction}
Cataclysmic Variables (CVs) are interacting binary systems, each consisting of a white dwarf accreting material from a
Roche-lobe-filling companion. The material flowing from the companion forms an accretion disk around the white dwarf before being dumped on its surface. 
In the case of a highly-magnetized white dwarf ($B > 10^{6}$ G), the magnetic field of the white dwarf truncates the inner regions of the disk or even completely prevents it from being formed. In this case the material follows the magnetic field lines onto the surface of the white dwarf. These are known as magnetic CVs (see \citealt{Warner_1995} for a review).
CVs experience several types of cataclysmic events and thermonuclear explosions, hence the name. One of these events is called a dwarf nova (DN) and is a viscosity-induced instability in the accretion disk, resulting in a temporary increase in mass transfer rate and heating the whole disk. DNe manifest as relatively low amplitude outbursts, typically $\Delta m \sim 2 - 5$ mag up to $\sim$ 9\,--\,10 mag in some extreme cases \citep{Kawash_etal_2021a}.

While material builds up on the surface of the white dwarf through secular accretion, the pressure and density increases in its surface layers. Once a critical density is reached, a thermonuclear runaway is triggered on the surface of the white dwarf, leading to an increase in the brightness of the system by up to 15\,mag or more in a matter of a few days \citep{Starrfield_etal72,Yaron_etal_2005}. These events are known as classical novae (see \citealt{Gallaher_etal_1978,Bode_etal_2008,Woudt_Ribeiro_2014,Della_Valle_Izzo_2020,Chomiuk_etal_2020} for reviews) and their recurrence timescale is typically $\gtrsim$ thousands of years \citep{Yaron_etal_2005}. In some cases, novae recur on shorter timescales -- short enough to be recorded more than once.
These are called recurrent novae and these systems are usually characterized by a high mass transfer rate, often due to the presence of an evolved secondary \citep{Schaefer_2010}.

After a classical nova event, the mass transfer rate is expected to be high enough to keep the disk in a hot, ionized state, temporarily preventing DNe in the system.
Theoretically, DNe are only expected hundreds/thousands of years later, when the mass transfer rate becomes low enough for the disk to cool and again become susceptible to disk instability events. This is implied by the hibernation scenario of CVs \citep{Shara_etal_1986}, which suggests that CVs go through cycles of low and high states of mass transfer rate. In this scenario, the states are mostly determined by the irradiation of the secondary by a nova event (which increases the mass transfer rate) and the growing separation between the two stars (which decreases the mass transfer rate). However, several CV systems have shown DN outbursts after classical nova eruptions;
e.g.,  GK~Per \citep{Sabbadin_Bianchini_1983,Bianchini_etal_1986,Zemko_etal_2017,ATel_11995}, V1017~Sgr \citep{Sekiguchi_1992,Salazar_etal_2017}, and V446~Her \citep{Honeycutt_etal_1995,Honeycutt_etal_2011} (see Table~\ref{table:cn_results} for a full list). 

\subsection{V1047~Cen -- the 2019 outburst}
\begin{figure*}
\begin{center}
  \includegraphics[width=0.49\textwidth]{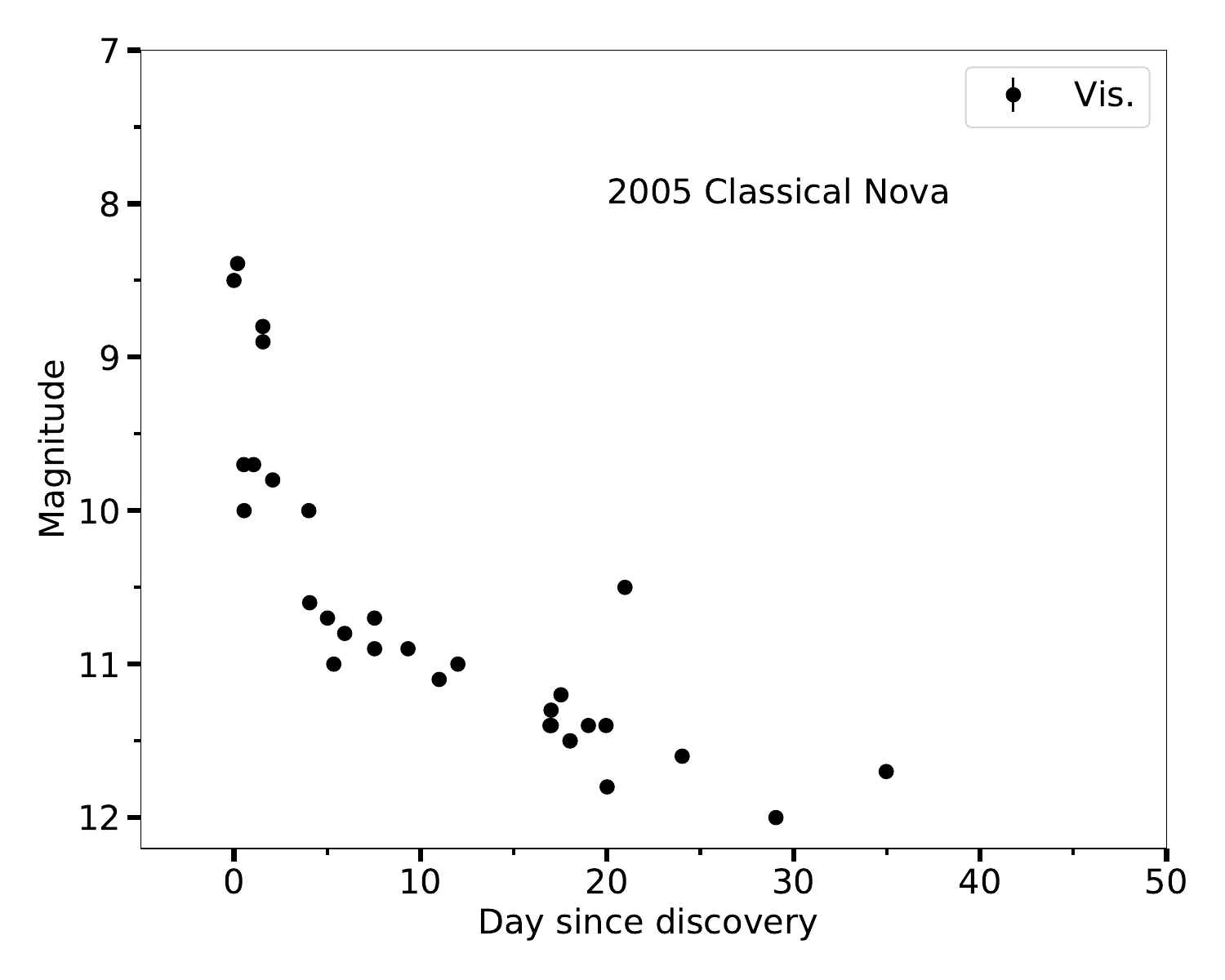}
  \includegraphics[width=0.49\textwidth]{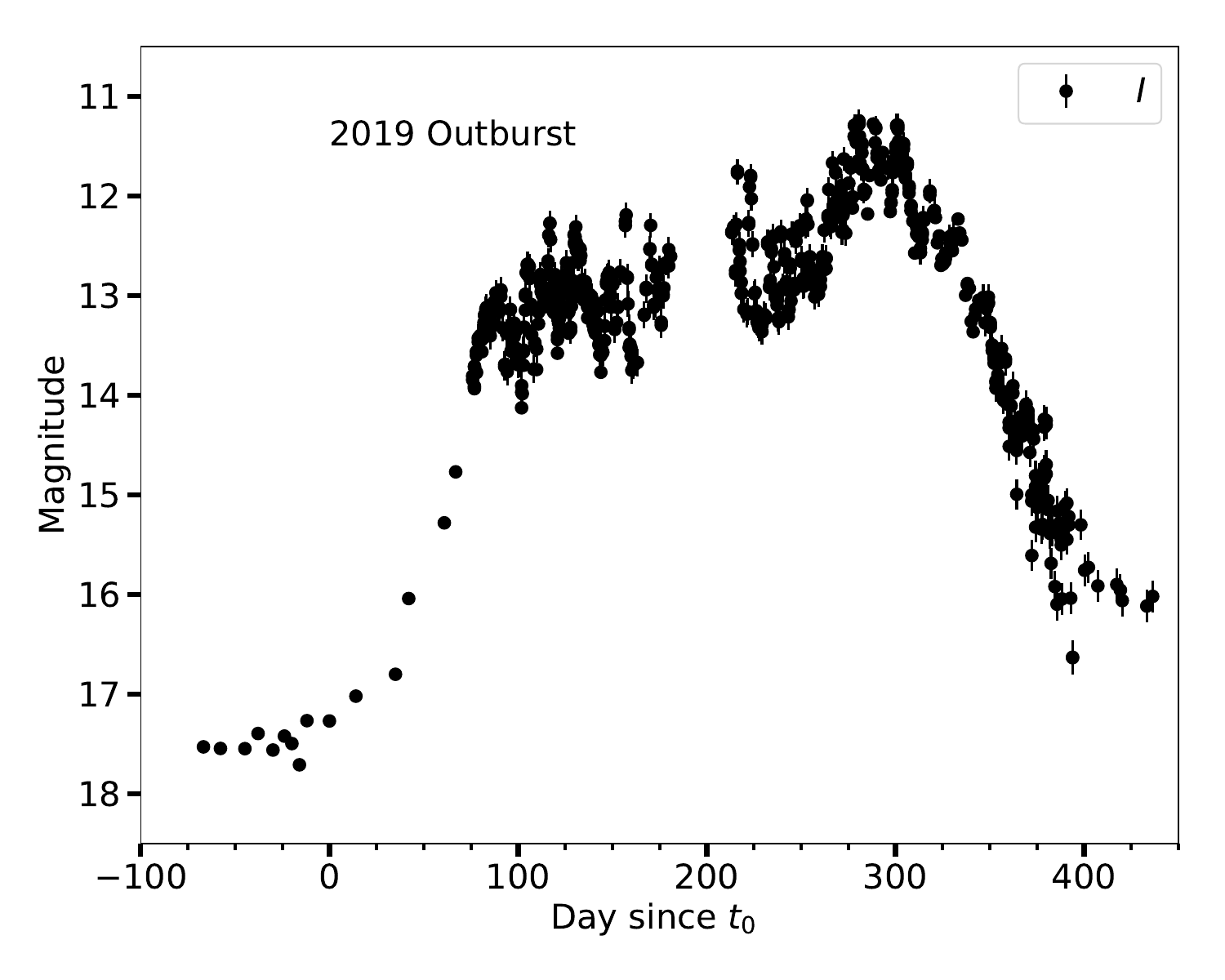}
\caption{A direct comparison between the visual light curve of the 2005 classical nova eruption from AAVSO (\textit{left}) and the OGLE $I$-band light curve of the 2019 outburst (\textit{right}). The remarkable difference between the two light curves indicates that the two events are of a completely different nature.}
\label{Fig:LC_comp}
\end{center}
\end{figure*}

V1047~Cen (Nova Cen 2005) was discovered as a Galactic transient on 2005 September 1.03 at a visual magnitude of around 8.5 (left panel of Figure~\ref{Fig:LC_comp}) and later classified spectroscopically as a classical nova eruption \citep{Liller_2005}. 
\citet{Walter_etal_2012} reported spectroscopic follow up taken approximately 5 and 7 days after discovery. The spectra showed
typical lines of Balmer, \eal{Fe}{II}, and \feal{O}{I}, which are characteristic of a classical nova near optical peak.
The Balmer lines showed multiple P Cygni absorption components with blueshifted velocities of around 750 and 1800\,km\,s$^{-1}$.

Otherwise, V1047 Cen was not extensively observed during the 2005 eruption, and
little else is known about the classical nova eruption. 
Archival observations of the system, taken by the Inner Galactic Disk with MIPS (MIPSGAL; \citealt{Carey_etal_2009}) Survey in the 24 and 70 micron bands, yield a 24-micron average magnitude of 0.72 $\pm$ 0.02\,mag \citep{Gutermuth_etal_2015}. 
The MIPSGAL survey data were collected between September 2005 and October 2006. The system was also observed by the Wide-field Infrared Survey Explorer (WISE) on February 2010 with IR magnitudes $w1$ = 11.2, $w2$ = 9.5, $w3$ = 4.2, and $w4$ = 2.2. The high IR brightness early after the outburst could indicate a dust formation event, which are common among classical novae (see, e.g., \citealt{Strope_etal_2010}), as well as, extinction along the line of sight. Additional archival observations by the VISTA Variables in the Via Lactea (VVV) survey between March 2010 and August 2011, show the nova fading from $K$=12.84 to $K$=13.11 \citep{Saito_etal_2012}.
The Neil Gehrels Swift Observatory (hereafter \textit{Swift}; \citealt{Gehrels_etal_2004}) observed the classical nova eruption between 2005 and 2008 \citep{Ness_etal_2007}. The observations obtained in November 2005 and January 2006 resulted in detections of hard X-rays from strongly absorbed shock emission, which are typically seen in novae during the early weeks/months of the eruption (see, e.g., \citealt{Mukai_etal_2008,Schwarz_etal_2011,Gordon_etal_2021}). The \textit{Swift} observations obtained in 2008 led to a non-detection in X-rays. 

Fourteen years after the 2005 nova eruption, \citet{2019TNSTR.985....1D} reported the discovery of an astronomical transient AT2019hik/Gaia19cfn possibly associated with V1047~Cen on 2019 June 11.6, with a discovery magnitude of 16.2 in the $G$-band. Based on regular monitoring by the Optical Gravitational Lensing Experiment (OGLE; \citealt{Udalski_etal_2015}) survey, \citet{ATel_12876} confirmed that the FK5 J2000 equatorial coordinates of the transient ($[\alpha, \delta]$ = [13$^{\mathrm{h}}$20$^{\mathrm{m}}$49$^{\mathrm{s}}$\!.78, --62$^{\circ}$37$'$50\arcsec\!\!.6]) were consistent with those of V1047~Cen. They found that the re-brightening of the system started as early as 2019 April 6.11 (HJD 2458579.61; right panel of Figure~\ref{Fig:LC_comp}). This date is considered as the outburst start ($t_0$) throughout this paper. \citet{ATel_12876} noted that the slow re-brightening of the system was inconsistent with a recurrent nova eruption. The re-brightening of the system triggered follow-up observations across the electromagnetic spectrum. \citet{ATel_12885} reported optical spectroscopy, which showed DN outburst spectral features superimposed on the spectral features of a classical nova nebula. \citet{Geballe_etal_2019} reported infrared spectroscopy of the then ongoing outburst, during its first 160 days, concluding that the event was possibly a DN outburst. However, upon further follow-up, the 2019 outburst of V1047~Cen seems to be a more exotic phenomenon. 

Here we report on multi-wavelength follow-up of the 2019 outburst of V1047~Cen spanning the spectrum from X-ray to radio. In Section~\ref{Obs} we present the observations and data reduction. In Section~\ref{Res} we show our results, while in Section~\ref{Disc} we offer discussion about the nature of the event and its peculiar observational features. Our conclusions are given in Section~\ref{Conc}.

\section{Observations and data reduction}
\label{Obs}
\subsection{Optical and near-IR photometry}
V1047~Cen has been observed by the OGLE sky survey \citep{Udalski_etal_2015} since May 2013, several years before the outburst, as part of the OGLE Galaxy Variability Survey (GVS). All data were taken in the $I$ band with an exposure time of 30~s, and they were reduced and calibrated using the standard OGLE pipeline \citep{Udalski_etal_2015}. A sample of the observations is listed in Table~\ref{table:OGLE_log}.

Optical photometry between days 76 and 437 was performed in the $BVRI$ bands by several observers from the American Association of Variable Star Observers (AAVSO; \citealt{Kafka_2020}). The bulk of the multi-band photometry comes from two observers, Gordon Meyer and Josch Hambsch. A sample of the observations is listed in Table~\ref{table:AAVSO_log}.  

We obtained SMARTS {\it Andicam} photometry in the $J, H$, and $K_s$
bands on 29 nights between days 80 and 117.
Data reduction is described in \cite{Walter_etal_2012}. A sample of the observations is listed in Table~\ref{table:SMARTS_log}.
The {\it Andicam} instrument was retired on 2019 August~1 (day 117). We also make use of IR photometry from the enhanced Wide-field Infrared Survey Explorer (NEOWISE; \citealt{Mainzer_etal_2011}) covering only three epochs during the outburst in the $W1$ (3.35\,$\mu$m) and $W2$ (4.60\,$\mu$m) bands. A log of the observations is listed in Table~\ref{table:NEOWISE_log}.

The field of V1047~Cen was observed by the Transiting Exoplanet Survey Satellite (\textit{TESS}; \citealt{Ricker_etal_2015}) during sectors 11, which covers the early rise of the 2019 outburst between days 17 and 44, and sector 38 covering the post-outburst period between days 753 and 781. We used the open-source tool ELEANOR \citep{Feinstein_etal_2019} to extract light curves from the \textit{TESS} full-frame images,
opting to utilize the corrected flux light curve and to only include data that are not associated
with a quality flag. 

All the data will be available as online material.
All multi-band photometry is calibrated using the
\textit{Vega} system zero points.

\begin{figure*}
\begin{center}
  \includegraphics[width=\textwidth]{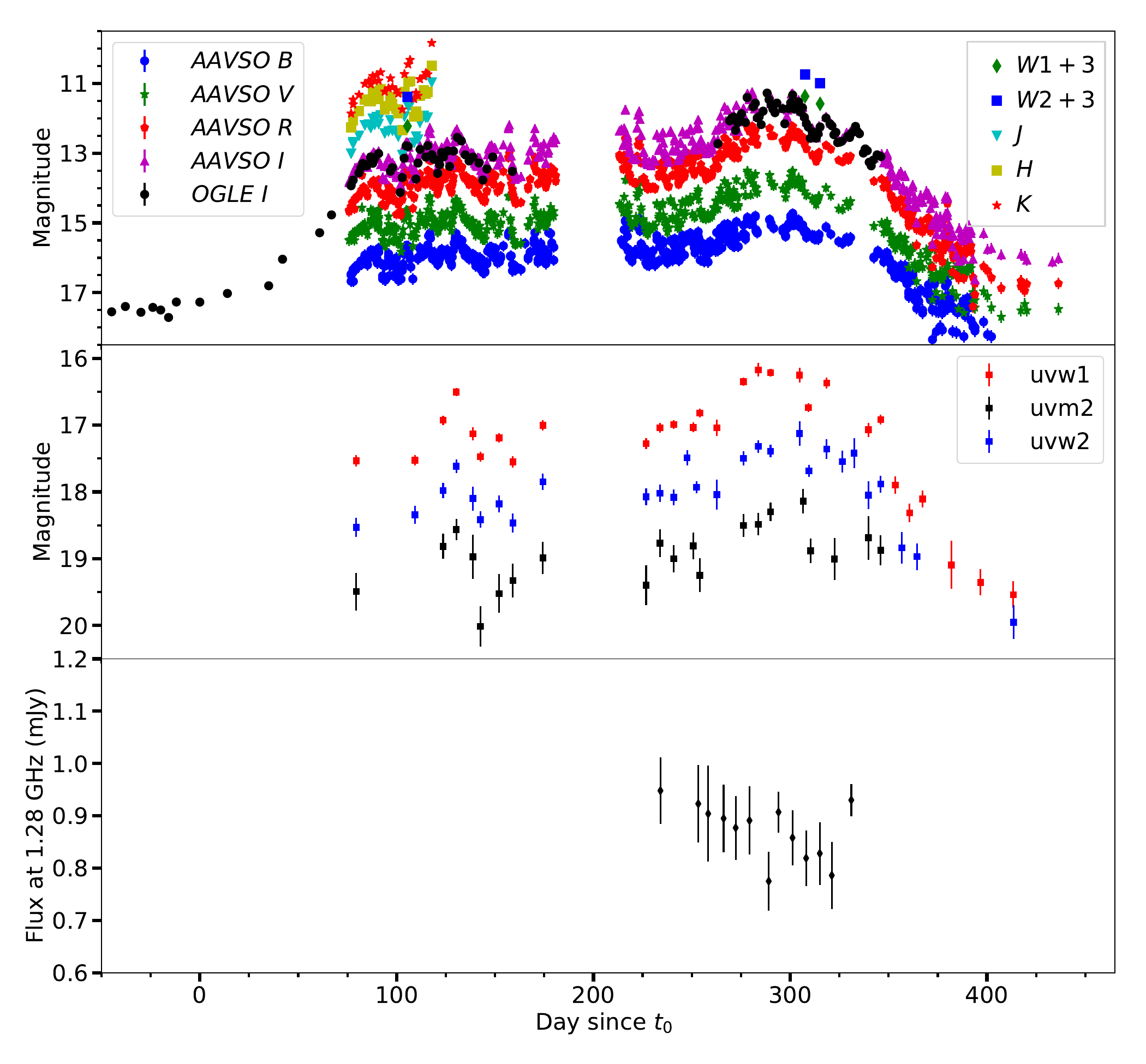}
\caption{Multi-band light curves of the 2019 outburst of V1047~Cen. The \textit{top panel} shows optical and IR photometry from AAVSO ($BVRI$), OGLE ($I$), SMARTS ($JHK$), and NEOWISE ($W1$ and $W2$) photometry. The \textit{middle panel} shows \textit{Swift} UVOT UV data, and the \textit{bottom panel} shows MeerKAT L-band radio flux at 1.28 GHz. Colors and symbols are coded as indicated in figure legends.}
\label{Fig:optical_UV}
\end{center}
\end{figure*}

\subsection{Optical and IR spectroscopy}

We obtained optical and IR spectroscopic observations of the 2019 outburst between days 74 and 643 using a diverse group of telescopes and instruments. A log of the spectral observations is presented in Table~\ref{table:spec_log}. 

We obtained spectra using the High Resolution Spectrograph (HRS; \citealt{Barnes_etal_2008,Bramall_etal_2010,Bramall_etal_2012,Crause_etal_2014}) and the Robert Stobie Spectrograph (RSS; \citealt{Burgh_etal_2003}; \citealt{Kobulnicky_etal_2003}) mounted on the Southern African Large Telescope (SALT; \citealt{Buckley_etal_2006,Odonoghue_etal_2006}) in Sutherland, South Africa. HRS was used in the low resolution LR mode, yielding a resolving power $R = \lambda/\Delta\lambda  \approx$ 14,000 over the range 4000\,--\,9000\,\AA. The primary reduction of the HRS spectroscopy was conducted using the SALT science pipeline \citep{Crawford_etal_2010}, which includes over-scan correction, bias subtraction, and gain correction. The rest of the reduction was done using the MIDAS FEROS \citep{Stahl_etal_1999} and $echelle$ \citep{Ballester_1992} packages. The reduction procedure is described by \citet{kniazev_etal_2016}. RSS was used in long-slit mode with the 1.5 arcsec slit and the PG900 grating, resulting in a resolving power $R \approx 1500$. The spectra were first reduced using the  PySALT pipeline \citep{Crawford_etal_2010}, which involves bias subtraction, cross-talk correction, scattered light removal, bad pixel masking, and flat-fielding. The wavelength calibration, background subtraction, and spectral extraction were done using the Image Reduction and Analysis Facility (IRAF; \citealt{Tody_1986}).

We also carried out low- and medium-resolution optical spectroscopy using the Goodman spectrograph \citep{Clemens_etal_2004} on the 4.1\,m Southern Astrophysical Research (SOAR) telescope located on Cerro Pach\'on, Chile. The observations were carried out in two setups: one setup using the 400~l\,mm$^{-1}$ grating and a 0.95\arcsec\ slit, yielding a resolving power $R \approx$ 1000 over the wavelength range 3820--7850\,\AA. Another setup was used with a 2100~l\,mm$^{-1}$ grating and a 0.95\arcsec\ slit, yielding a resolving power $R \approx$ 5000 in a region centered on H$\alpha$ that is 570\,\AA\ wide. The spectra were reduced and optimally extracted using the \textsc{apall} package in IRAF. 

Four high-resolution spectra were also obtained using the Chiron fiber-fed echelle spectrograph \citep{Tokovinin_etal_2013} mounted on the CTIO 1.5m telescope.
Integration times were 20 minutes, with three integrations per night summed for 1 hour net exposure time. All spectra were taken in ``fiber mode'', with
4x4 on-chip binning yielding a resolving power $R \approx$27,800. The data were reduced using a pipeline coded in IDL (Walter 2017)\footnote{\url{http://www.astro.sunysb.edu/fwalter/SMARTS/CHIRON/ch_reduce.pdf}}. 

\subsection{Infrared spectroscopy}

Near-infrared spectra of V1047 Cen were obtained at the Gemini South Telescope on Cerro
Pachon in Chile on days 223 and 230 using the facility instrument FLAMINGOS-2, for program GS-2020A-Q-201. The
observations are summarized in Table~\ref{table:IR_spec},. The 0.36'' wide slit and R3000 grism were employed to
obtain spectra at resolving powers, $R$, of up to 3000 in portions of the $J, H,$ and $K$ windows. The JH R1200 grism was used with the same slit to obtain a spectrum covering 0.89-1.75 $\mu$m. Note
that for each grism there is considerable variation in $R$ across each wavelength interval. (see Ref.\footnote{\url{http://www.gemini.edu/instrumentation/flamingos-2/components\#Grisms}}). Data reduction
employed standard near-infrared techniques utilizing both Gemini/IRAF and Figaro commands \citep{Shortridge_etal_1992}. Flux calibrations, derived from spectra of the standard stars listed in
Table~\ref{table:IR_spec} and are accurate to $\pm$30\%.

NASA SOFIA \citep[Stratospheric Observatory for Infrared Astronomy;][]{Young_etal_2012} airborne observations of V1047~Cen were obtained with the the Faint Object InfraRed  CAmera \cite[FORCAST;][]{Herter_etal_2018}, the dual-channel mid-infrared imager and grism spectrometer operating from 5 to 40~\micron{} on two separate, consecutive flight series on 2019 July 02.542 UT 
(F0589) 03.559 UT 
(F0590), days 88 and 89, using the G111 and G227 grisms with the instrument configured using a long-slit (4\farcs7 $\times$ 191\arcsec), yielding a spectral resolving power $R\sim$ 140--300. the G111 grating was used to observe V1047~Cen both nights. Standard pipeline processed and flux calibrated data \citep[for details of the reduction processes see][]{Clarke_etal_2015} were retrieved for science analysis from the Infrared Processing and Analysis Center (IPAC) Infrared Science Archives (IRSA). The data products contain a computed atmospheric transmission model appropriate for the flight altitudes which were used to mask-out spectral points in the observed spectral energy distributions where the atmospheric transmission was $\lesssim$ 70\%. Within the statistical errors, no variations between the G111 data sets obtained on the two different nights were noted, hence these data were averaged into a single resultant spectrum.

\subsection{MeerKAT observations}
We observed the field of V1047 Cen with the MeerKAT radio interferometer \citep{2016mks..confE...1J} 15 times between days 277 and 700.
Observations were taken with the 64 antenna array at a central frequency of $1.28\,\rm{GHz}$ with a $856\,\rm{MHz}$ bandwidth. Each observation consisted of 15 minutes integration on the field of V1047 Cen with two minutes on the phase calibrator J1424--4913 before and after. J1939--6342 was used to set the flux scale and calibrate the bandpass response of the instrument. Data were reduced using the OxKAT (see \citealt{Heywood_2020oxkat} for details) reduction scripts, which include recipes for both phase reference and amplitude and phase self-calibration. The typical noise in a region of our images without obvious emission is 35\,$\mu\rm{Jy}$/beam. A log of the MeerKAT observations is given in Table~\ref{table:radio_log}

\subsection{\textit{Swift} observations}
Observations of V1047 Cen with \textit{Swift} commenced on 2019 June 24, 79 days after the re-brightening start. The UV/Optical
Telescope (UVOT; \citealt{Roming_etal_2005}) detected emission across all three UV filters ($uvw1$: central wavelength
of 2600 \AA; $uvm2$: 2246 \AA; $uvw2$: 1928 \AA; a sample of the observations is listed in Table~\ref{table:Swift_log}). No X-ray emission was
detected with the X-ray Telescope (XRT; \citealt{Burrows_etal_2005}), however. A second observation was performed a month later (day 109), followed by approximately weekly observations from day 123 until day 420 after the start of the outburst. Throughout this time, a variable UV source was detected, while no individual observations
showed significant X-ray emission. Co-adding the full $\sim$50 ks of XRT data, a faint X-ray source with a count rate of 2.8$^{+1.3}_{-1.1}$~$\times$~10$^{-4}$ count~s$^{-1}$ was detected.
However, with only 16 counts in the source extraction region, no spectral
analysis can be sensibly performed.

\section{Results}
\label{Res}
\subsection{Optical/UV light curves}

The AAVSO ($BVRI$), OGLE $I$, and \textit{Swift}  UV light curves of V1047 Cen are presented in Figure~\ref{Fig:optical_UV}. The data 
reveal peculiar behavior,
with the outburst lasting for around 400 days. There is a general trend of increased brightness between days 0 and $\sim$ 310, before the brightness of the system starts decreasing. Throughout the 400 day outburst, the light curves show variability on timescales of a few days and with amplitude of $\lesssim$ 1\,magnitude. 
Based on the OGLE data the amplitude of the outburst reached 6.2 magnitudes (Figure~\ref{Fig:LC_comp}).

\begin{figure}
\begin{center}
  \includegraphics[width=\columnwidth]{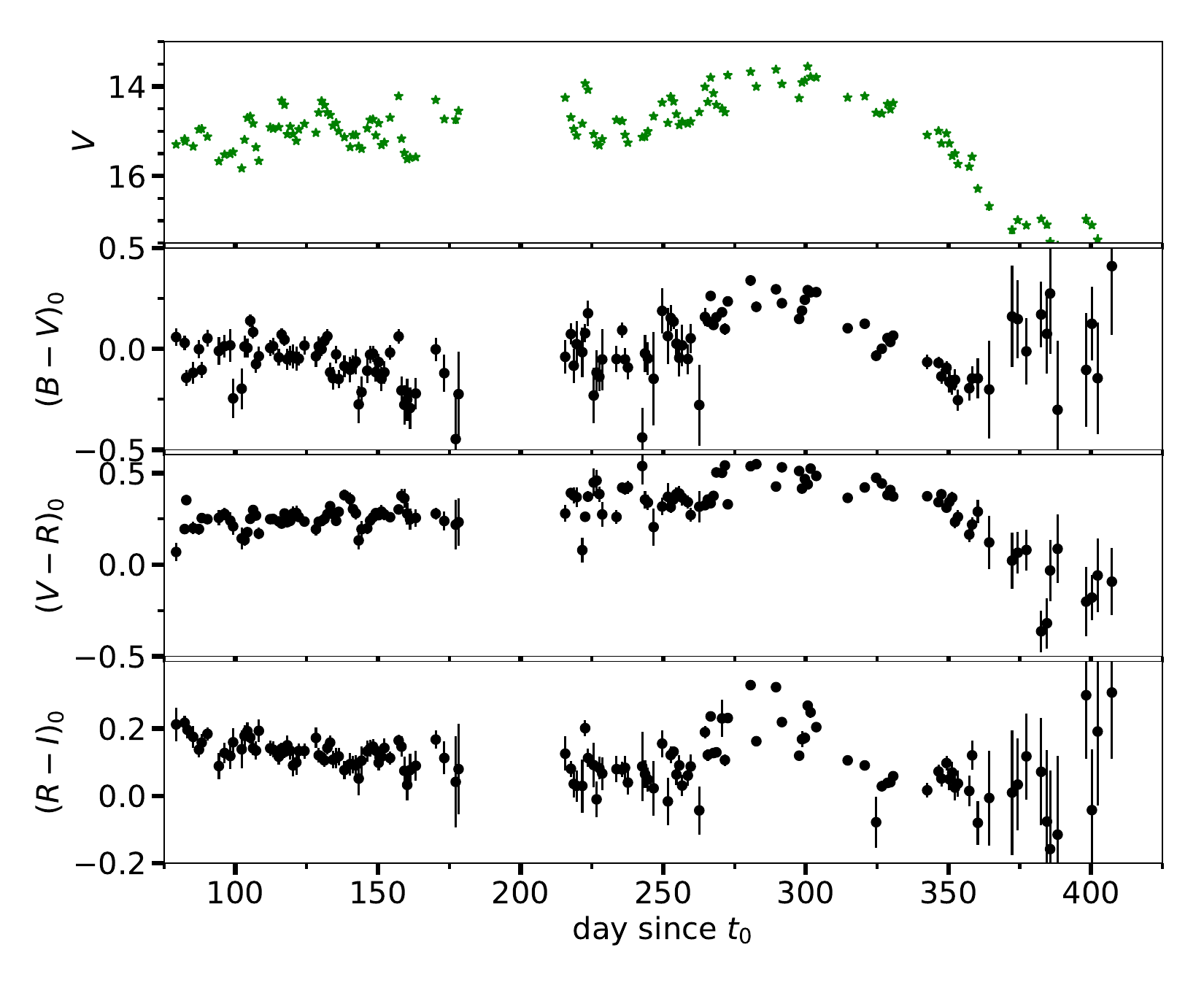}
\caption{The evolution of the extinction-corrected broadband colors $(B-V)_0$, $(V-R)_0$, and $(R-I)_0$, throughout the outburst.}
\label{Fig:color_evolution}
\end{center}
\end{figure}

The evolution of the optical broadband colors $(B-V)_0$, $(V-R)_0$, and $(R-I)_0$ is presented in Figure~\ref{Fig:color_evolution}. We applied a reddening correction based on the $E(B-V)$ extinction derived in Section~\ref{red_dist}. The colors show random fluctuations around a mean value throughout the majority of the outburst. However, on day 250, the three colors show a noticeable redward trend, coincident with a bump in the optical light curves. This is followed by a blueward trend after day 300, as the outburst ends.

\begin{figure}
\begin{center}
  \includegraphics[width=\columnwidth]{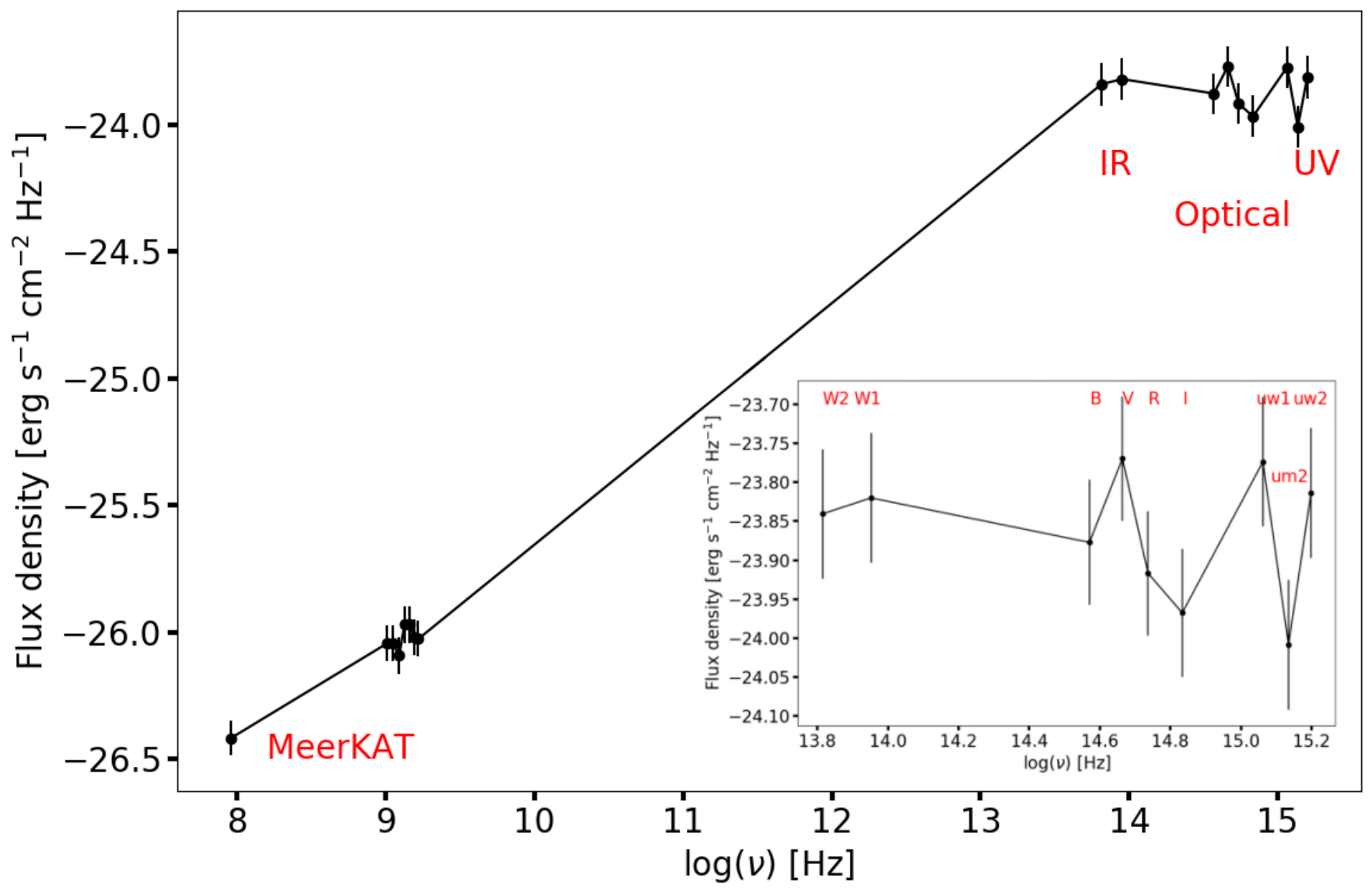}
\caption{Extinction-corrected SED plot, showing the UV to radio SED of V1047 Cen on day 308. The error bars are 1$\sigma$ uncertainties and they include contributions from the photometric and extinction uncertainties. A We also present a zoom-in on the IR/optical/UV data.} 
\label{Fig:SED_extended}
\end{center}
\end{figure}

We have near-IR photometry only for a short period near the start of the plateau.
Aside from a trend toward redder colors as the source brightened (between days 76 and 82, corresponding to late part of the rising phase), the
near-IR colors were fairly stable, at $I-K=2.25 \pm 0.03$,
$J-K=1.24 \pm 0.02$, and $H-K=0.51 \pm 0.03$. We use our data to create multi-wavelength SED of V1047~Cen during the 2019 outburst. Despite our extensive coverage of the outburst, we do not have simultaneous optical, IR, UV, and radio data except for one day (day $\approx$ 308), where we have simultaneous optical, UV, and NEOWISE NIR photometry, along with MeerKAT radio data (see Figure~\ref{Fig:SED_extended}). Since the brightness of the system shows variability, up to 1\,mag in a couple of days, we avoid any interpolation of the light curve to create SEDs on other days. We attempted blackbody fit to the multi-wavelength SED, but no blackbody model can result in a reasonable fit to the data and we end up with poor fitting even in the case of multiple blackbody components. This indicates complex emission, possibly from multiple sources (e.g, the nebula of the 2005 nova event, the white dwarf surface, the accretion disk, and ejected material/outflow during the 2019 outburst). Similarly, we attempted disk model fitting to the multi-wavelength SED, but no disk model with reasonable parameters (e.g., mass accretion rate and disk radius) could fit the data, particularly due to the high brightness of V1047~Cen during outburst relative to accreting white dwarf systems. Whether we include the MeerKAT data or not, the mass accretion rate resulting from the best fit disk model to the SED is of the order of $10^{-5}$\,M$_{\odot}$\,yr$^{-1}$, which is not physically reasonable. This shows that accretion alone is not enough to explain the emission during the outburst, again raising questions about its nature.

\begin{figure*}
\begin{center}
  \includegraphics[width=0.49\textwidth]{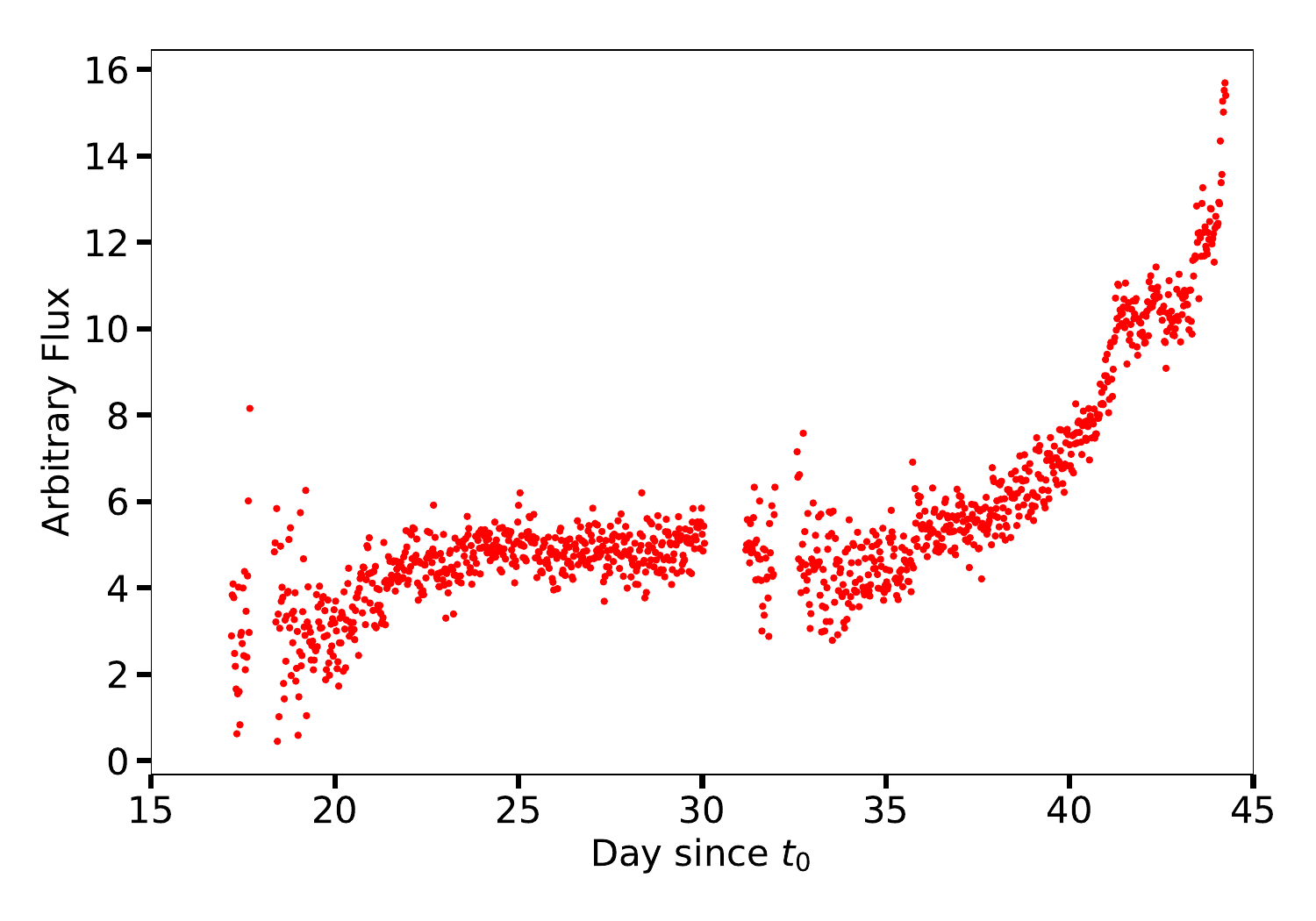}
  \includegraphics[width=0.49\textwidth]{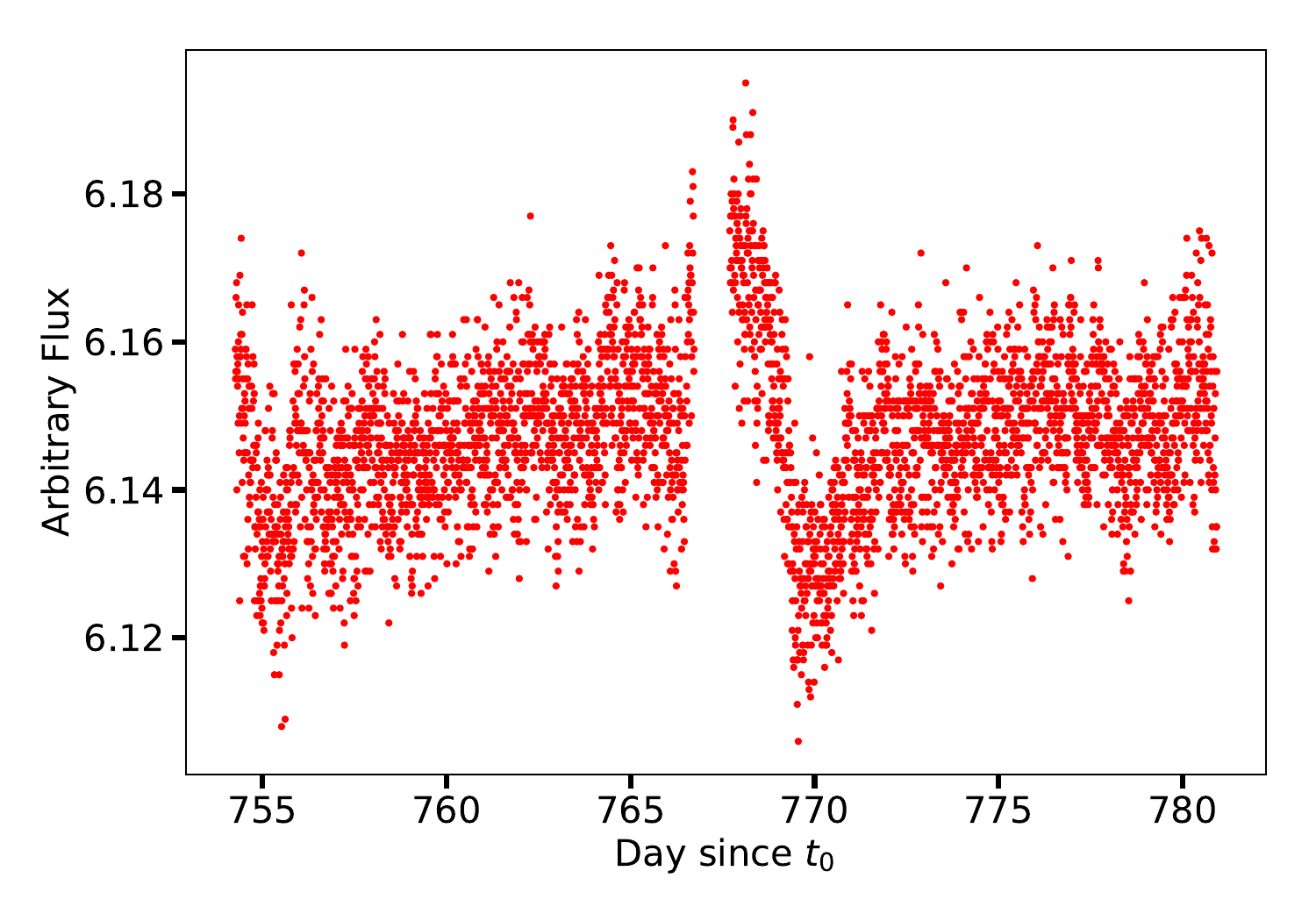}
\caption{The \textit{TESS} light curve of the 2019 outburst of V1047~Cen during sectors 11 (\textit{left}) and 28 (\textit{right}).}
\label{Fig:TESS_LC}
\end{center}
\end{figure*}

The sector 11 \textit{TESS} light curve of V1047~Cen, presented in Figure~\ref{Fig:TESS_LC} (left panel), shows the early rise of the outburst between days 17 and 44. \citet{ATel_12889} reported a period of 0.36 days in the \textit{TESS} light curve. However, after extracting the data using different techniques and performing period analysis on the different light curves, we failed to find the period in some of these light curves. This indicates that the 0.36 days period is possibly an artifact introduced by the extraction technique or it is caused by emission leaking into the aperture from background, unresolved sources. This period is also not present in the \textit{TESS} data from sector 38 (Figure~\ref{Fig:TESS_LC}, right panel), which covers the post-outburst period between days 753 and 781.

\subsection{Radio light curve and spectral indices}

\begin{figure*}
\begin{center}
  \includegraphics[width= 0.48\textwidth]{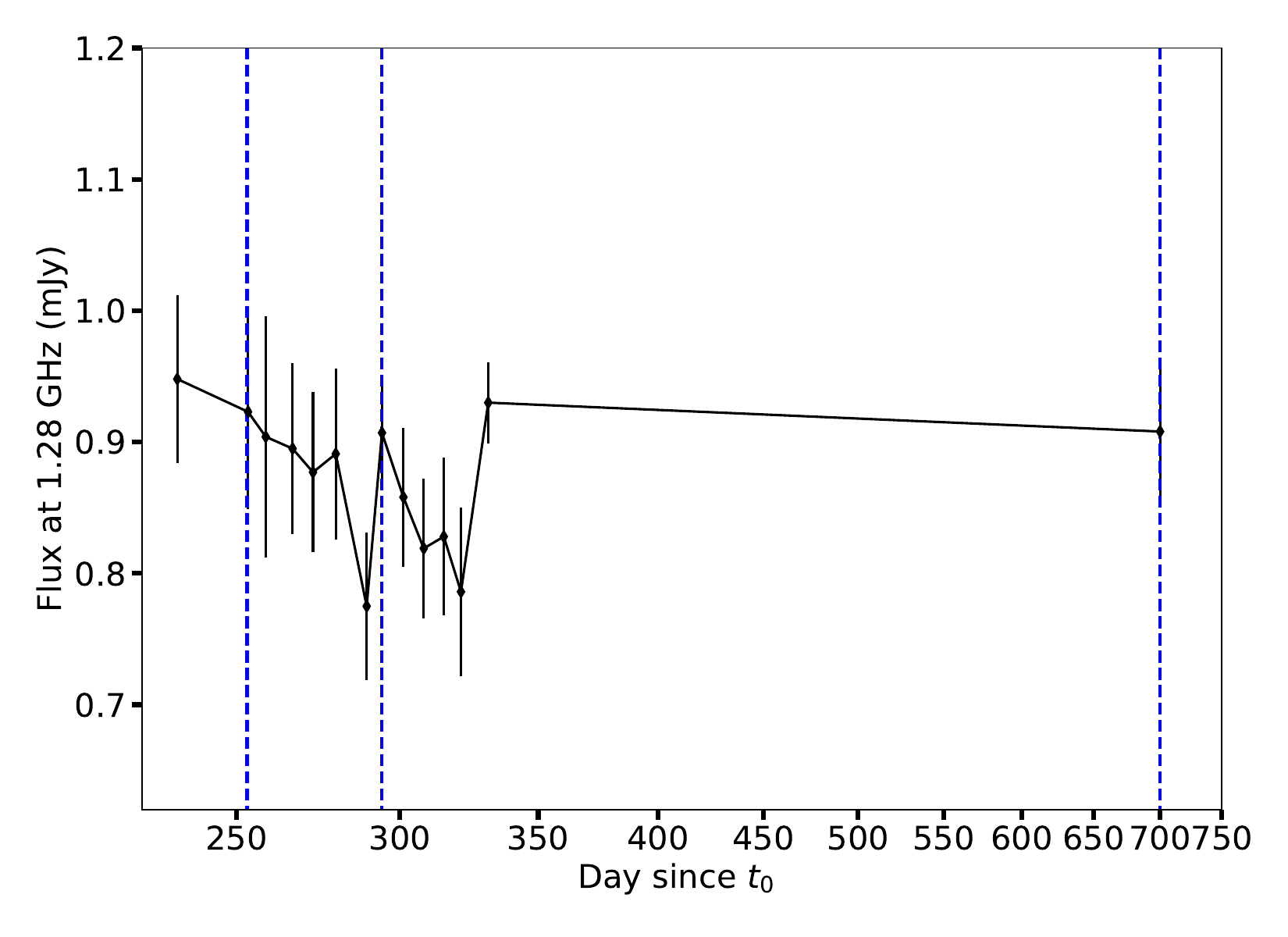}
  \includegraphics[width= 0.48\textwidth]{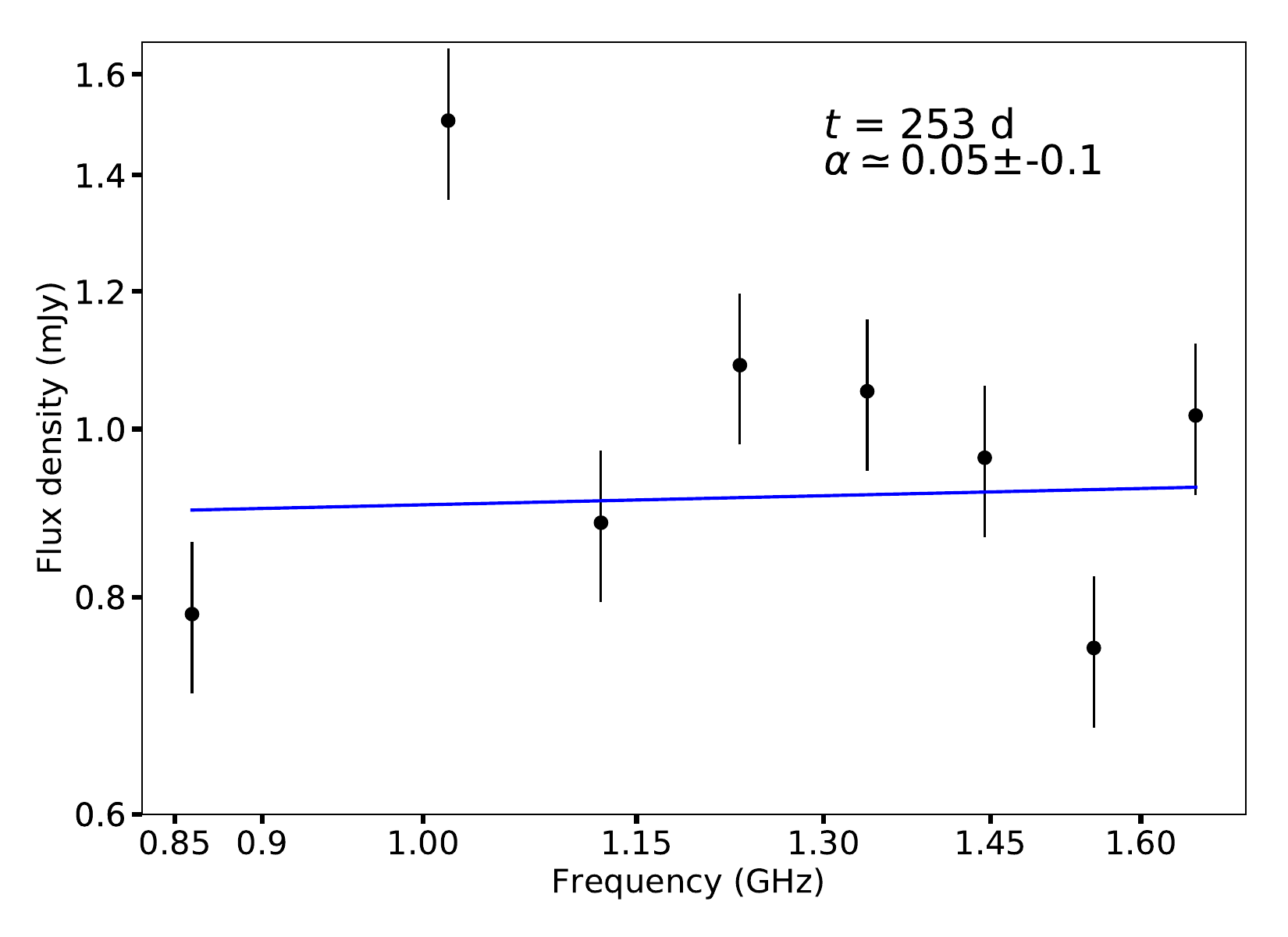}
  \includegraphics[width= 0.48\textwidth]{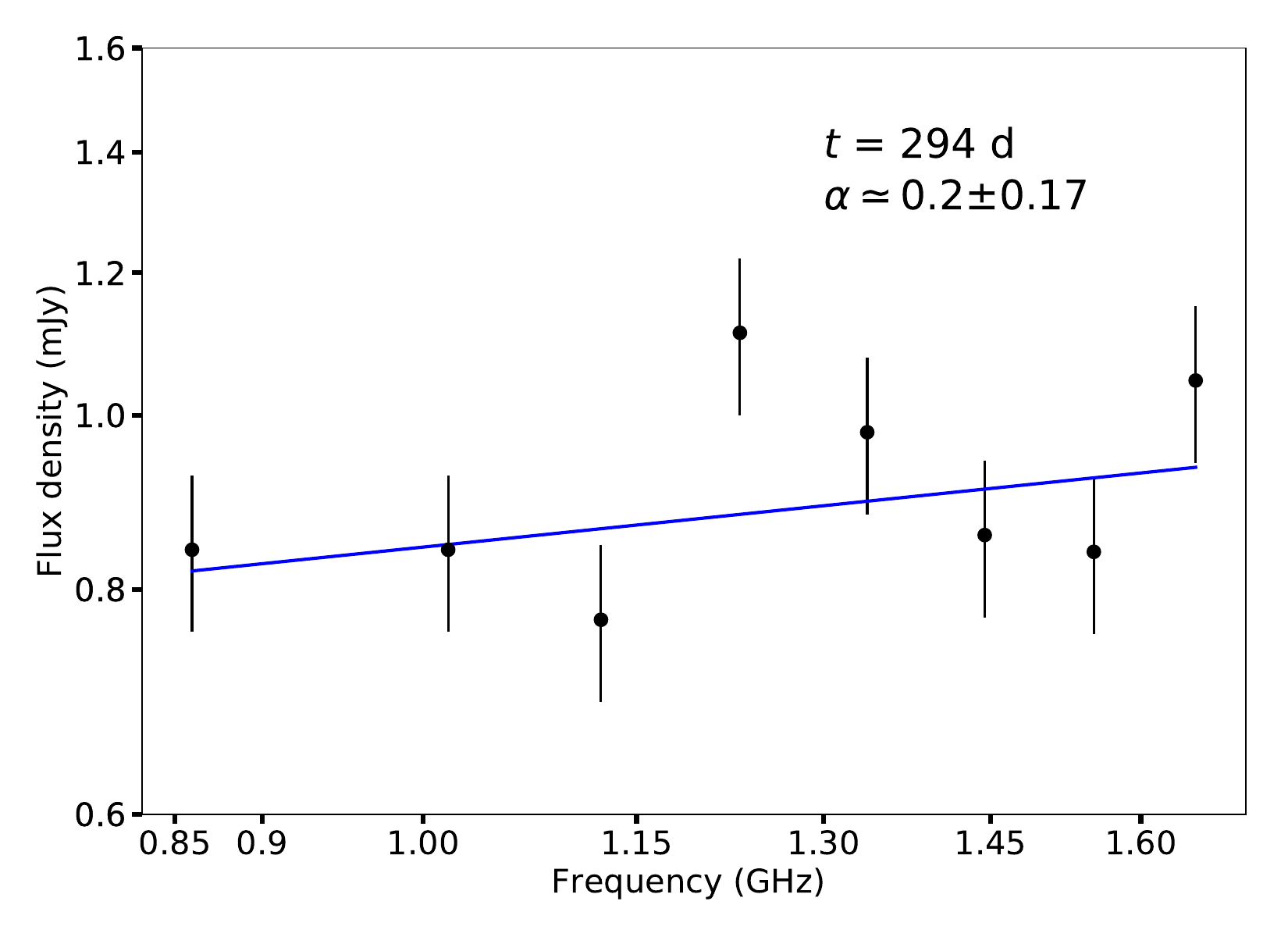}
  \includegraphics[width= 0.48\textwidth]{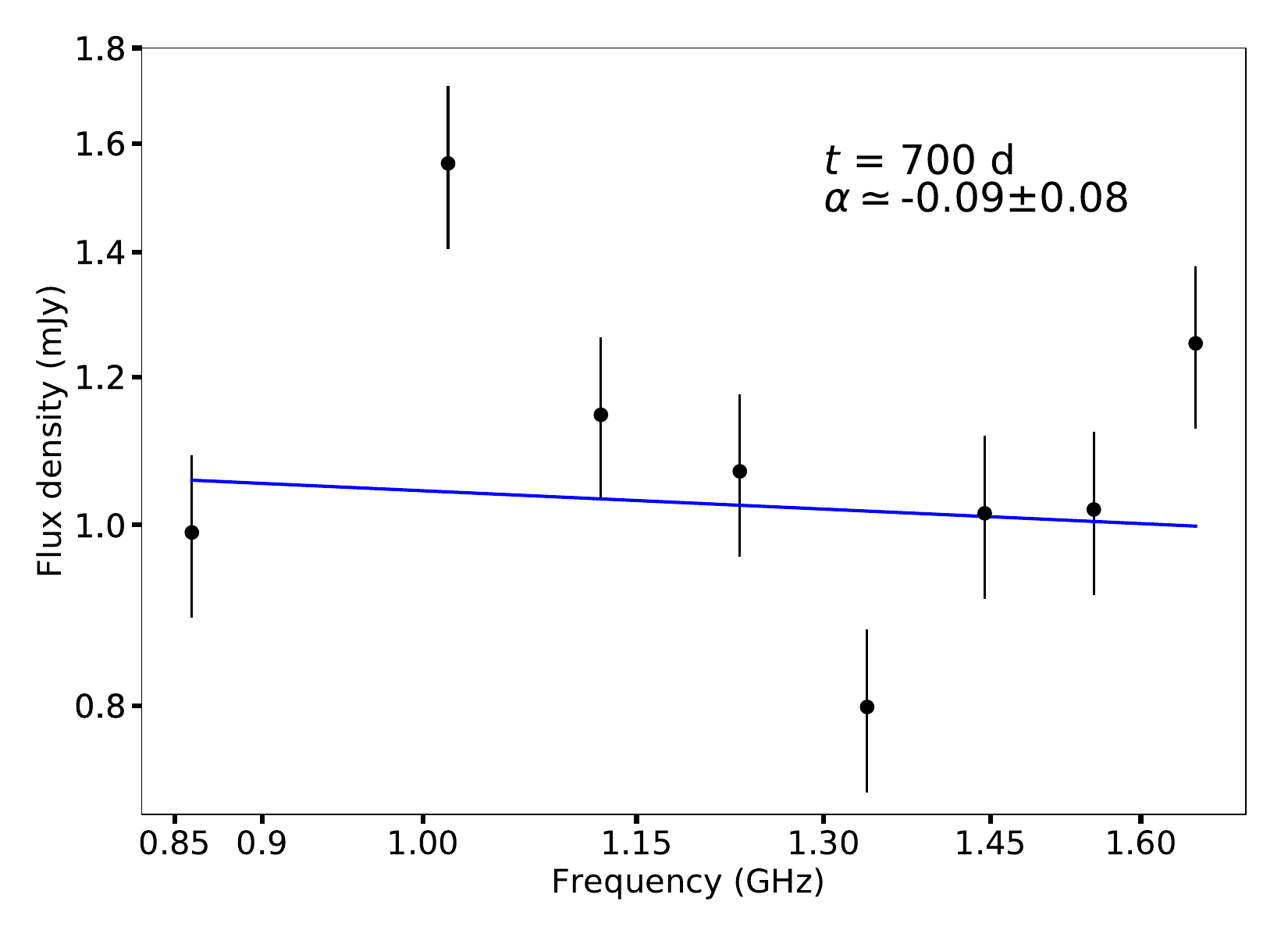}
\caption{The MeerKAT radio light curve measured at 1.28 GHz is plotted in the \textit{top left panel}, for comparison with three radio spectra measured on days 256, 294, and 700. Spectral indices are listed in each panel. The vertical dashed lines on the light curve plot signify the epochs when spectra are measured.} 
\label{Fig:spec_fitting}
\end{center}
\end{figure*}

The MeerKAT radio light curve is plotted in Figure~\ref{Fig:optical_UV}. The $L$-band (900 - 1670 MHz) flux shows variability between $\sim$ 0.7 and 1.0\,mJy during the optical outburst. There is no obvious correlation between the optical and radio emission. 
At day 700, ten months after the end of the optical outburst, the radio emission from the system is still bright at 0.91\,mJy (Figure~\ref{Fig:spec_fitting}).

We derived the spectral indices of the MeerKAT observations
on days 258, 301, and 421 (Figure~\ref{Fig:spec_fitting}). To do that, we divided the data into eight frequency intervals centered at 859 MHz, 1.016 GHz, 1.123 GHz, 1.230 GHz, 1.337 GHz, 1.444 GHz, 1.551 GHz, and 1.658 GHz, 
allowing us to measure the flux density of the source at each sub-band. Consequently we calculated the spectral index, $\alpha$, by fitting a 
single power law to the radio spectra that is represented as: 
 $ S \propto \nu^{\alpha}$. 
We note that the sub-band calibration for MeerKAT is still an active area of exploration, and hence the systematic uncertainties on the sub-band flux densities may yet be underestimated. While the spectrum is mostly flat, given the large uncertainties 
in the spectral indices, caution is required when interpreting these indices. Overall, it is hard to draw strong conclusions from these values, other than that the radio emission is not optically thick. We elaborate more on the origin of the radio emission in Section~\ref{Disc}.

\begin{figure*}
\begin{center}
  \includegraphics[width=0.9\textwidth]{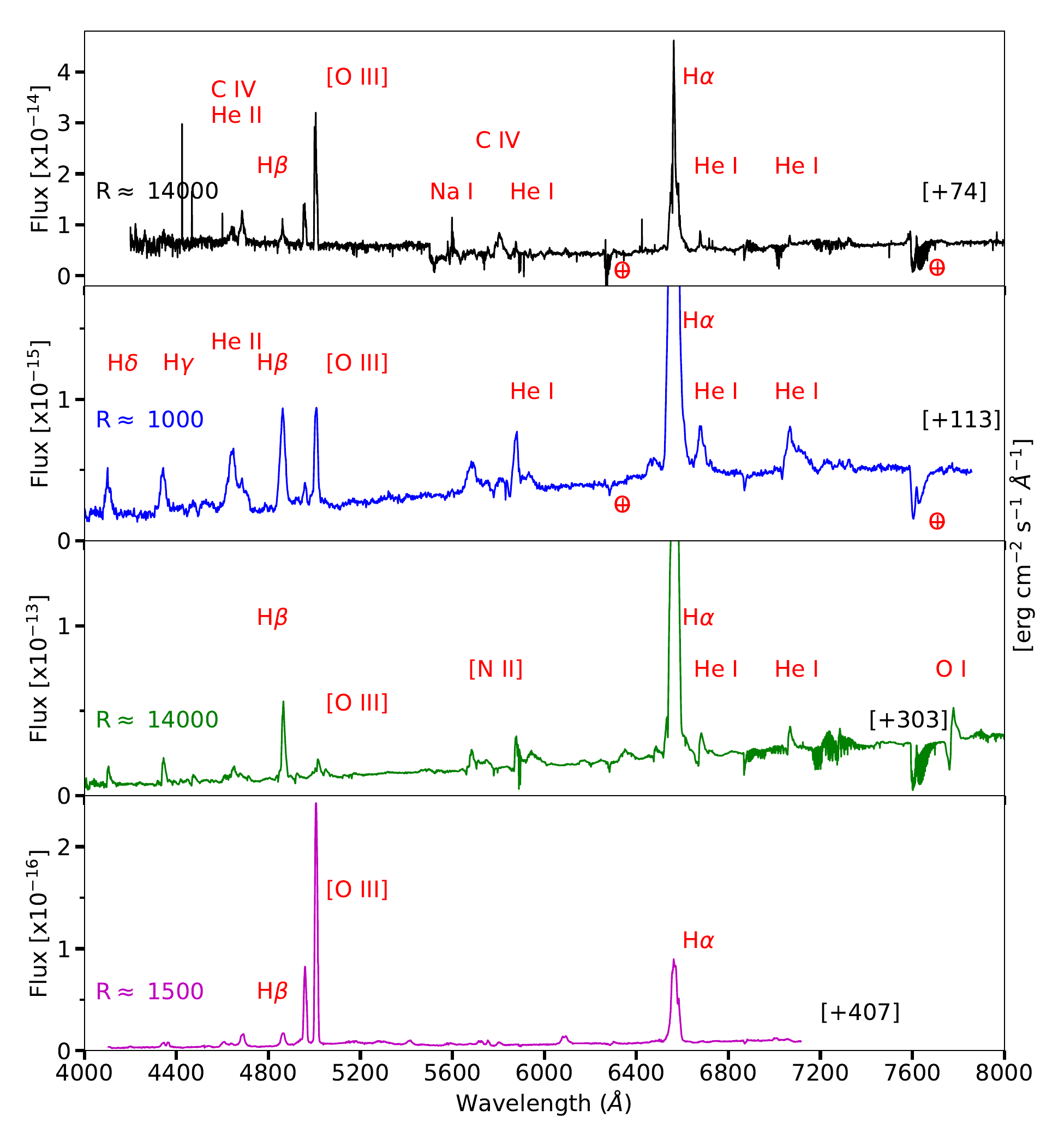}
\caption{The optical spectral evolution of V1047~Cen throughout different stages of its outburst. The numbers in brackets are days since $t_0$. We also quote the resolving power of the spectra on the plot. Note that the calibration pipeline introduced artifacts in the continuum of the spectrum of day 74, particularly below 5000\,$\mathrm{\AA}$.} 
\label{Fig:main_spec}
\end{center}
\end{figure*}

\subsection{Optical spectral evolution}
\begin{figure*}
\begin{center}
  \includegraphics[width=0.7\textwidth]{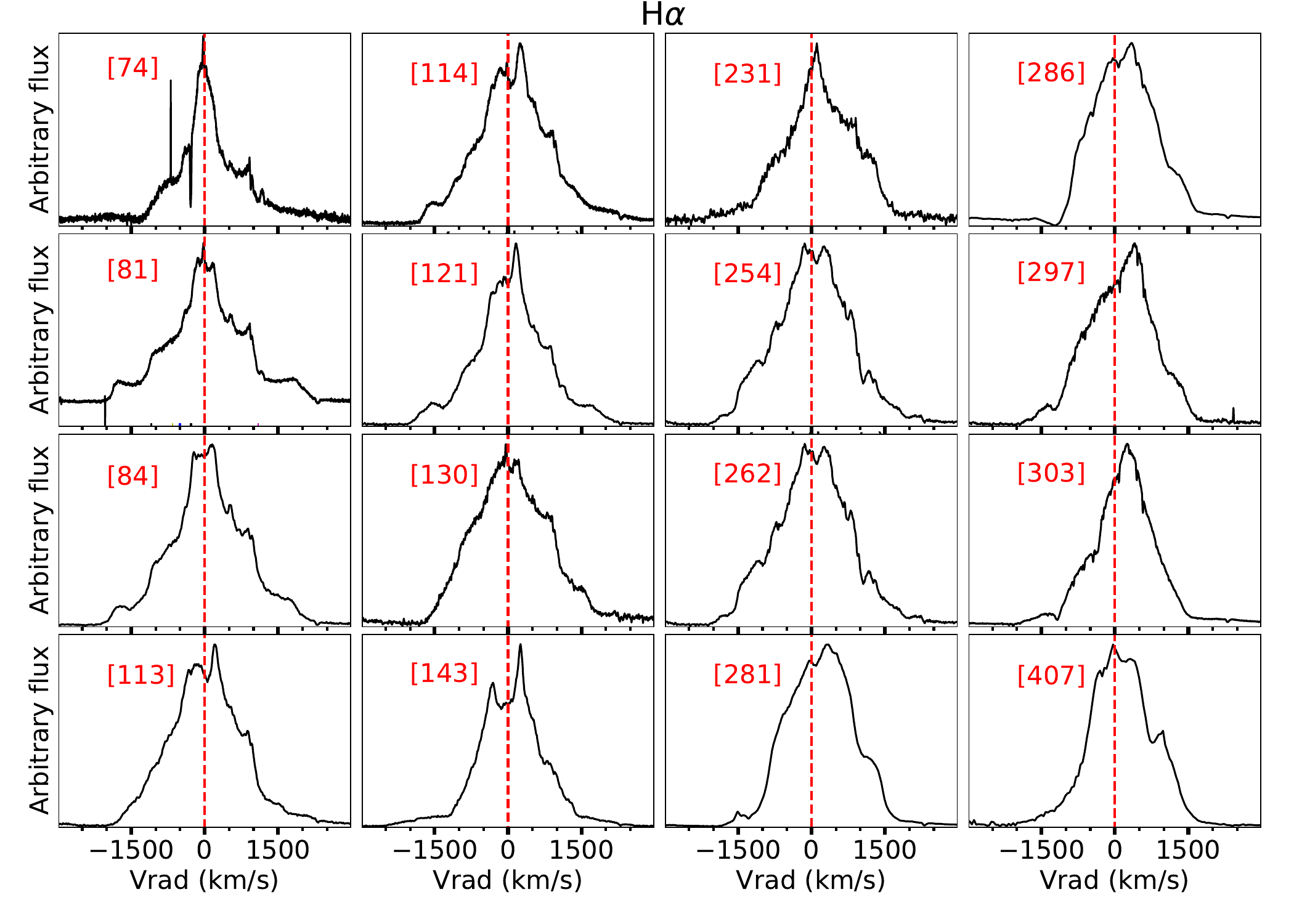}
\caption{The line profile evolution of H$\alpha$ throughout the outburst of V1047~Cen. The numbers between brackets are days after outburst. The red dashed lines represents the rest velocity ($V_{\mathrm{vrad}}$ = 0\,km\,s$^{-1}$). A heliocentric correction is applied to the radial velocities.} 
\label{Fig:Halpha}
\end{center}
\end{figure*}

In Figure~\ref{Fig:main_spec} we present a sample of the optical spectra, evolving throughout the outburst of V1047 Cen and covering three stages: the rise (day 74), the plateau phase (days 113 and 303), and the post-outburst (day 407). The full spectral evolution is presented in Figures~\ref{Fig:spec_1} to~\ref{Fig:spec_5}. The first spectrum, obtained 74 days after the start of the outburst, shows relatively narrow emission lines of H (Balmer), \eal{He}{I} and \eal{He}{II}, with Full Widths at Half Maximum (FWHMs) $<$ 300\,km\,s$^{-1}$ and Full Widths at Zero intensity (FWZIs) of around 500\,km\,s$^{-1}$, which are typical features of a DN outburst (e.g., \citealt{Morales-Rueda_Marsh_2002}) and low for classical nova eruptions. These features co-exist and are superimposed on top of broader emission lines of H, \feal{O}{III}, and \eal{C}{VI}, with FWZIs of around 2500\,km\,s$^{-1}$ for the Balmer lines and 1100\,km\,s$^{-1}$ for the \feal{O}{III} lines, which are characteristic of a classical nova nebula (likely arising in the remnant of the 2005 nova eruption). We also identify an even narrower feature in the Balmer lines, during the first two spectral epochs (days 74 and 81), characterized by FWHMs of $<$ 40\,km\,s$^{-1}$ (see Figures~\ref{Fig:Halpha} and~\ref{Fig:beta}). We do not have a definite explanation for the origin of these narrow features.

\begin{figure*}
\begin{center}
  \includegraphics[width=0.77\textwidth]{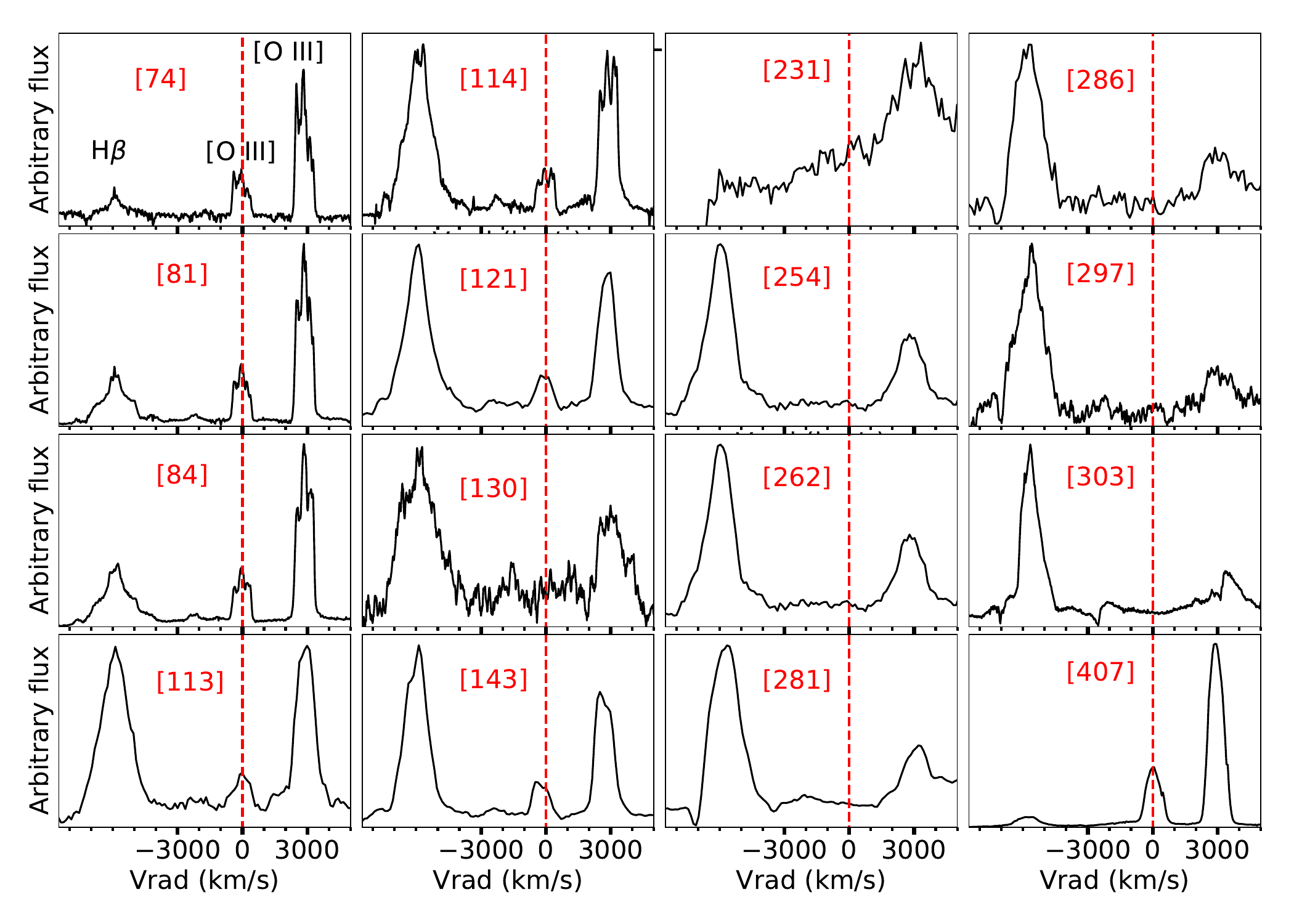}
\caption{The evolution of the profiles of the \feal{O}{III} lines at 4959 and 5007\,$\mathrm{\AA}$ in comparison to H$\beta$, throughout the outburst of V1047~Cen. The numbers between brackets are days after outburst. The red dashed lines represents the rest velocity ($V_{\mathrm{vrad}}$ = 0\,km\,s$^{-1}$) of \feal{O}{III} 4959\,$\mathrm{\AA}$. A heliocentric correction is applied to the radial velocities. Note that on day 231 H$\beta$ falls near the end of the spectral range.} 
\label{Fig:beta_OIII}
\end{center}
\end{figure*}

By day 113, the fluxes of the Balmer lines had increased significantly relative to the \feal{O}{III} lines; compare
$F_{H\alpha}/F_{5007} =1.56$ on day 74 with $\sim$15 on day 113 
(e.g., $F_{H\beta}/F_{5007}$ increased from $\sim$0.35 on day 74 to $\sim$3.1 on day 113; see Figure \ref{Fig:beta_OIII} for a direct comparison between the evolution of H$\beta$ and the \feal{O}{III}lines). The high-resolution spectra show that the Balmer and \feal{O}{III} line profiles are very different from one another, with the \feal{O}{III} lines (in the high-resolution spectra) having rectangular shapes and jagged tops, characteristic of nova nebular lines. In contrast, H$\beta$ shows complex profiles which vary throughout the outburst. The Balmer lines also had significantly broadened from day 113, with the FWZIs increasing by a factor of $\sim$2 compared to day 74, reaching 4000\,--\,4500\,km\,s$^{-1}$ (Figures \ref{Fig:Halpha} and~\ref{Fig:beta_OIII}). Note that a broad base in H$\alpha$ can be observed as early as day 81 but it becomes prominent from day 113 onwards. 

From day 130, some of the \eal{He}{I} emission lines show P Cygni profiles with absorption troughs at blueshifted velocities of around 1400\,km\,s$^{-1}$ (Figures~\ref{Fig:spec_2} and~\ref{Fig:spec_3}). 

\begin{figure}
\begin{center}
  \includegraphics[width=\columnwidth]{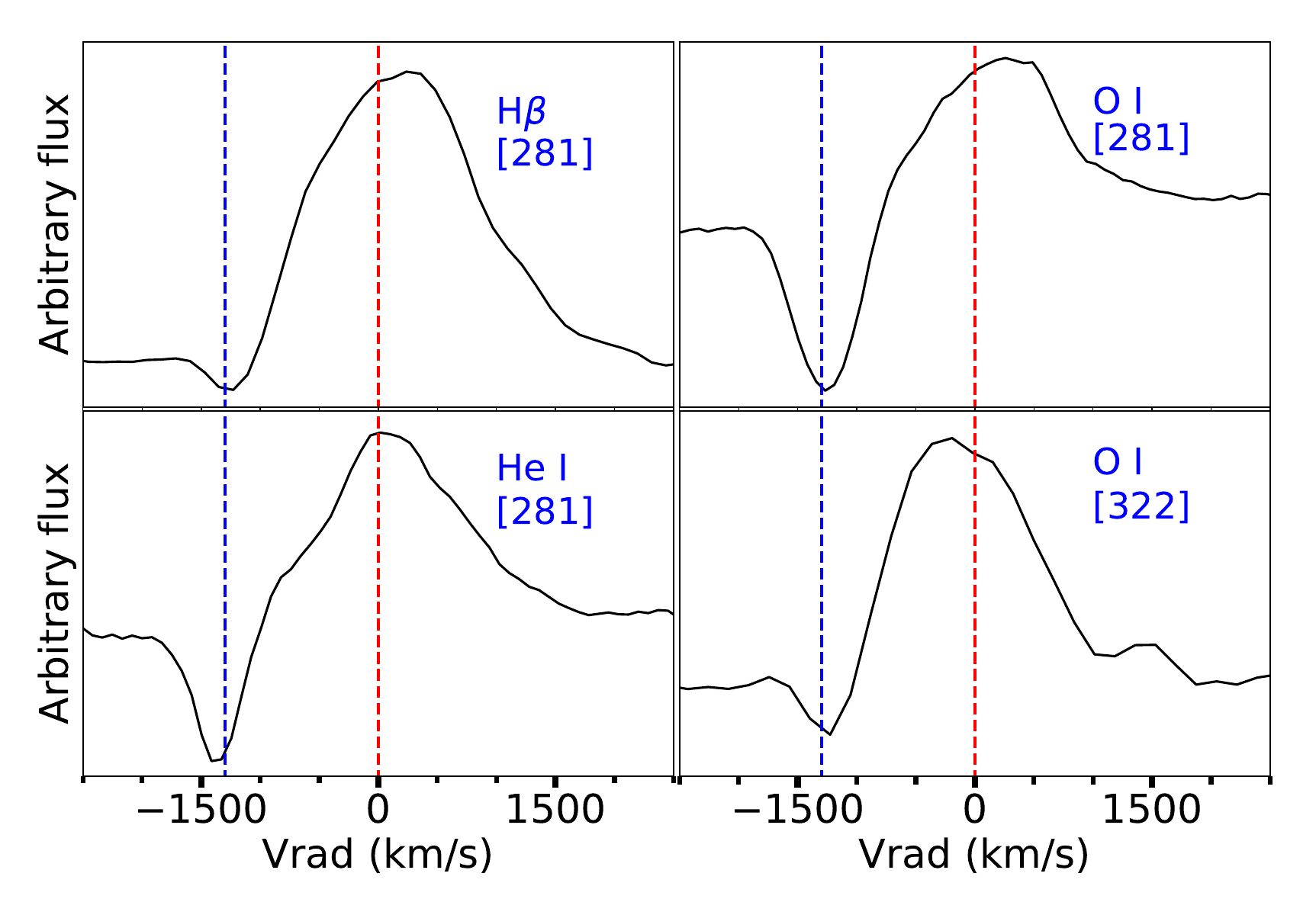}
  \caption{The P Cygni line profiles of H$\beta$, \eal{He}{I} 7065$\mathrm{\AA}$, and \eal{O}{I} 7773\,$\mathrm{\AA}$ taken on day 281 and \eal{O}{I} 1.128 $\mu$m taken on day 323 (Numbers between brackets are days since $t_0$). The red and blue dashed lines represent $v_r$ = 0\,km\,$^{-1}$ and $v_r$ = $-1300$\,km\,$^{-1}$, respectively.} 
\label{Fig:PCygni}
\end{center}
\end{figure}

Between days 262 and 303 \eal{O}{I} and \eal{N}{II} emission lines emerge, while the \feal{O}{III} nebular lines fade relative to the other spectral lines. At this stage, in addition to the \eal{He}{I} emission lines, Balmer, \eal{N}{II} and \eal{O}{I} emission lines also show P Cygni profiles with absorption troughs at velocities of around $-1200$ to $-1400$\,km\,s$^{-1}$ (Figure~\ref{Fig:PCygni}). During the same period, particularly on days 281 and 286, the optical spectra which extend above 8000\,$\mathrm{\AA}$ show broad double-peaked emission of \eal{O}{I} at 8446\,$\mathrm{\AA}$ with
FWZIs of around 3600\,km\,s$^{-1}$ (Figure~\ref{Fig:spec_6}). The \eal{O}{I} P Cygni profile at 7773\,$\mathrm{\AA}$ and the double-peaked \eal{O}{I} 8446\,$\mathrm{\AA}$ are not typical features of DN outbursts \citep{Morales-Rueda_Marsh_2002}. 

\begin{figure}
\begin{center}
  \includegraphics[width=\columnwidth]{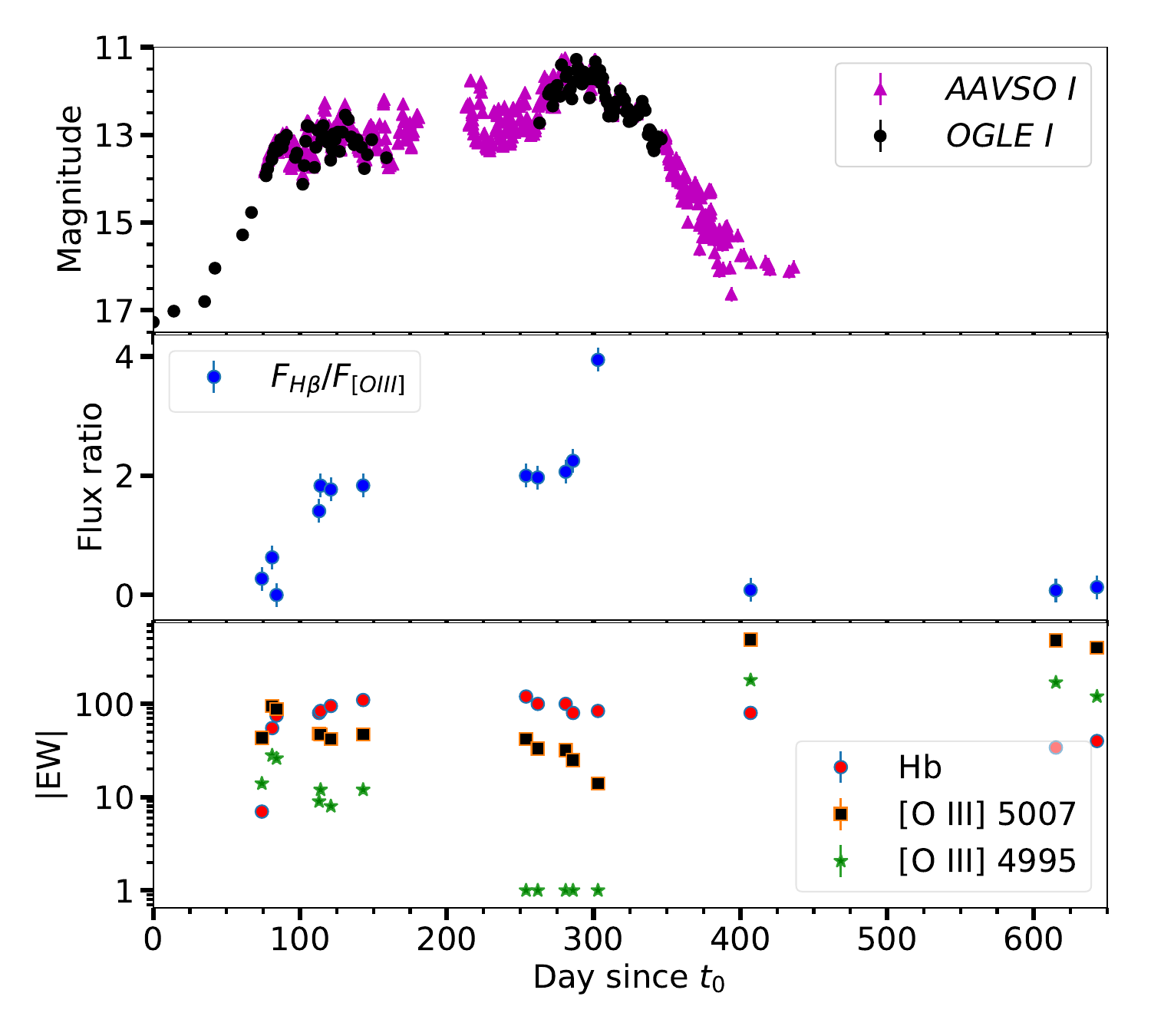}
\caption{\textit{Top}: the optical light curve in the $I$-band. \textit{Middle}: The evolution of the flux ratio between H$\beta$ and \feal{O}{III} 5007\,$\mathrm{\AA}$. \textit{Bottom}: The evolution of the of the equivalent width (EW) of the H$\beta$, \feal{O}{III} 5007\,$\mathrm{\AA}$, and \feal{O}{III} 4995\,$\mathrm{\AA}$ emission lines. } 
\label{Fig:line_ratio}
\end{center}
\end{figure}

After the end of the optical outburst, the spectrum obtained on day 407 shows substantial changes (Figure~\ref{Fig:main_spec}), with the \feal{O}{III} emission lines now dominating the spectrum relative to the Balmer lines. In Figure~\ref{Fig:line_ratio} we show the evolution of the line ratio between the \feal{O}{III} line at 5007\,$\mathrm{\AA}$ and H$\beta$ and the evolution of the equivalent width of H$\beta$, \feal{O}{III} 5007\,$\mathrm{\AA}$, and \feal{O}{III} 4995\,$\mathrm{\AA}$ emission lines. It is remarkable how the \feal{O}{III} lines were still relatively strong during the rise of the 2019 outburst, but fade throughout the outburst, before strengthening again relative to the Balmer lines by the end of the outburst. 

Half a year later, our spectra on days 615 and 643 still show strong \feal{O}{III} lines, in addition to other lines, which likely originate in the 2005 classical nova nebula, such as \feal{O}{II} 7320\,$\mathrm{\AA}$, \feal{N}{II} 5755\,$\mathrm{\AA}$, and high ionization \feal{Fe}{VII} lines (Figure~\ref{Fig:spec_5}).

Our last spectrum taken on day 774, more than 2 years after the start of the 2019 outburst, shows in addition to the 2005 nova nebular lines, weak lines from high ionization transitions of He, N, O, and C, such as the \eal{He}{II} lines at 4686 and 5412\,$\mathrm{\AA}$, \eal{N}{V} 4603\,$\mathrm{\AA}$, \eal{O}{V} 5920\,$\mathrm{\AA}$, \eal{C}{IV} 5802\,$\mathrm{\AA}$, and \eal{O}{IV} 7713\,$\mathrm{\AA}$ or \eal{Ne}{IV} 7716\,$\mathrm{\AA}$ (Figure~\ref{Fig:spec_7}). Such lines have been observed in systems like V617~Sgr and V~Sge and are associated with nuclear shell burning (e.g., \citealt{Herbig_etal_1965,Cieslinski_etal_1999,Steiner_etal_1999}).

\begin{figure}
\begin{center}
  \includegraphics[width=\columnwidth]{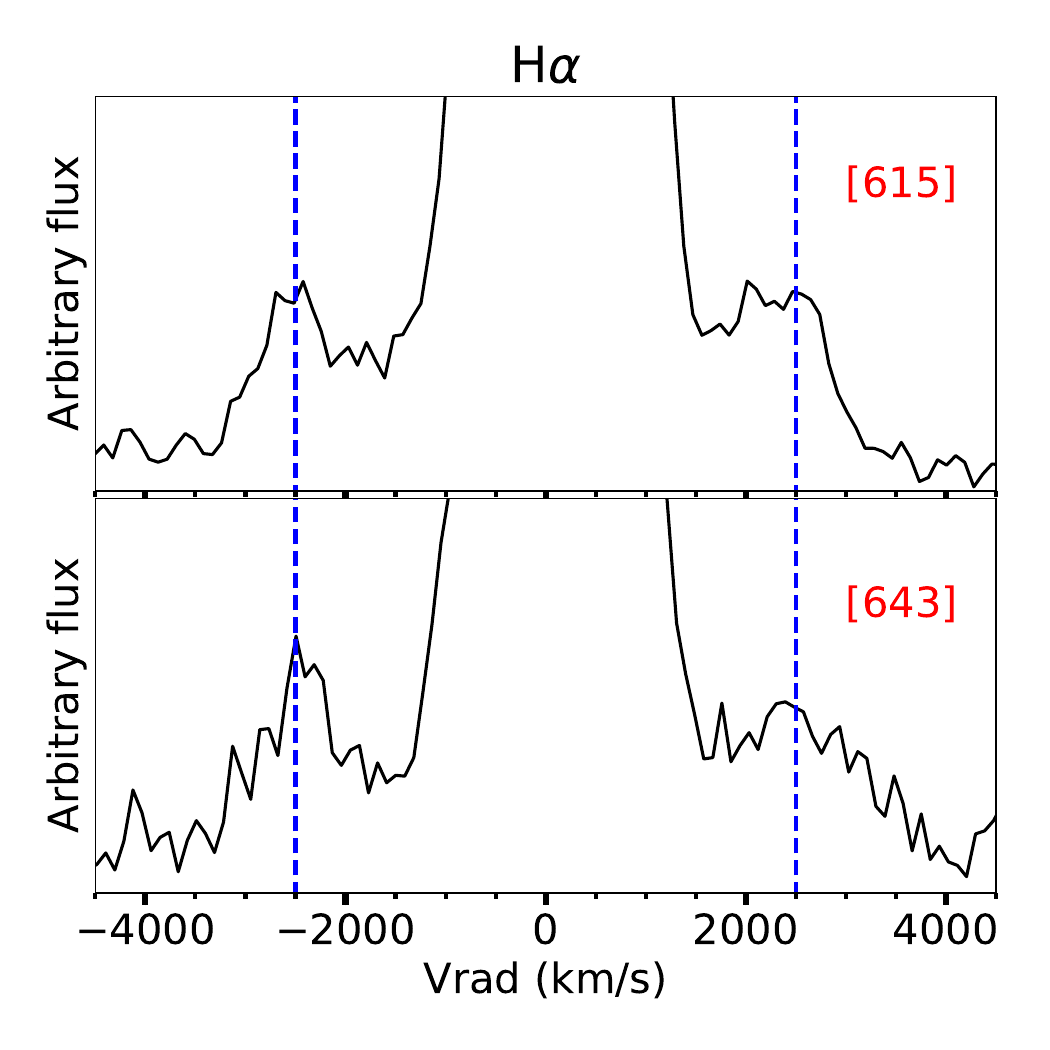}
\caption{High-velocity ``satellite" features in the late H$\alpha$ line profiles, taken on days 615 and 643. The origin of these features is considered in the Discussion.} 
\label{Fig:jet_lines}
\end{center}
\end{figure}

\begin{figure*}
\begin{center}
  \includegraphics[width=1.01\textwidth]{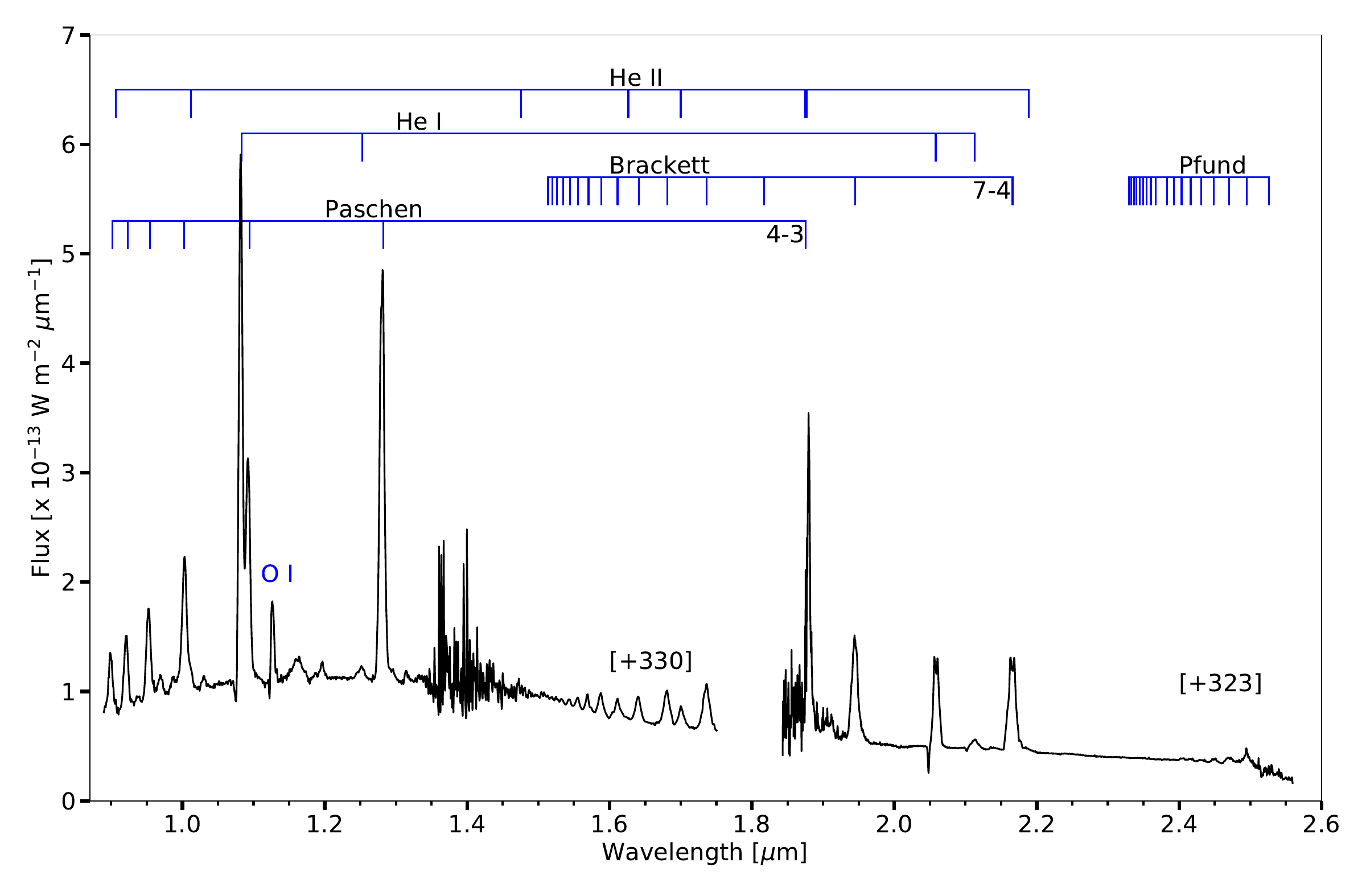}
\caption{The Gemini-S 0.9\,--\,2.5\,$\mu$m spectra obtained on days 323 and 330. We identify hydrogen recombination and other He and O lines. For the recombination lines, we show the transition of the longest wavelength member.} 
\label{Fig:Gemini_spec}
\end{center}
\end{figure*}

Several months after the end of the optical outburst on days 615 and 643, H$\alpha$ shows satellite emission components at $\pm$ 2500\,km\,s$^{-1}$ (Figures~\ref{Fig:jet_lines} and \ref{Fig:spec_5}). We elaborate on the origin of these components in the discussion.

\subsection{IR spectral evolution}
\label{sec_IR_spec}
\begin{figure}
\begin{center}
\includegraphics[width=0.51\textwidth, height = 7.4cm]{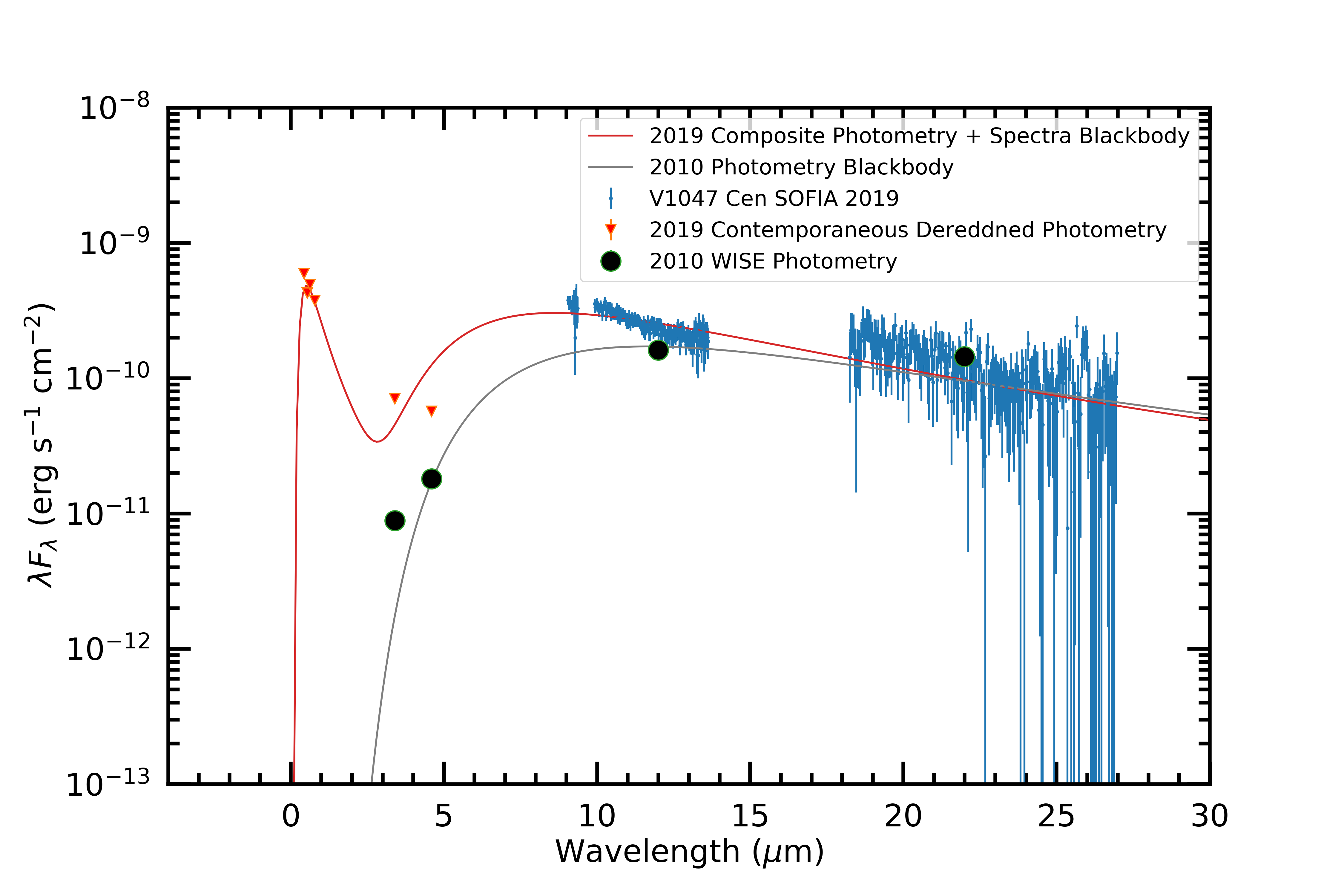}
\caption{The 2019 SOFIA mid-infrared spectra of V1047 Cen obtained on 
day 87. The blue symbols are the SOFIA spectra (combination of two 
grating settings G111 and G227) which includes all spectral points 
where the atmospheric transmission at the flight altitudes are greater 
than 70\% (hence the gap near $\sim 9.5$~\micron{}). The wavelength 
coverage of the two gratings is not continuous. Superposed is the 
best-fit composite blackbody (red solid line), i.e., the sum
of a T$_{\rm bb} = 7070\pm950$~K from the 2019 photometry
data  plus a T$_{\rm bb} = 425\pm12$~K, the latter is interpreted as 
thermal emission from heated circumstellar material (dust) from
the blast of the 2019 eruption. The inverted orange 
triangles are contemporaneous (2019) dereddened $BVRI$ and NEOWISE 
photometry, while the filled-black circles are 2010 WISE 
photometry showing that the infrared spectral energy distribution 
peak was represented by a cooler blackbody (solid gray line) 
of T$_{\rm bb} = 315\pm30$~K.}
\label{Fig:SOFIA_spec}
\end{center}
\end{figure}

The Gemini and SOFIA IR spectra are represented in Figures~\ref{Fig:Gemini_spec} and \ref{Fig:SOFIA_spec}, respectively. A detailed description and analysis of the IR spectral evolution
during the first 120 days of the outburst is presented in \citet{Geballe_etal_2019}. The Gemini $JHK$ spectra obtained on day 323 show emission lines of \eal{H}{I} (Paschen, Brackett, and Pfund series), \eal{He}{I}, \eal{He}{II}, and \eal{O}{I}. The FWZIs of the lines are around 4000\,km\,s$^{-1}$, similar to the ones measured from the optical spectral lines. The \eal{O}{I} line at  1.1289$\mu$m and some of the \eal{H}{I} and \eal{He}{I} lines show P Cygni profiles with absorption troughs at blue-shifted velocities of 1100 to 1800\,km\,s$^{-1}$, also comparable to the ones measured for optical lines (Figure~\ref{Fig:PCygni}).

Figure~\ref{Fig:SOFIA_spec} presents the SOFIA composite
spectra obtained on day 87, including contemporaneous, dereddened 
(see \S~\ref{red_dist}) $BVRI$ 
photometry obtained from the AAVSO database, as well as NEOWISE 
photometry obtained 18 days later. A blackbody fit to the 2019 
SOFIA spectra yields T$_{\rm bb} = 425 \pm 12$~K, which we
interpret as thermal emission from circumstellar material heated
by the outburst event.  WISE photometry obtained 
in 2010 (prior to the eruption) is also plotted and a blackbody
fit to this photometry yields a cooler temperature of 
T$_{\rm bb} = 315 \pm 30$~K. Thus the circumstellar
material (dust, likely from the 2005 nova event) has been heated as a result of processes related to the 2019 outburst. The SOFIA spectra show no evidence for strong H or \eal{He}{I} emission lines on day 87. However the [\ion{O}{4}] 25.91~\micron{} fine structure line, frequently seen in 
other novae \citep{Gehrz_etal_2015,Evans_etal_2012,Helton_etal_2012}
is marginally detected. A Gaussian fit gives a line flux of 
$6.9 \pm 1.6~\rm{W}~\rm{m}^{-2}$. The upper level of this line 
is collisionally de-excited at electron densities ($n_{e}$) in excess of
$9.9 \times 10^{3}$~cm$^{-3}$ (for an electron temperatures of  $10^{4}$~K); 
the presence of the line therefore indicates that the electron density 
in the region where the line is produced is less than this value.

\section{Discussion}
\label{Disc}

\subsection{Reddening and distance}
\label{red_dist}
In order to derive the reddening towards the system, we use the equivalent widths of several absorption lines from diffuse interstellar bands, in combination with the empirical relations from \citet{Friedman_etal_2011}. We derive an average $E(B-V) \approx 1.0 \pm 0.1$ mag and $A_V = 3.1 \pm 0.3$ mag for $R_V = 3.1$, in good agreement with the results of \citet{Geballe_etal_2019}. 
We avoid using the interstellar \eal{Na}{I}~D absorption doublet due to saturation. This is much smaller than the values derived from the Galactic reddening maps of \citet{Schlafly_etal_2011}, which estimate $A_V \approx 35$\,mag towards the system. However, the measurements from \citet{Schlafly_etal_2011}'s reddening maps should be considered an upper limit on the reddening along the line of sight. In addition, at Galactic latitude $<$ $\vert 5\vert$ deg, which is the case for V1047~Cen, the reddening estimates of the maps are not reliable.

The Gaia parallax measurement of V1047~Cen suffers from large uncertainties. The latest Gaia EDR3 parallax of the system is 0.338$\pm$0.249\,mas \citep{Gaia_EDR3}. With a flat prior (Galactic latitude-based priors are not appropriate for novae), this implies a distance of $2.7^{+3.9}_{-1.0}$ kpc, which is not a useful constraint. 

Due to the high uncertainty in
the Gaia parallax, we use the three dimensional Galactic reddening maps from \citet{Chen_etal_2019} and our measured reddening value to estimate the distance to V1047 Cen. The reddening map uses measurements from the Gaia DR2, 2MASS and WISE surveys. Therefore, we converted the previously derived $E(B-V)$ measurement to reddening values in the 2MASS $JHK$ filters and the Gaia DR2 $G$, $G_{Bp}$, and $G_{Rp}$ bands, using the extinction law from \citet{Wang_etal_2019} and \citet{Chen_etal_2019}.
We derive $E(G-K_s) = 2.2 \pm 0.1$, $E(G_{Bp}-G_{Rp}) = 1.3 \pm 0.1$, and $E(H-K_s) = 0.2 \pm 0.1$ mag. Using these reddening values, we derive an average distance of $3.2 \pm 0.2$\,kpc, consistent with the Gaia parallax distance within its large uncertainties.
The above uncertainty for the reddening-based distance likely underestimates the systematic uncertainties in this calculation.

\subsection{Evolutionary stage of the secondary}
\label{sec:secondary}

The average $V$- and $I$-band magnitudes as measured by OGLE long after the nova, between 2013 and 2018, are $I$ = 17.1 and $V$ = 17.5 mag. 
The post-nova color index is then $(V-I)$ = 0.38$\pm$0.07, implying $(V-I)_0 = -1.19$, which is  bluer for a typical CV system, implying a potential contribution from ongoing nuclear burning on the white dwarf. In Figure~\ref{Fig:charts} we show the OGLE light curve during the six years prior to the 2019 outburst. Adopting a distance of 3.2\,kpc, $A_V$ = 3.1  mag, and $A_I$ = 1.46 mag, we derive $M_V \approx 1.9$ mag and  $M_I \approx 3.5$ mag during the post-nova period. In addition to the companion star, we also expect contributions from the accretion disk, hot spot, the 2005 nova nebula, and any possible ongoing nuclear burning. Therefore, using the optical magnitudes/colors between 2005 and 2019 to constrain the evolutionary stage of the companion star is not straightforward. 

The system is not listed in the USNO-B1.0 catalog \citep{Monet_etal_2003}, and is not apparent in visual inspection of blue or red plates of the Digitized Sky Survey (DSS; Figure~\ref{Fig:charts}). Since the completeness level of USNO-B1.0 is roughly $V \sim 20.5$--21 mag, the absolute magnitude of the system during quiescence (pre-2005) should be $M_V >$ 5\,mag (at least 3 magnitudes fainter than the post-2005 nova brightness). This rules out anything more luminous than a main-sequence secondary and favors a typical CV system with an orbital period of the order of a few hours \citep{Darnley_etal_2012}. However, if the distance and extinction we derive are underestimated, the secondary star might be more evolved; If we assume a distance $\approx$ 6\,kpc (closer to the upper limit of the distance estimated from the Gaia parallax), and $A_V$ 5\,mag (note that the \citealt{Schlafly_etal_2011} maps estimates $A_V$ $\approx$ 35\,mag in the direction of the system), we derive an absolute quiescent magnitude brighter than 2 magnitudes in the $V$-band -- implying a more evolved companion star (e.g, a sub-giant).

\subsection{The 2005 eruption vs the 2019 outburst}
\label{sec_2005vs2019}

The 2005 nova eruption reached an apparent magnitude of around 8\,mag in the visual, compared to a peak $V$-band magnitude of 13.5 for the 2019 outburst (Figures~\ref{Fig:LC_comp} and~\ref{Fig:optical_UV}). These translate to an absolute visual magnitude of $\approx -8$
for the 2005 eruption and $\approx -2$
for the 2019 outburst, using $A_V = 3.03$ and a distance of 3.2\,kpc (see Section~\ref{red_dist}).

The energy radiated during the 2005 nova eruption can be estimated by integrating the Eddington luminosity of the WD over the period of the eruption. The Eddington luminosity of a 1\,M$_{\odot}$ WD is of the order of 10$^{38}$\,erg\,s$^{-1}$. Novae emit close to Eddington (and in many cases super-Eddington) luminosity for months up to years \citep{Starrfield_etal_2008}. As a conservative estimate, we will assume emission at Eddington luminosity over 100 days. Note that the \textit{Swift} follow up of the 2005 nova event, 5 months after the discovery, showed hard X-ray emission which is typically detected prior to the X-ray supersoft phase (see, e.g., \citealt{Gordon_etal_2021}). Therefore, it is likely that the nova radiated at near Eddington luminosity for longer than 100 days. Over 100 days, the amount of energy radiated by the 2005 nova event is of the order of $\sim$ 10$^{45}$\,erg. This is only the energy radiated, but a considerable portion of the energy during a nova goes into kinetic energy of the ejecta (e.g., \citealt{Gallagher_Starrfield_1976}). Therefore, the energy output during the 2005 nova event is at least a few times 10$^{45}$\,erg --- a very conservative estimate. For the 2019 outburst, determining the energy output is not straightforward, given that the event might be a combination of mechanisms, starting with a disk re-brightening, due to enhanced mass transfer or disk instability, leading to ejection of material. 

There is no model that describes well the emission during the 2019 event, hence, given the lack of appropriate models, we will assume blackbody emission, which should be a fair assumption for the purpose of the rough estimates we are trying to derive. At a distance of 3.2 kpc and $A_V= 3.1$ mag, a blackbody with $T$ $\approx 18,000$\,K and  $R \approx$ 1.5 $\times 10^{11}$\,cm should have an apparent $V$-magnitude of $\approx$ 15.0. This is comparable to the average maximum $V$-band magnitude throughout the 2019 outburst. Such a blackbody has a bolometric luminosity of a few times 10$^{36}$\,erg\,s$^{-1}$. Assuming that this blackbody emission represents most of the luminosity, the total energy radiated during the 2019 outburst would be $\sim$10$^{44}$\,erg in 400 days. This could be an overestimate given that the peak average magnitude of $\approx$ 15.0 lasted for 200 days only and the outburst showed 100 days of rise and 100 days of decline. So, the actual energy radiated is possibly less than 10$^{44}$\,erg, based on the blackbody emission. But given the uncertainty on the Bolometric Correction and given that we do not take into account the kinetic energy of potential material ejection, we will assume 10$^{44}$\,erg to be a rough estimate for the total energy output of the 2019 outburst. This shows that the 2005 nova event is at least an order of magnitude more energetic than the 2019 outburst, proving that the two events are of distinct natures.

The morphology of the optical light curves are also distinctively different, with the 2005 light curve typical of a fast classical nova, while the light curve of the 2019 event plateaued at peak for around a year. The velocities measured from the spectra taken during the 2005 eruption range from 750 to 1800\,km\,s$^{-1}$. These velocities are slow to moderate in comparison to the velocities observed in 
classical novae. During the rise of the 2019 outburst, the velocities measured from some of the lines were around a few hundreds km\,s$^{-1}$ (typical of CVs), but later in the outburst, the spectral lines showed velocities of $\approx$ 2000 km\,s$^{-1}$, raising more questions about the nature of this event. 

\subsection{The 2005--2019 post-nova period}
\label{sec:2005--2009}

\begin{figure*}
\begin{center}
  \includegraphics[width= 8.3cm, height = 7.3cm]{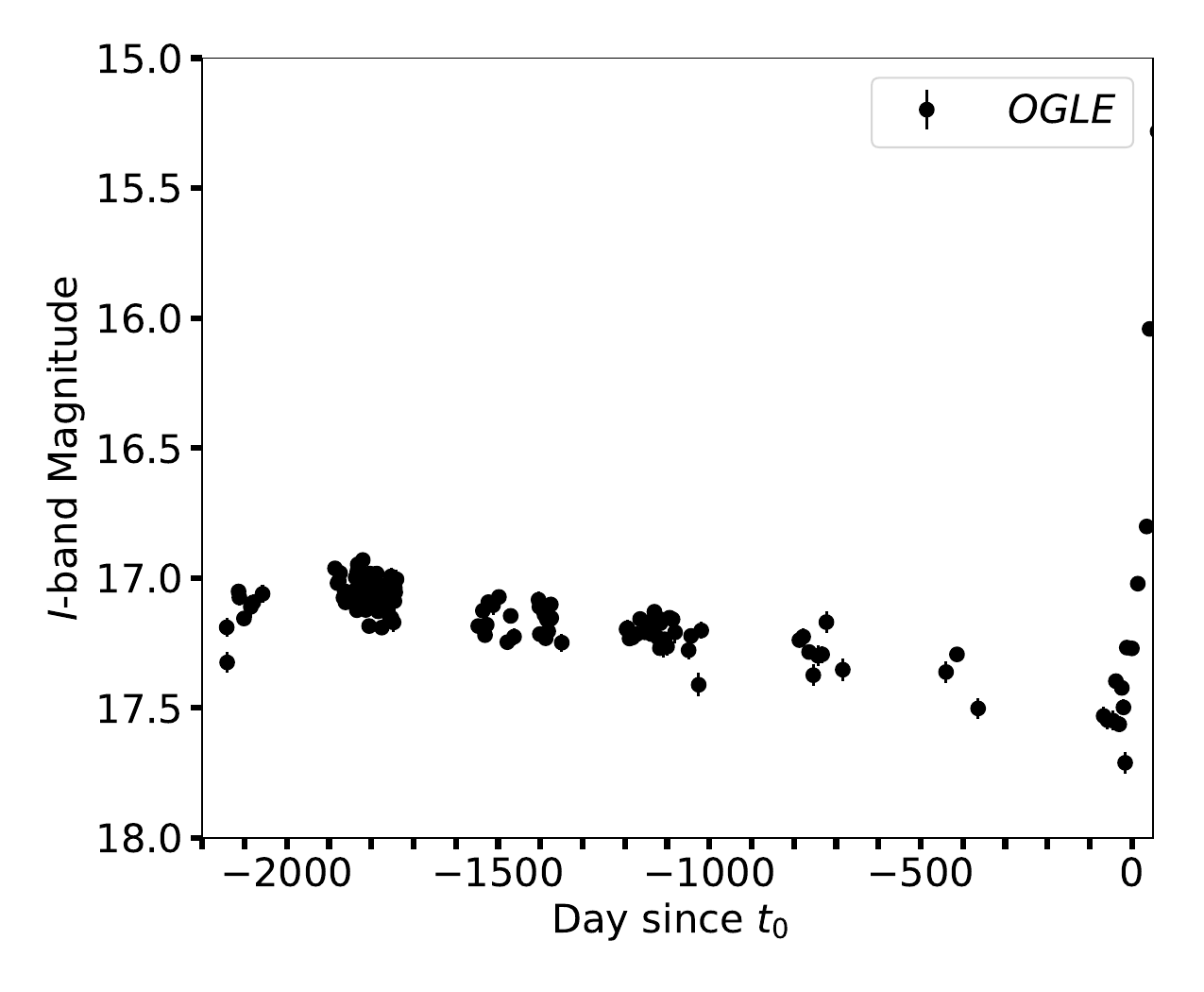}
  \includegraphics[width= 8.3cm, height = 7.3cm]{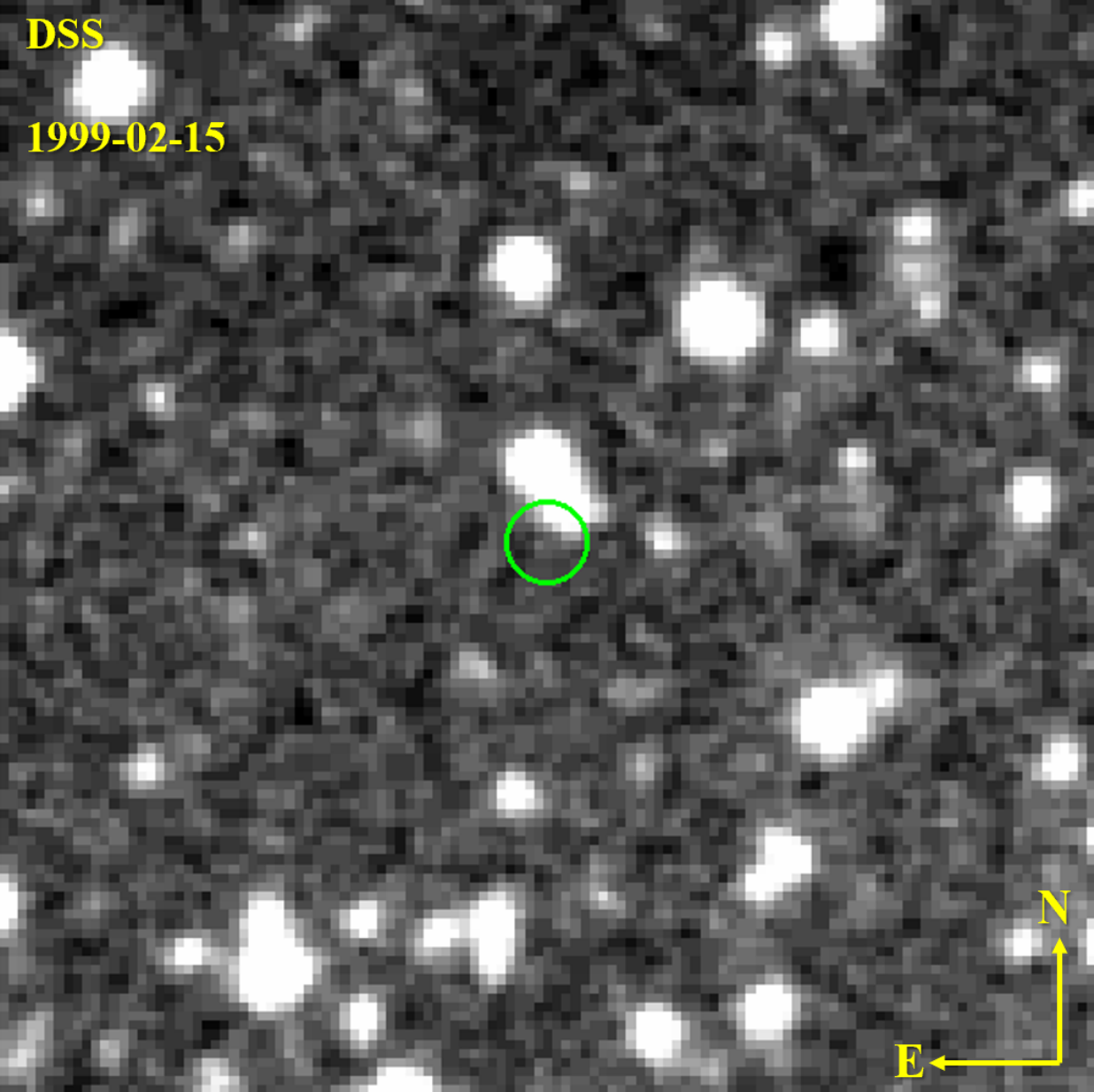}
  \includegraphics[width= 8.3cm, height = 7.3cm]{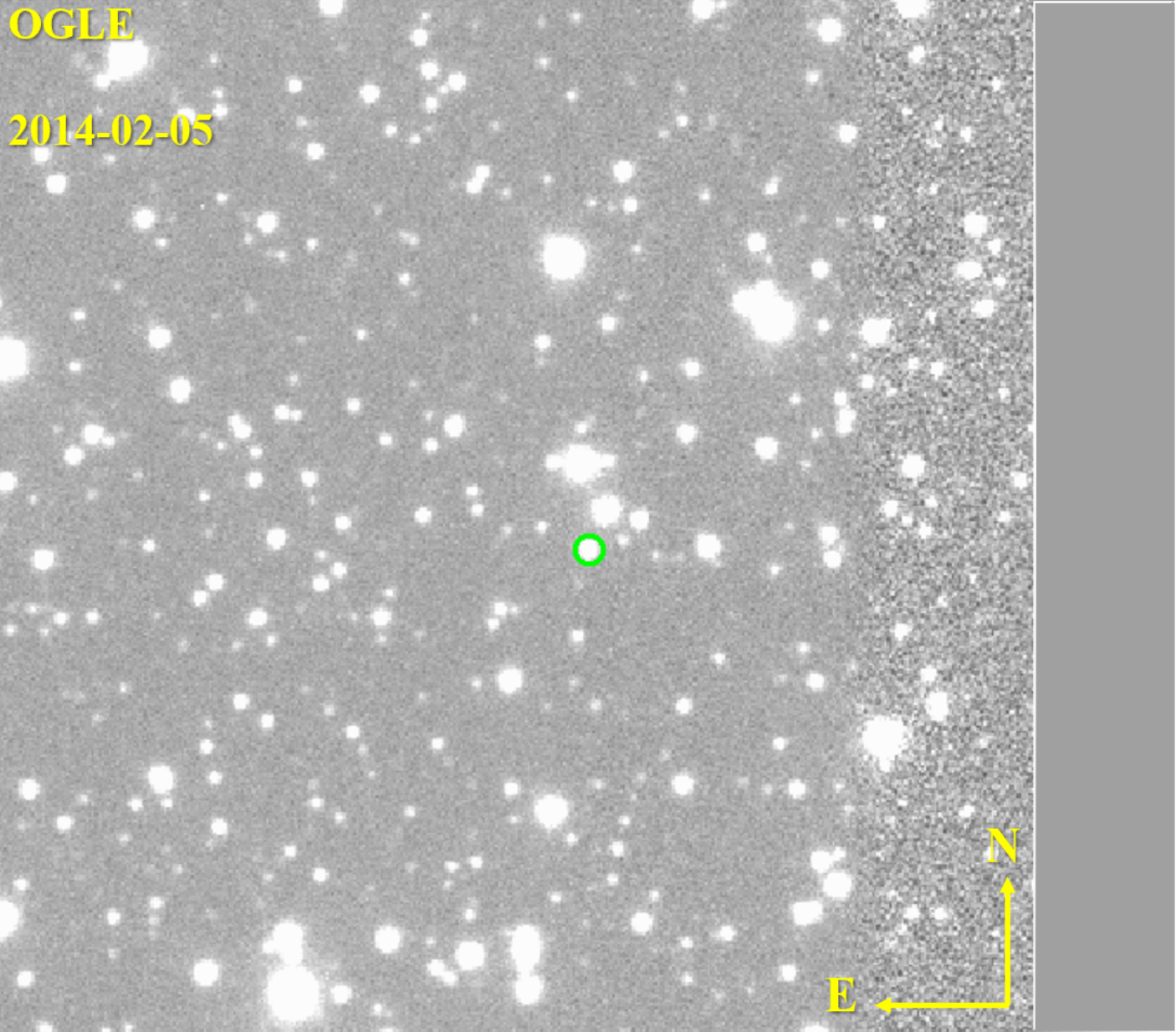}
  \includegraphics[width= 8.3cm, height = 7.3cm]{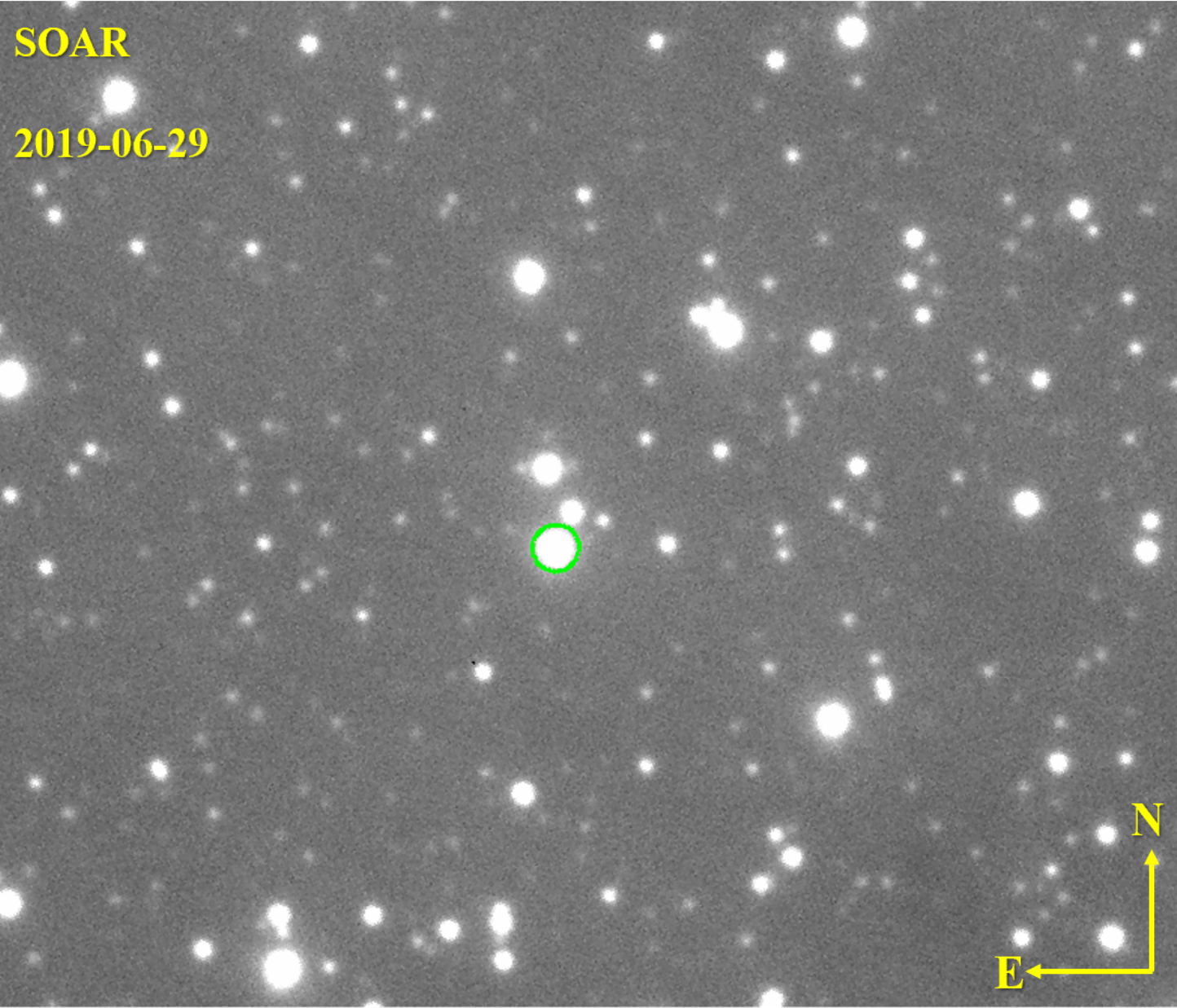}
\caption{\textit{Top-left:} the OGLE light curve between 2013 and 2019. Days are relative to the 2019 outburst start. \textit{Top-right:} A red plate of the field of V1047~Cen from the Digitized Sky Survey (DSS) obtained in February 1999 (position of V1047~Cen indicated by a green circle). \textit{Bottom-left:} combined OGLE $I$-band images of V1047~Cen (circled in green) taken between in February 2014. \textit{Bottom-right:} SOAR unfiltered acquisition image of V1047~Cen (circled in green) taken in June 2019, during the 2019 outburst. In all the charts the field is 2 $\times$ 2 arcmin. V1047 Cen lies on the edge of the OGLE images, which is the reason why the chart is cropped on the west edge.} 
\label{Fig:charts}
\end{center}
\end{figure*}

In Figure~\ref{Fig:charts}, we show charts of the field of V1047~Cen from DSS (red plates taken in Feb. 1999), OGLE (taken in Feb 2014), and SOAR (taken in June 2019). Clearly, there is an excess in brightness of the system in 2014, nine years after the 2005 nova eruption, compared to the pre-nova. This indicates that the system did not return to the pre-nova brightness.

The OGLE $(V-I)$ colors between 2010 and 2019 indicate a blue source with high temperatures, in excess of $10^5$\,K. These substantially blue colors could be due to the contribution of emission lines to the spectra---particularly the [\eal{O}{III}] lines from the nova nebula, whose fluxes contribute to the $V$-band. These lines were relatively strong compared to the other lines during the early spectra of the 2019 outburst and after the end of the outburst (Figure~\ref{Fig:main_spec}). Unlike the $V$-band, no strong nebular lines contribute to the flux in the $I$-band (Figure~\ref{Fig:spec_1}). 

Moreover, the blue colors (implying high temperatures) are likely an indication of ongoing nuclear burning on the surface of the WD. Many nova systems have shown extended supersoft X-ray emission and continued remnant thermonuclear burning on the surface of the WD several years after nova eruptions (e.g., \citealt{Schaefer_Collazi_2010,Zemko_etal_2015,Zemko_etal_2016}). 
This could be remnant burning from the accreted nova envelope or due to enhanced mass transfer from an irradiated secondary post-nova \cite{Ginzburg_etal_2021}. Stable nuclear burning from enhanced accretion onto the white dwarf has also been studied by \citet{Wolf_etal_2013}.
They derived the mass accretion rate needed for stable burning on a white dwarf and found that, for white dwarf masses between 0.6 and 1.3\,M$_{\odot}$, the mass accretion rate should be of the order of $\sim 10^{-8}\,$--\,$10^{-7}$\,M$_{\odot}$\,yr$^{-1}$. 

We do not have definite estimates of the  WD mass in V1047~Cen. \citet{Hachisu_Kato_2007} derived a low  WD mass of around 0.7\,M$_{\odot}$ for V1047~Cen, based on the light curve of the 2005 nova eruption.
However, the values derived by \citet{Hachisu_Kato_2007} are uncertain, given that the light curve of V1047~Cen does not follow their ``universal decline law''. On the other hand, the 2005 nova eruption of V1047~Cen showed a rapid decline in its optical light curve, which is indicative of an ejection of a low-mass envelope with large ejecta velocities \citep{Starrfield_etal_2020}; low mass ejecta in turn tend to be associated with eruptions occurring on massive WDs \citep{Yaron_etal_2005}.
\citet{Shara_etal_2018} combined simulations of nova eruptions with optical photometric data from \citet{Strope_etal_2010} and \citet{Schaefer_2010} to estimate the masses and accretion rates of white dwarf stars in novae. Based on their relations and the parameters of the 2005 classical nova eruption of V1047~Cen --- an eruption amplitude of $\approx$ $21-8.0$ = 13\,mag and a time to decline from peak by 2\,mag of $t_2 \approx$ 5 days --- we estimate a quiescent mass accretion rate of the order of 10$^{-10}$\,--\,10$^{-9}$\,M$_{\odot}$\,yr$^{-1}$ and a white dwarf mass of around 1.2\,--\,1.4\,M$_{\odot}$. This implies that V1047~Cen possibly has a massive white dwarf. However, the ejecta velocities measured from the optical
spectral lines during the 2005 classical nova are moderate (FWHM $<$ 2000\,km\,s$^{-1}$; \citep{Liller_2005}, which argues against a massive WD \citep{shafter_etal_2011}. 

For a WD mass in the range of 1.2 to 1.4\,M$_{\odot}$, the accretion rate needed for stable burning is a few times $10^{-7}$\,M$_{\odot}$\,yr$^{-1}$ \citep{Wolf_etal_2013}. This is relatively high for a typical accretion rate on a WD in a CV system. Even during a post-nova event, the mass transfer rate is usually of the order of $10^{-8}$\,M$_{\odot}$\,yr$^{-1}$ (e.g., \citealt{Kovetz_Prialnik_1985,Hillman_etal_2020}). However, recent work by \citet{Ginzburg_etal_2021} showed that the post-nova mass-accretion rate could be as high as $10^{-7}$\,M$_{\odot}$\,yr$^{-1}$ for several centuries. At such mass-transfer rate onto a massive WD (1.2--1.4\,M$_{\odot}$), the system would be a bright supersoft source \citep{Wolf_etal_2013,Page_etal_2020}, easily detectable with \textit{Swift} at a distance of $\sim$ 3--4\,kpc and with $A_V$ = 3.1 (implying interstellar $N(H)$ = 8.7 $\times 10^{21}$\,cm$^{-2}$ ; \citealt{Bahramian_etal_2015}).
However, there was no X-ray detection with \textit{Swift} in 2008 nor throughout the 2019 outburst.

Another alternative is that the stable nuclear burning is taking place on a low mass WD (0.6--0.8\,M$_{\odot}$) with emission too soft to be detected with \textit{Swift}. Stable burning on such a relatively low mass WD would require a mass accretion rate of the order of $10^{-8}$\,M$_{\odot}$\,yr$^{-1}$ \citep{Wolf_etal_2013}, which is a more typical post-nova mass transfer rate. Similarly, the lack of X-ray detection during the 2019 outburst is an indication that---if nuclear burning is present during the 2019 outburst---it must be taking place on a low mass white dwarf rather than a high mass white dwarf.

If the ongoing nuclear burning is from residual material from the nova eruption, this also means that the WD in nova V1047~Cen is of a low mass (0.6--0.8\,M$_{\odot}$), for it to lasts for several years. The residual nuclear burning on a high mass WD after a nova eruption typically lasts for a few weeks/months only (e.g., \citealt{MacDonald_1996,Schwarz_etal_2011}).

Regardless of the mass of the white dwarf, stable nuclear burning is likely taking place on the surface of the white dwarf between 2005 and 2019, and is responsible for the post-2005-nova brightness excess compared to the pre-nova brightness and the substantially blue colors in the OGLE data between 2013 and 2019.

\subsection{The 2019 outburst of V1047~Cen - a potential record breaker?}

Only a few classical novae are known to have shown DN outbursts after a nova eruption, such as GK Per (Nova Persei 1901; \citealt{Sabbadin_Bianchini_1983,Bianchini_etal_1986}), V1017~Sgr (Nova Sagittarii 1919; \citealt{Sekiguchi_1992,Salazar_etal_2017}) and V446 Her (Nova Herculis 1960;
\citealt{Honeycutt_etal_1995,Honeycutt_etal_2011}; see Table~\ref{table:cn_results} for a full list). The first DN outbursts of these novae occurred several decades after the nova eruption, with the shortest gaps being 54 years (V1017~Sgr), 47 years (GK Per), and 30 years (V446~Her). The 14 year gap between the classical nova eruption and the 2019 outburst of V1047~Cen is the shortest ever recorded, if it is indeed a DN outburst. Note that the older novae such as GK~Per and V1017~Sgr might have had earlier DN outbursts that were missed due to monitoring gaps.

During the 2019 outburst of V1047 Cen, the rise to optical peak, plateau phase, and decline lasted 100, 210, and 100 days, respectively. This 400 day outburst is twice as long as the longest known DN outburst, previously recorded for V1017~Sgr and four times longer than the longest outburst recorded for GK Per (e.g., the 2006 outburst; \citealt{Evans_etal_2009}).

The peak absolute magnitude at $V$-band during the 2019 outburst reached $\approx -2$\,mag. This 
is higher than the typical absolute magnitudes of dwarf novae during outburst, which range between 3.8 and 2.6 mag \citep{Ramsay_etal_2017}. Some systems with long periods (of the order of days) such as V1017~Sgr and V630~Cas, have extremely luminous DN outbursts, with absolute magnitudes reaching $-0.3$ and 1.4 mag, respectively. However, even these are still substantially less luminous compared to V1047~Cen. Therefore, if the 2019 outburst of V1047~Cen is a DN, it would be the most luminous outburst of a DN observed to date. Note that \citet{Kawash_etal_2021a} showed that some DN outbursts could reach absolute magnitudes brighter than zero, but accurate distances and extinction values were unavailable for these systems in their study. As mentioned in Section~\ref{sec_2005vs2019} the peak luminosity of the 2019 outburst is more than a few times 10$^{36}$\,erg\,s, which is orders of magnitude larger than the typical luminosity of DNe ($\sim$10$^{34}$\,erg\,s; \citealt{Warner_1995}). At a few times 10$^{36}$\,erg\,s, the outburst is too energetic to be powered by accretion alone.

All the above indicate that the 2019 outburst of V1047 Cen is either a record breaking DN outburst or---more likely---a more energetic phenomenon than a disk instability event, as we discuss below.

\subsection{Origin of the spectral features}
\label{spec_features}
During the rise phase (days 0 to 90), the spectrum of V1047 Cen resembled that of a DN superimposed on a spectrum of an old classical nova shell, implying an origin in a bright accretion disk. However, the subsequent evolution, particularly of the Balmer lines, differed considerably from that of DNe. The widths of the Balmer lines reached velocities much larger than those of the nova's nebular lines, with FWZIs of around 4000\,km\,s$^{-1}$ (compared to FWZIs $\approx$ 1100\,km\,s$^{-1}$ for the latter; see Figures~\ref{Fig:Halpha} and~\ref{Fig:beta_OIII}). These velocities are also much larger than those measured for the Balmer lines during GK Per's recent outburst (FWHM $\sim$ 500-650\,km\,s$^{-1}$; e.g., \citealt{ATel_7217}) or other DNe in outburst \citep{Morales-Rueda_Marsh_2002}.
The Balmer lines also showed substantial increases in their line fluxes relative to the nova nebular lines more than 100 days into the 2019 outburst (Figure~\ref{Fig:beta_OIII}). The cause of this dramatic flux increase and broadening is not clear, but it likely due to a fast, low density outflow (given that these velocities are higher than the escape velocities of a 0.8--1.3\,M$_{\odot}$ WD). All this indicates that, in addition to the early disk brightening, there are other mechanisms shaping the electromagnetic signatures of V1047~Cen.

Between days 262 and 310, most of the lines in the spectrum (e.g., Balmer, \eal{He}{I}, and \eal{O}{I}) showed P Cygni-like absorption features at velocities of around $-1400$\,km\,s$^{-1}$. Such features are not characteristic of a DN outburst. They probably originate in an optically thick wind/outflow. The onset of these features coincides with the start of the brightness increase between days $\sim$ 260 and 310 (Figure~\ref{Fig:optical_UV}), supporting the possibility of a wind or shell ejection during this phase of the outburst.  

The satellite emission components at $\pm$ 2500\,km\,s$^{-1}$ in H$\alpha$ around 7 months after the end of the outburst (Figure \ref{Fig:jet_lines}), are reminiscent of the features observed in Z And-like classical symbiotic outbursts. 
They are associated with collimated bipolar flows (see e.g., \citealt{Burmeister_etal_2007,Skopal_etal_2013,Tomov_Tomova_2013}). 

\subsection{The origin of the bright radio emission}
\label{radio_disc}
In the past decade, CVs have been established as radio emitters during both quiescent and outburst phases (e.g., \citealt{Cordova_etal_1983,Coppejans_etal_2015,Coppejans_etal_2016,Barrett_etal_2017}). Their radio properties are diverse, and the emission mechanisms powering CV radio luminosity are still being established.

\citet{Coppejans_etal_2015} observed several nova-like CVs and found that they are significant radio emitters. Nova-like CVs are non-magnetic CVs characterized by a 
sufficiently high mass transfer rate
to maintain the accretion disk in a constant hot state, unlike CV systems which undergo DN outbursts. For the individual systems in their sample, \citet{Coppejans_etal_2015} found that the emission in these nova-like CVs is consistent with optically thick synchrotron, gyro-synchrotron, or cyclotron maser emission. Their 6 GHz spectral radio luminosities are $\sim10^{15}-10^{17}$ erg\,s$^{-1}$\,Hz$^{-1}$ \citep{Coppejans_etal_2015}.

Non-magnetic CVs which undergo DN outbursts are also known to be radio sources during outburst \citep{Coppejans_etal_2016}. The most famous of these systems is SS~Cyg, which shows a radio flare during the early days of its optical outbursts, peaking at $\sim$ 1\,mJy before fading gradually, dropping below radio detectability by the end of the optical outburst. This flaring radio emission resembles that of X-ray binaries and is suggested to be the result of synchrotron emission from a transient jet (e.g., \citealt{Kording_etal_2008,Miller-Jones_etal_2011,Russell_etal_2016,Coppejans_etal_2016,Coppejans_etal_2020}). \citet{Coppejans_etal_2016} found that the radio spectral luminosities of DNe in outburst range between 10$^{14}$ and 10$^{16}$\,erg\,s$^{-1}$\,Hz$^{-1}$ at 10\,GHz. 

Based on a radio survey targeting a large sample of magnetic CVs, \citet{Barrett_etal_2017} suggest that they are also radio emitters, dominated by weakly polarized gyro-synchrotron emission or highly polarized electron-cyclotron maser emission. Most of these are nearby (less than a kpc) sources and are characterized by flux densities, $\sim$20--400 $\mu$Jy; the implication is radio spectral luminosities in the range $\sim10^{14}-10^{17}$ erg\,s$^{-1}$\,Hz$^{-1}$ \citep{Barrett_etal_2020}. Only one of them, AE~Aqr, shows substantially higher flux density ($\sim$5\,mJy), but it is located remarkably nearby, at a distance of $\approx$ 90\,pc \citep{Ramsay_etal_2017}.

Similar to the other observational features of V1047~Cen across the spectrum, the radio emission from this system is record breaking and puzzling.  Assuming a distance of 3.2\,kpc, the spectral luminosity of V1047~Cen at 2 GHz is $\sim$ 10$^{19}$\,erg\,s$^{-1}$\,Hz$^{-1}$.
This means that V1047~Cen is at least two orders of magnitude brighter at radio wavelengths than all other cataclysmic variables.

If the radio emission from a DN outburst is optically thick thermal emission from ionized gas at a typical brightness temperature of 10$^4$\,--\,10$^5$\,K, then the emitting regions should have radii of $\sim$ 10$^{14}$\,--\,10$^{15}$\,cm (assuming circular sources as projected on the sky) to produce the observed flux densities in V1047~Cen. During a DN outburst, it is reasonable to assume that the emitting gas has a size roughly equivalent to the size of the binary. CVs with orbital periods of the order of a few hours have orbital radii of $\sim$ 10$^{11}$\,cm, which is three orders of magnitude smaller in size than the optically thick emitting gas responsible for the observed fluxes. This implies that if the outburst is a DN, any emission of order the size of the orbit must be non-thermal (brightness temperature $> 10^5$ K).

\begin{figure*}
\begin{center}
  \includegraphics[width=0.8\textwidth]{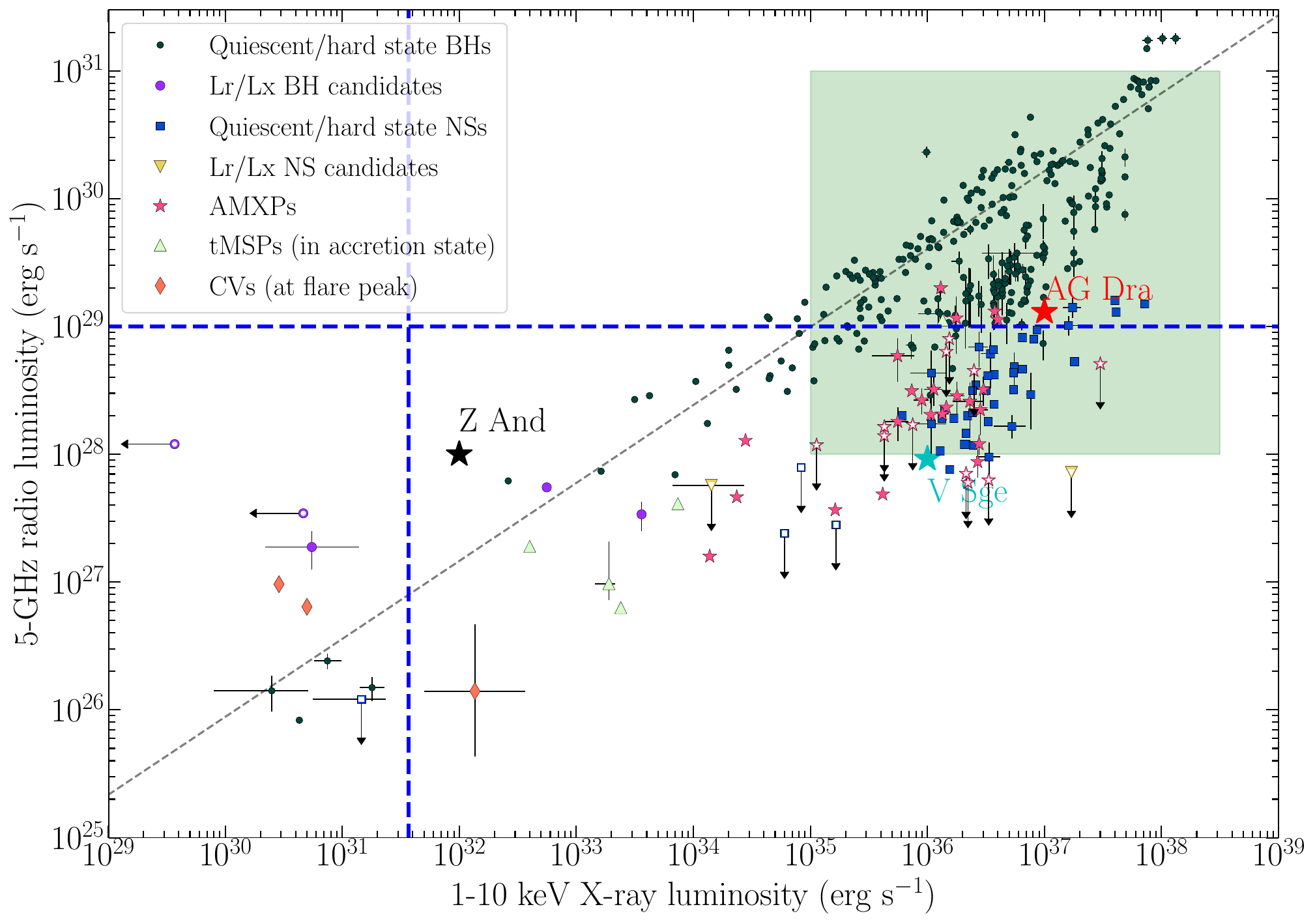}
\caption{The plane of 5 GHz radio luminosity ($L_{\mathrm{radio}}$) vs. 1--10 keV X-ray luminosity ($L_X$), with points representing X-ray binaries (accreting neutron stars, black holes), accreting and transitional millisecond pulsars, flaring CVs, classical novae, and classical symbiotics in outburst. The green shaded box represents the range of luminosities of classical novae. The classical symbiotics Z~And and AG~Dra are represented with black and red stars, respectively, while the X-ray binary V Sge is represented with a Cyan star. The radio luminosity of V1047 Cen at 5 GHz is plotted with dashed blue horizontal line and is calculated assuming a flat spectrum. 
Adopted from \citet{Bahramian_etal_2018_1252036} \url{https://zenodo.org/record/1252036/export/hx}.}
\label{Fig:LxLr}
\end{center}
\end{figure*}

One possibility is that the radio emission is non-thermal synchrotron emission from a transient radio jet, similar to the case of other DN outbursts. While the flux density of V1047~Cen equals that of SS~Cyg during its short-lasting flare, the luminosity of V1047~Cen is several orders of magnitude brighter than that of SS~Cyg (the distance to SS~Cyg is 114\,pc; \citealt{Miller-Jones_etal_2013}).  
In Figure~\ref{Fig:LxLr} we plot V1047~Cen on the $L_{\mathrm{radio}}$ vs $L_X$ diagram in comparison with jet emission in other CVs, classical novae, Z~And, V~Sge, and other astrophysical objects such as X-ray binaries, accreting millisecond X-ray pulsars, and transitional millisecond pulsars (we multiply the spectral luminosity by 5 GHz to estimate the radio luminosity at 5 GHz assuming a flat spectrum). 
V1047 Cen's 5\,GHz luminosity, $\sim 1 \times 10^{29}$\,erg\,s$^{-1}$, is much higher than that observed in CVs and is comparable to (or even brighter in some cases than) X-ray binaries and pulsars with compact primaries like neutron stars and black holes. Such an energetic jet should also be a bright X-ray source. We used WebPIMMS \footnote{\url{https://heasarc.gsfc.nasa.gov/cgi-bin/Tools/w3pimms/w3pimms.pl}} to derive the X-ray unabsorbed flux using the stacked \textit{Swift} X-ray detection of the 2019 outburst and assuming a 5\,keV thermal bremsstrahlung model and $N(H)$ = 8.7 $\times$ 10$^{21}$\,cm$^{-2}$ (we use $N(H)$ = 2.81 $\times$ 10$^{21}$ $A_V$; \citealt{Bahramian_etal_2015}). This translates to an X-ray luminosity of around 3.7$\times$10$^{31}$\,erg\,s$^{-1}$ at a distance of 3.2\,kpc, which is comparable to the X-ray luminosity of other CVs in outburst. 

It is not possible to determine if V1047~Cen showed flaring radio emission during the early days
of the outburst like SS~Cyg and was an even brighter radio source at that time, or if the $\sim$ 1\,mJy emission remained more or less constant throughout the optical outburst. Nevertheless, what is really puzzling is that at the end of the optical outburst, the radio flux did not drop. This is in contrast to the trends shown by SS~Cyg and other DN outbursts, where the radio flux drops below the detection limit by the end of the optical outburst \citep{Russell_etal_2016}. This disparity between the radio behavior and energetics of V1047~Cen 
and other DN outbursts raises additional questions about the origin of the radio emission in V1047~Cen and if a transient synchrotron jet is an appropriate explanation.

\citet{Bode_etal_1987} conducted a radio survey of 26 classical novae observed with the VLA less than 10 years after eruption and detected only two in the sample. Both novae were detected at 5\,GHz, 2 years (V4077~Sgr) and 8 years (NQ~Vul) after the nova eruptions. The emission from these two novae was suggested to be thermal and had radio spectral luminosities $\sim10^{18}-10^{19}$ erg\,s$^{-1}$\,Hz$^{-1}$. 
The higher-end radio spectral luminosity of V4077~Sgr is due to the fact that the system was observed only 2 years after the nova eruption; radio emission from classical novae often peaks on timescales $\sim$1--3 yr after eruption \citep{2021arXiv210706251C}. In addition, given that radio emission from decade-old or older nova ejecta originates in the nova's extended thermal remnant, it can vary only slowly, not on the weeks--months timescales observed in V1047~Cen \citep{2021arXiv210706251C}. Therefore, it is very unlikely that the radio emission from V1047~Cen originates in the ejecta of the 2005 classical nova eruption.

The long-lasting, luminous radio emission in V1047~Cen resembles the radio emission from Z And-like classical symbiotic, accretion-powered, outbursts (e.g., \citealt{Crocker_2001,Mikolajewska_2002,Brocksopp_2004,Sokoloski_etal_2006}). These systems are luminous radio sources ($L \approx 10^{18}-10^{19}$ erg\,s$^{-1}$\,Hz$^{-1}$; \citealt{Ogley_etal_2002, Brocksopp_2004}),
where the emission is usually suggested to be thermal, originating from bipolar collimated jets/outflows. Classical symbiotic systems consist of giant stars, which in most cases do not fill their Roche Lobes, transferring material onto WDs;
accretion can proceed via a disk or not, depending on the relative velocity of the red giant wind and the accreting WD \citep{Livio_1992, Mohamed_Podsiadlowski_2007}. Even out of outburst, symbiotic stars can be luminous radio emitters ($L \approx 10^{17}-10^{20}$ erg\,s$^{-1}$\,Hz$^{-1}$), as the hot accreting white dwarf ionizes the substantial circumstellar material expelled by the giant companion \citep{Seaquist_etal_1984, Seaquist_Taylor_1990,Mikolajewska_2012}. 

We showed earlier that if the radio emission is optically thick thermal emission from ionized gas at a typical brightness temperature of 10$^4$\,--\,10$^5$\,K, the emitting regions should have a radius of $\sim$ 10$^{14}$\,--\,10$^{15}$\,cm to produce the observed flux densities. 
If the radio emission observed from V1047~Cen proves to be transient and associated with the 2019 outburst, it might be attributed to outflows moving at velocities $\gtrsim$2000\,km\,s$^{-1}$ (based on the optical line profiles), which will cover a radius of more than 10$^{15}$\,cm in less than 2 months. The variability of the radio source can then be explained as variability in the fast outflow. 

\subsection{The nature of the outburst}

The spectroscopic follow up of V1047~Cen during the rise to peak (first 90 days) and the low-amplitude of the outburst are fairly consistent with a DN outburst; however, the 400 day long outburst, the 14 year gap between the 2005 nova eruption and the 2019 outburst, the luminous radio and optical emission, the dramatic spectral changes and high velocity spectral components are not consistent with a DN and raise questions about the nature of this event. In addition, the amplitude of the outburst ($\approx$ 6\,mag), the absolute peak magnitude, and the spectral evolution are not consistent with a classical nova event. Moreover, there is a striking difference between the 2005 classical nova and the 2019 outburst. All this suggests that the 2019 outburst of V1047~Cen is unique and is not consistent with common phenomena observed in CV systems, such as DNe and classical novae. A definitive explanation for what happened during the 2019 outburst has not yet arisen based on the current dataset, but we offer below several potential scenarios, while listing their pros and cons.

Many of the features observed in V1047~Cen resemble those of Z And-like classical symbiotic outbursts \citep{Sokoloski_etal_2006}. This type of events starts with a disk instability outburst, leading to the accretion of a massive accretion disk onto the white dwarf. The accretion burst is large enough to trigger enhanced nuclear burning on the surface of the white dwarf and the ejection of an optically thick outflow/shell (see also \citealt{Munari_2019}). 

While classical symbiotic systems are characterized by orbital periods of the order of hundreds of days (\citealt{Mikolajewska_2003}; much longer than the potential few hours orbital period in V1047~Cen), the observational features of classical symbiotic outbursts have similarities with those of V1047~Cen (the long-lasting optical outburst, the long-lasting and luminous radio emission, the P Cygni spectral features with velocities $\approx$ 1500--2000\,km\,s$^{-1}$, the high ionization lines in the optical spectra, and the late satellite spectral features around the Balmer lines indicative of bipolar flows).

Therefore, one of the possibilities is that the 2019 outburst of V1047~Cen is similar in nature to classical symbiotic outbursts, starting with a DN disk instability type outburst (based on the early optical spectra), leading to the accretion of the entire accretion disk onto the white dwarf. The accretion of the disk might have led to enhanced nuclear shell burning on the surface of the white dwarf, which launched a radiation driven wind/outflow \citep{Sokoloski_etal_2006}. This scenario could explain the 400 day outburst duration in the optical (e.g., the 2000's outburst of Z~And lasted for 2 years) and the dramatic brightening and broadening of the Balmer lines.
The light curve bump or re-brightening that occurred $\sim$250 days after the outburst start could mark the stage at which an optically thick outflow started. This is consistent with the development of P Cygni-like absorption features in the optical spectral lines around the same period (Figure~\ref{Fig:PCygni}). The absence of the P Cygni profiles in the spectra obtained on day 407 after the end of the optical outburst (Figure~\ref{Fig:main_spec}; note that the P Cygni profiles were present in the previous epoch on day 303) indicates that the optically thick wind/outflow 
dissipated rapidly and became optically thin, which is very similar to the behavior observed in the outburst of
Z~And \citep{Sokoloski_etal_2006}. The presence of high ionization lines of He, C, O, N, and potentially forbidden Fe in the late spectra of V1047~Cen (Figure~\ref{Fig:spec_7}), obtained after the end of the optical outburst could indicate ongoing nuclear shell burning, which would strengthen the case for this scenario, which transitions from an accretion-powered outburst to a nuclear-powered outburst. 
The same high ionization lines of \eal{N}{V} and \eal{O}{VI} are typically observed in V~Sge like stars (the Milky-Way counterparts of the supersoft sources in the Magellanic Clouds), and typically indicate nuclear shell burning (e.g., \citealt{Herbig_etal_1965,Cieslinski_etal_1999,Steiner_etal_1999,starrfield_etal_2004}). \citet{Sokoloski_etal_2006} derived the luminosity of Z~And a good part of the 2006 outburst to be of the order of 10$^{36}$\,erg\,s, which is close to the luminosity of V1047~Cen during the flat peak phase of the 2019 outburst.

This scenario could also explain the long-lasting, luminous radio emission. We showed in Section~\ref{radio_disc} that the flux densities observed in V1047~Cen could be explained as thermal emission from an outflow of ionized gas. Such radio emitting outflows/jets are characteristics of classical symbiotic outbursts (e.g, \citealt{Brocksopp_2004} and references therein). In addition, there is strong evidence for the presence of such collimated outflows or jets in V1047~Cen based on the satellite emission components observed in H$\alpha$ (Figure~\ref{Fig:jet_lines}), which are similar to those observed in Z And-like classical symbiotic outbursts (e.g., \citealt{Munari_etal_2005,Burmeister_etal_2007,Skopal_etal_2013,Tomov_Tomova_2013}). The bright radio emission in Z~And like systems, is mostly due to the presence of dense circumstellar material from the evolved secondary, ionized by the outburst on the white dwarf. However, for a system like V1047~Cen, which likely hosts a dwarf companion, we do not expect a dense circumstellar environment. Unless, the medium around the system has been enriched by material from a swelled companion after the 2005 nova, during a phase of enhanced mass-transfer, coupled with mass-loss from the system. While there is no explanation for why this happens in a system like V1047~Cen specifically, it is within our attempts to explain the uniqueness of this outburst.

If strong internal shocks are not present within the outflow, bright X-ray emission might not be present, because the X-ray emission associated with the nuclear-burning on a low mass WD (see Section~\ref{sec:2005--2009}) would be hidden by the interstellar column density along with the puffed up photosphere and outflow during the outburst. This could explain the lack of X-ray detection in the \textit{Swift} individual observations. 

The energy radiated during the 2019 outburst is likely powered by different mechanisms (e.g., disk brightening and nuclear burning), making it difficult to estimate (see Section~\ref{sec_2005vs2019}). However, if we assume that all ($\sim$ 10$^{44}$\,erg) the energy radiated is powered by nuclear burning, the mass needed to be accreted between 2005 and 2019 to power the 10$^{44}$\,erg would be around 1.6 $\times 10^{-8}$\,M$_{\odot}$ of hydrogen, assuming a 0.7\% nuclear burning efficiency \citep{Clayton_1968}. This amount of material could be accreted onto the WD with a moderate post-nova mass transfer rate ($\sim$ $10^{-9}$\,--$10^{-8}$\,M$_{\odot}$\,yr$^{-1}$.)

While a scenario similar to a classical symbiotic outburst could indeed explain some of the observational features during the 2019 outburst, this scenario suffers from some weaknesses.
\citet{Sokoloski_etal_2006} suggest that pre-existing quasi-steady shell burning on the WD surface is necessary to trigger a classical symbiotic nova. 
As mentioned earlier, the OGLE photometry between 2013 and 2019 
indicates that quasi-stable nuclear burning on the WD surface could indeed have been ongoing prior to the 2019 outburst, with a mass accretion rate of the order of $10^{-8}$\,M$_{\odot}$\,yr$^{-1}$, or remnant mass from the 2005 nova event. However, this mass accretion rate is high for a disk instability event to take place at the beginning of the 2019 outburst of V1047~Cen. While some systems like GK~Per and V1017~Sgr showed DN outbursts, just a few decades after a nova eruption when the accretion rate is expected to be of the order of $10^{-8}$\,M$_{\odot}$\,yr$^{-1}$, these systems are characterized by long orbital periods ($>$ 1\,d) and therefore potentially massive accretion disks, where a disk instability event could take place despite the high accretion rate. Similarly, Z And-like systems are characterized by high mass accretion rates ($\sim10^{-8}$\,M$_{\odot}$\,yr$^{-1}$) in long orbital period systems, harboring massive disks. However, if V1047~Cen is characterized by an orbital period of less than 10 hours, a disk instability would be more difficult to take place at this high mass accretion rate.

Moreover, the potential ongoing nuclear burning on the WD surface will heat up the disk to temperatures where a disk instability is not possible to occur. If the system brightness ($M_V \approx 1.9$\,mag) prior to the 2019 outburst is mostly contributed by the nuclear burning irradiating the disk, this means that the disk temperature could reach over 10,000\,K, implying that disk instabilities are not possible. Therefore, despite the early spectra's consistency with a DN outburst (Balmer and \eal{He}{I} lines with FWHM of a few 100\,km\,s$^{-1}$ and strong \eal{He}{II} at 4868$\mathrm{\AA}$), a disk instability may not be the trigger of the 2019 outburst. In addition, the optical light curve of V1047~Cen is of higher-amplitude compared to typical light curves of classical symbiotic outbursts. For example, the 2006 outburst of Z~And reached an amplitude of 2--3\,mag compared to more than 6 magnitudes in V1047~Cen. However, it is worth noting that the amplitude of the outburst is not only related to the outburst radiated energy, but also the magnitude of the companion star. Since the companion star in Z~And is a bright, giant star, the amplitudes of the outbursts in such systems are expect to be low. 

If the distance and reddening estimate towards V1047~Cen are larger than the ones we derive, this would imply that the companion star is an evolved one (e.g., a sub-giant; see Section~\ref{sec:secondary}). If so, this would place V1047~Cen in the same category of GK~Per and V1017~Sgr, i.e., systems characterized by a large disk and an orbital period longer than 1-2\,days. This implies that a disk instability could still take place despite a high mass transfer rate. Moreover, even if the inner parts of the disk are characterized by temperatures in excess of 10$^4$\,k, the outer parts of the large disk could be cold enough for instability to happen.

Based on the early spectral features, a disk brightening likely took place during the early weeks of the outburst. If this did not happen due to an instability in the disk, an alternative possibility is that the disk brightened due to enhanced mass transfer rate from the companion. \citet{Bollimpalli_etal_2018} suggested that for the 2000-2002 outburst of Z And, a disk instability could not be the trigger of the outburst. This is mainly due to the irradiation of the disk by the hot WD, implying that a dwarf nova would only be significant at very high mass transfer rates (10$^{-6}$\,M$_{\odot}$\,yr$^{-1}$), which is higher than the typical mass transfer rate in such systems. \citet{Bollimpalli_etal_2018} suggest that the outburst of Z And between 2000 and 2002 might have been triggered by a mass-transfer enhancement from the giant companion, leading to an increase in nuclear burning on the WD surface. It could be that something similar took place during the early stages of the 2019 outburst of V1047~Cen. However, based on the scolors of V1047~Cen prior to 2019, there is evidence for ongoing nuclear burning on the WD surface, likely caused by post-nova enhanced mass transfer rate. Therefore, we do not have a definitive explanation for what could lead to further enhancing the mass transfer rate from the secondary into the surface of the WD.

One other possible explanation of the 2019 outburst is that the system experienced a non-ejection nova event. 
The extended grid of nova models by \citet{Yaron_etal_2005} shows that some combinations of parameters might lead to a thermonuclear runaway without ejecting material 
(see also \citealt{Fujimoto_1982,Shara_etal_1977}). \citet{Yaron_etal_2005} suggest that such outbursts cause only a slow increase in luminosity, 
followed by a slow decay, particularly in the case of low mass WDs where the timescales of the rise and decay could be thousands of days.
The supersoft X-ray transient ASASSN-16oh has been suggested to be a thermonuclear runaway event without mass ejection \citep{Hillman_etal_2019}. The system showed a slowly rising (1585 days) and declining (268 days) optical light curve, with an amplitude of less than 4 magnitude. While there are similarities between the optical light curves of ASASSN-16oh and V1047~Cen, the former has been detected as a supersoft X-ray source at the distance of the Small Magellanic Cloud. This is very different from the case of V1047~Cen, which was not detected in individual epochs by \textit{Swift} during the 2019 outburst despite our extensive monitoring. 
In the case of a thermonuclear runaway, the luminosity of the supersoft source is expected to be of the order of 10$^{38}$\,erg\,s$^{-1}$, which would have been easily observed at a distance of 3.2\,kpc by \textit{Swift}, if the whitw dwarf is of higher mass ($>1$M$_{\odot}$). The supersoft emission of ASASSN-16oh was characterized by a luminosity of around 10$^{37}$\,erg\,s$^{-1}$, which is lower than that expected 
from the models of non-ejecting thermonuclear runaways. This low luminosity has been attributed to an optically thick accretion disk hiding most of the WD surface. While a massive accretion disk could also be blocking some of the emission from the WD surface in case of V1047~Cen, it is a less likely possibility given the \textit{Swift} non-individual-detections over several months. Moreover, the optical spectral evolution is consistent with the presence of an optically thick ejection/outflow. Therefore, a non-ejection nova scenario is less likely.

In conclusion, the 2019 outburst of V1047~Cen does not resemble the common events that take place in CV systems (e.g., DNe and classical novae). Based on the multi-wavelength observational features, the outburst is likely the combination of multiple mechanisms, starting with a disk brightening, followed by the generation of an outflow. Such a unique outburst has never before been observed in a CV system that has experienced a recent classical nova eruption, indicating the possibility that we have witnessed a new astronomical phenomenon.

\section{Conclusions}
\label{Conc}
We have presented multi-wavelength observations of the 2019 outburst in the 2005 classical nova V1047~Cen. The outburst amplitude reached around 6 magnitudes in the optical and lasted for more than 400 days. We derive a distance of 3.2$\pm$0.2\,kpc to the system and a peak absolute magnitude of the outburst $M_V = -2$, placing it between novae and dwarf novae. The first spectra we obtained of the system around 74 days after the start of the outburst were consistent with a disk instability dwarf nova event in a classical nova system. If V1047 Cen is a DN, the event would be a record breaker, making it the longest DN outburst on record, the shortest gap between a nova and DN in a CV system, and the most luminous optical and radio DN outburst. 
This, along with the different observational features across the electromagnetic spectrum point towards a phenomenon more exotic than just a DN outburst.

We therefore suggest that the event is a combination of multiple mechanisms, starting with a brightening in the disk (due to enhanced mass-transfer or less likely an instability in the disk), which then triggered enhanced nuclear shell burning on the surface of the white dwarf and eventually led to an optically thick, radiation driven wind/outflow. This scenario fits well the 400 day outburst duration, the dramatic changes in the optical line profiles and the $>$ 2000\,km\,s$^{-1}$ velocities inferred from these profiles, the P Cygni line profiles which appeared several months after the outburst start, the high peak optical brightness, and most importantly the long-lasting, superluminous radio emission, which likely originates from ionized gas (thermal emission) in collimated bipolar flows, which are characteristic of classical symbiotic outbursts. Strong evidence for such outflows can be seen in the late optical spectral line profiles. Mid-infrared observations also indicate that pre-existing dust --- likely formed during the 2005 classical nova eruption --- has been heated by the radiation from the 2019 outburst.

Therefore, the 2019 outburst of V1047~Cen  is a unique phenomenon observed for the first time in a typical CV system, characterized by a short ($\sim$ hours) orbital period and which has undergone a recent classical nova eruption.

\section*{Acknowledgments}

We thank B. Schaefer for useful discussion. We thank the AAVSO observers from around the world who contributed their magnitude measurements to the AAVSO International Database used in this work.

Support for this work was provided by NASA through the NASA Hubble Fellowship grant HST-HF2-51501.001-A awarded by the Space Telescope Science Institute, which is operated by the Association of Universities for Research in Astronomy, Inc., for NASA, under contract NAS5-26555. EA, LC, and KVS acknowledge NSF award AST-1751874, NASA award 11-Fermi 80NSSC18K1746, and a Cottrell fellowship of the Research Corporation. JS was supported by the Packard Foundation. DAHB gratefully acknowledges the receipt of research grants from the National Research Foundation (NRF) of South Africa. PAW kindly acknowledges the National Research Foundation and the University of Cape Town. KLP acknowledges funding from the UK Space Agency. Nova research at Stony Brook University has been made possible by NSF award AST-1611443. MG is supported by the EU Horizon 2020 research and innovation programme under grant agreement No 101004719.
A part of this work is based on observations made with the Southern African Large Telescope (SALT), with the Large Science Programme on transients 2018-2-LSP-001 (PI: DAHB). Polish participation in SALT is funded by grant no. MNiSW DIR/WK/2016/07. This paper was partially based on observations obtained at the Southern Astrophysical Research (SOAR) telescope, which is a joint project of the Minist\'{e}rio da Ci\^{e}ncia, Tecnologia e Inova\c{c}\~{o}es (MCTI/LNA) do Brasil, the US National Science Foundation's NOIRLab, the University of North Carolina at Chapel Hill (UNC), and Michigan State University (MSU).
The OGLE project has received funding from the National Science Centre, Poland, grant MAESTRO 2014/14/A/ST9/00121 to AU. 
The CHIRON and $Andicam$ instruments are managed by the Todd Henry and the SMARTS Consortium. The MeerKAT telescope is operated by the South African Radio Astronomy Observatory, which is a facility
of the National Research Foundation, an agency of the Department of Science and Innovation. We acknowledge the use of the ilifu cloud computing facility - \url{www.ilifu.ac.za}, a partnership between the University of Cape Town, the University of the Western Cape, the University of Stellenbosch, Sol Plaatje University, the Cape Peninsula University of Technology and the South African Radio Astronomy Observatory. The ilifu facility is supported by contributions from the Inter-University Institute for Data Intensive Astronomy (IDIA - a partnership between the University of Cape Town, the University of Pretoria and the University of the Western Cape), the Computational Biology division at UCT and the Data Intensive Research Initiative of South Africa (DIRISA).
This research is based in part on observations obtained at the international Gemini Observatory, a program of NSF's NOIRLab, which is managed by the Association of Universities for Research in Astronomy (AURA) under a cooperative agreement with the National Science Foundation. on behalf of the Gemini Observatory partnership: the National Science Foundation (United States), National Research Council (Canada), Agencia Nacional de Investigaci\'{o}n y Desarrollo (Chile), Ministerio de Ciencia, Tecnolog\'{i}a e Innovaci\'{o}n (Argentina), Minist\'{e}rio da Ci\^{e}ncia, Tecnologia, Inova\c{c}\~{o}es e Comunica\c{c}\~{o}es (Brazil), and Korea Astronomy and Space Science Institute (Republic of Korea). DPKB is supported by a CSIR Emeritus Scientist grant-in-aid, which is being hosted by the Physical Research Laboratory, Ahmedabad.
This work is based in part on observations made with the 
NASA/DLR Stratospheric Observatory for Infrared Astronomy (SOFIA). SOFIA is jointly operated by the Universities Space Re- search Association, Inc. (USRA), under NASA contract NNA17BF53C, and the Deutsches SOFIA Institut (DSI) under DLR contract 50 OK 0901 to the University of Stuttgart. Financial support for CEW/RDG 
related to this work was provided by NASA through award SOF07-0005 issued by USRA to the University of Minnesota. This publication makes use of data products from the Near-Earth Object Wide-field Infrared Survey Explorer (NEOWISE), which is a joint project of the Jet Propulsion Laboratory/California Institute of Technology and the University of Arizona. NEOWISE is funded by the National Aeronautics and Space Administration. VARMR acknowledges financial support from the Funda\c{c}\~{a}o para a Ci\^encia e a Tecnologia (FCT) in the form of an exploratory project of reference IF/00498/2015/CP1302/CT0001, and from the Minist\'erio da Ci\^encia, Tecnologia e Ensino Superior (MCTES) through national funds and when applicable co-funded EU funds under the project UIDB/EEA/50008/2020, and supported by Enabling Green E-science for the Square Kilometre Array Research Infrastructure (ENGAGE-SKA), POCI-01-0145- FEDER-022217, and PHOBOS, POCI-01-0145-FEDER029932, funded by Programa Operacional Competitividade e Internacionaliza\c{c}\~ao (COMPETE 2020) and FCT, Portugal. Analysis made significant use of \textsc{python} 3.7.4, and the associated packages \textsc{numpy}, \textsc{matplotlib},, \textsc{seaborn}, \textsc{scipy}. 
Data reduction made significant use of \software{MIDAS FEROS \citep{Stahl_etal_1999}, echelle \citep{Ballester_1992}, PySALT \citep{Crawford_etal_2010}, IRAF \citep{Tody_1986,Tody_1993}, ELEANOR \citep{Feinstein_etal_2019}, OGLE pipeline \citep{Udalski_etal_2015}}.

\bibliography{biblio}
\bibliographystyle{aasjournal}



\appendix
 
\renewcommand\thetable{\thesection.\arabic{table}}    
\renewcommand\thefigure{\thesection.\arabic{figure}}   
\setcounter{figure}{0}

\section{Supplementary plots and tables}
\label{appB}
In this Appendix we present supplementary plots and tables.

\begin{figure*}
\begin{center}
  \includegraphics[width=0.95\textwidth]{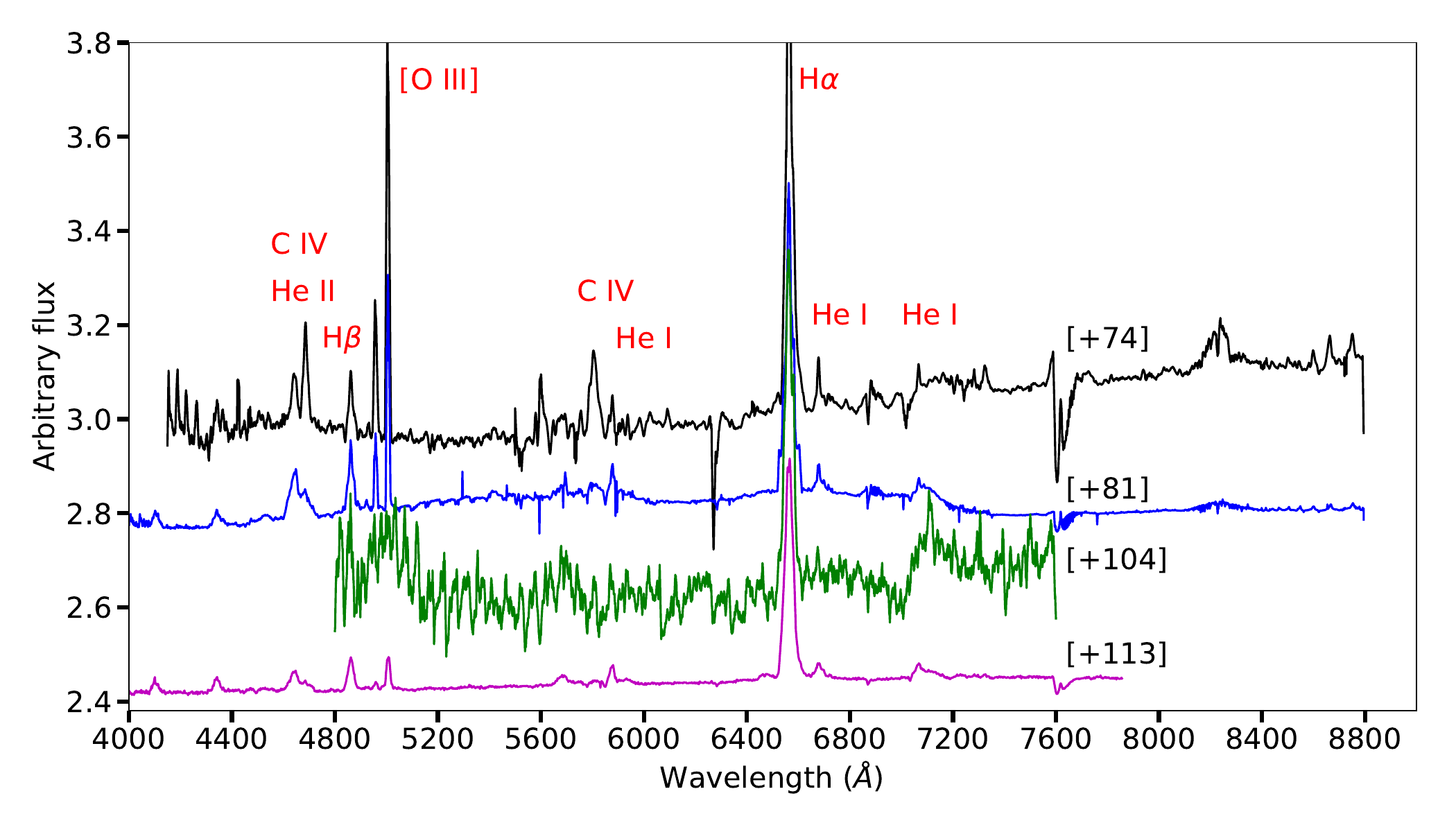}
\caption{The optical spectral evolution of V1047~Cen. The numbers in brackets are days since $t_0$.} 
\label{Fig:spec_1}
\end{center}
\end{figure*}

\begin{figure*}
\begin{center}
  \includegraphics[width=0.95\textwidth]{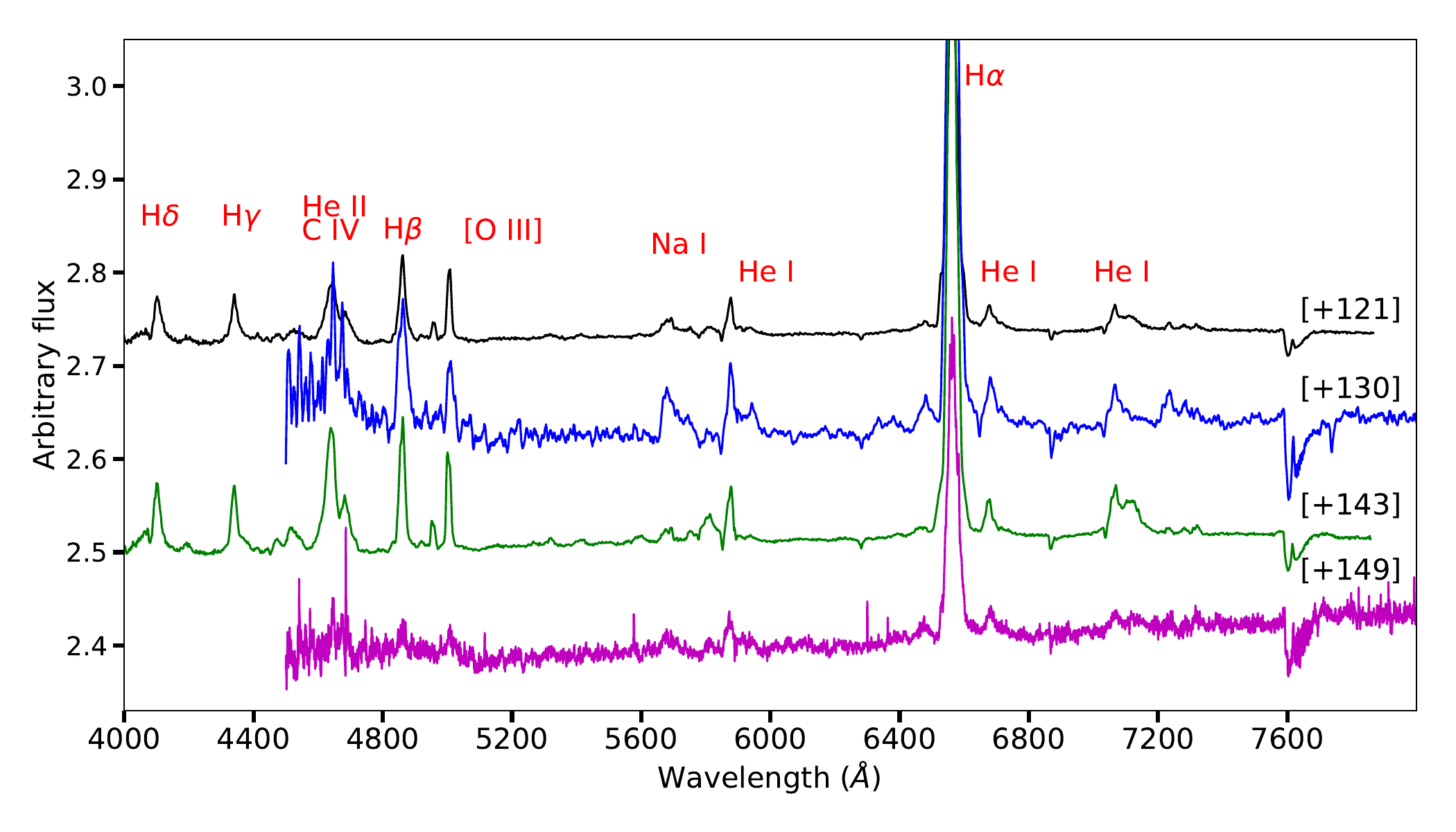}
\caption{The optical spectral evolution of V1047~Cen. The numbers in brackets are days since $t_0$.} 
\label{Fig:spec_2}
\end{center}
\end{figure*}

\begin{figure*}
\begin{center}
  \includegraphics[width=0.95\textwidth]{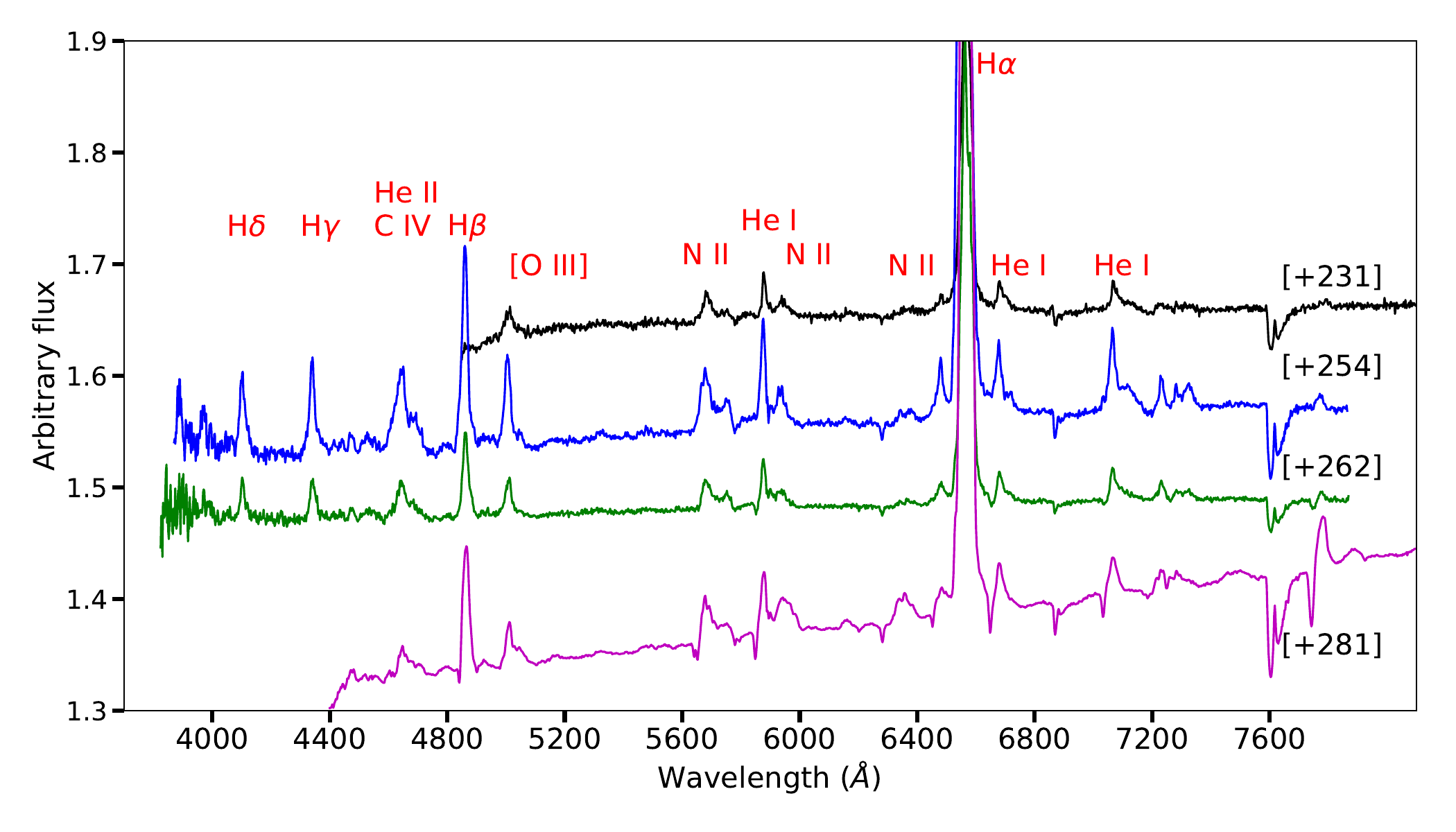}
\caption{The optical spectral evolution of V1047~Cen. The numbers in brackets are days since $t_0$.} 
\label{Fig:spec_3}
\end{center}
\end{figure*}

\begin{figure*}
\begin{center}
  \includegraphics[width=0.95\textwidth]{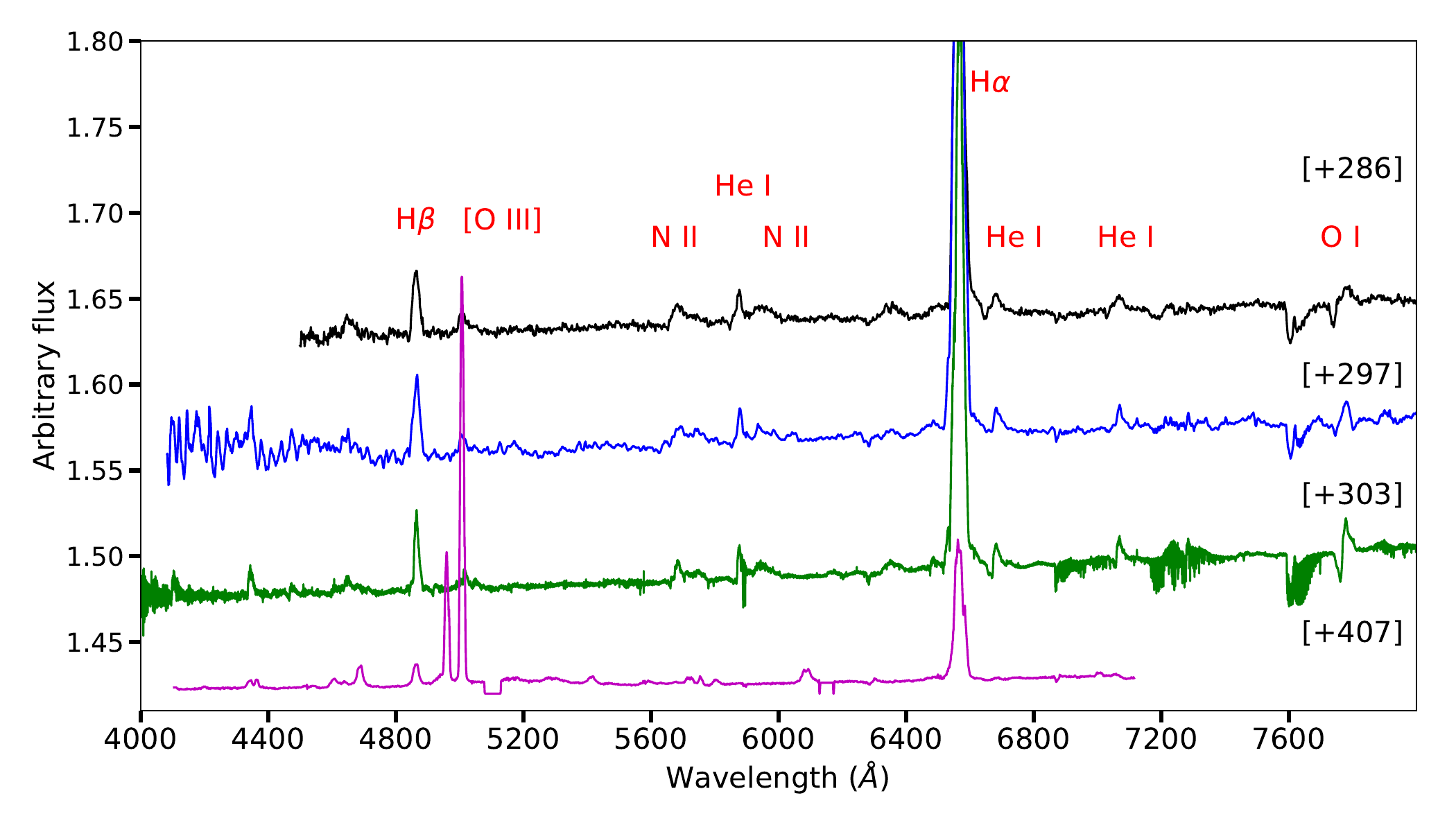}
\caption{The optical spectral evolution of V1047~Cen. The numbers in brackets are days since $t_0$.} 
\label{Fig:spec_4}
\end{center}
\end{figure*}

\begin{figure*}
\begin{center}
  \includegraphics[width=0.95\textwidth]{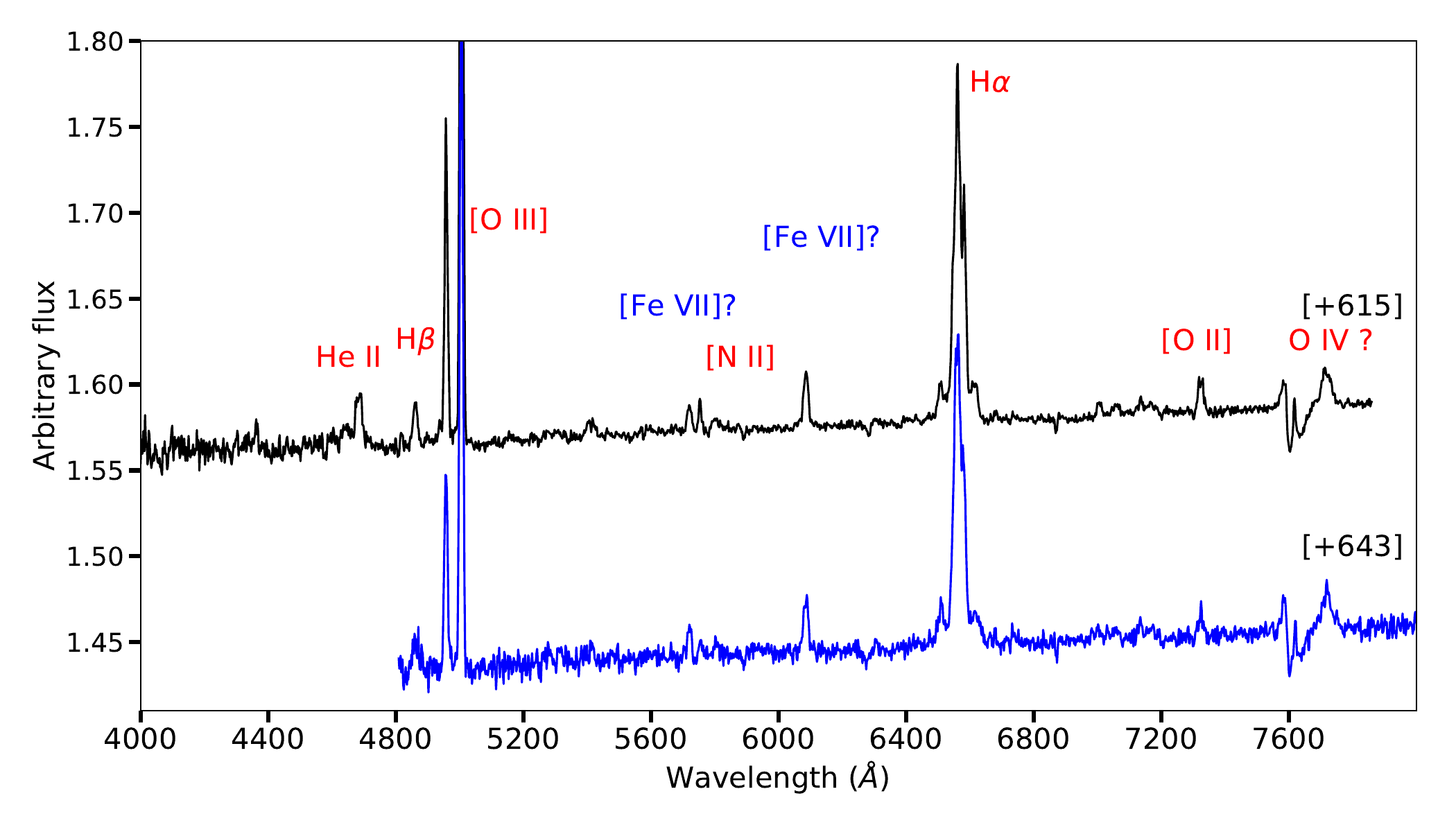}
\caption{The optical spectral evolution of V1047~Cen. The numbers in brackets are days since $t_0$.} 
\label{Fig:spec_5}
\end{center}
\end{figure*}

\begin{figure*}
\begin{center}
  \includegraphics[width=0.95\textwidth]{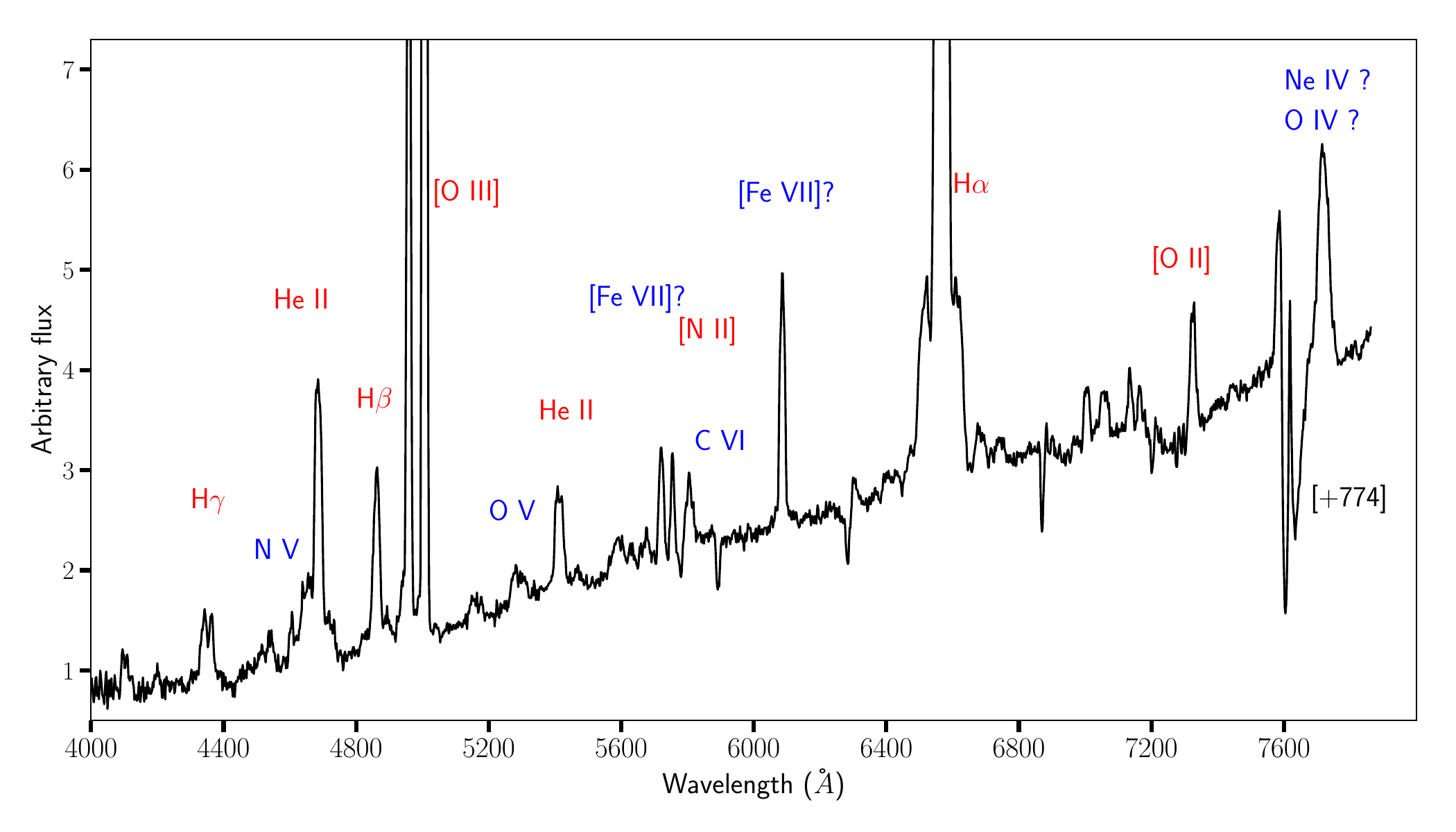}
\caption{The optical spectral evolution of V1047~Cen. The numbers in brackets are days since $t_0$.} 
\label{Fig:spec_7}
\end{center}
\end{figure*}

\begin{figure*}
\begin{center}
  \includegraphics[width=0.95\textwidth]{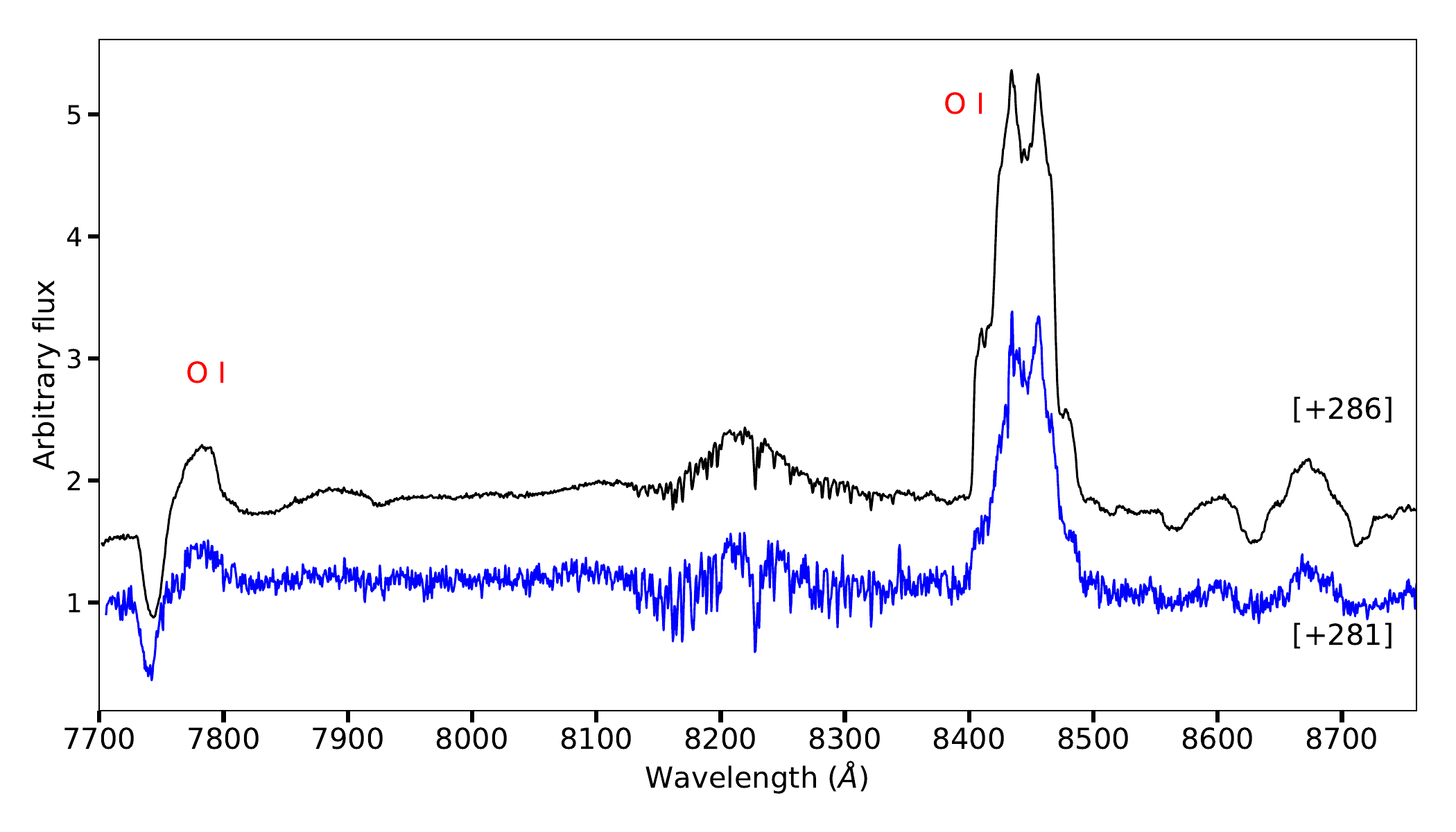}
\caption{The optical spectral evolution of V1047~Cen. The numbers in brackets are days since $t_0$.} 
\label{Fig:spec_6}
\end{center}
\end{figure*}

\begin{figure*}
\begin{center}
  \includegraphics[width=0.7\textwidth]{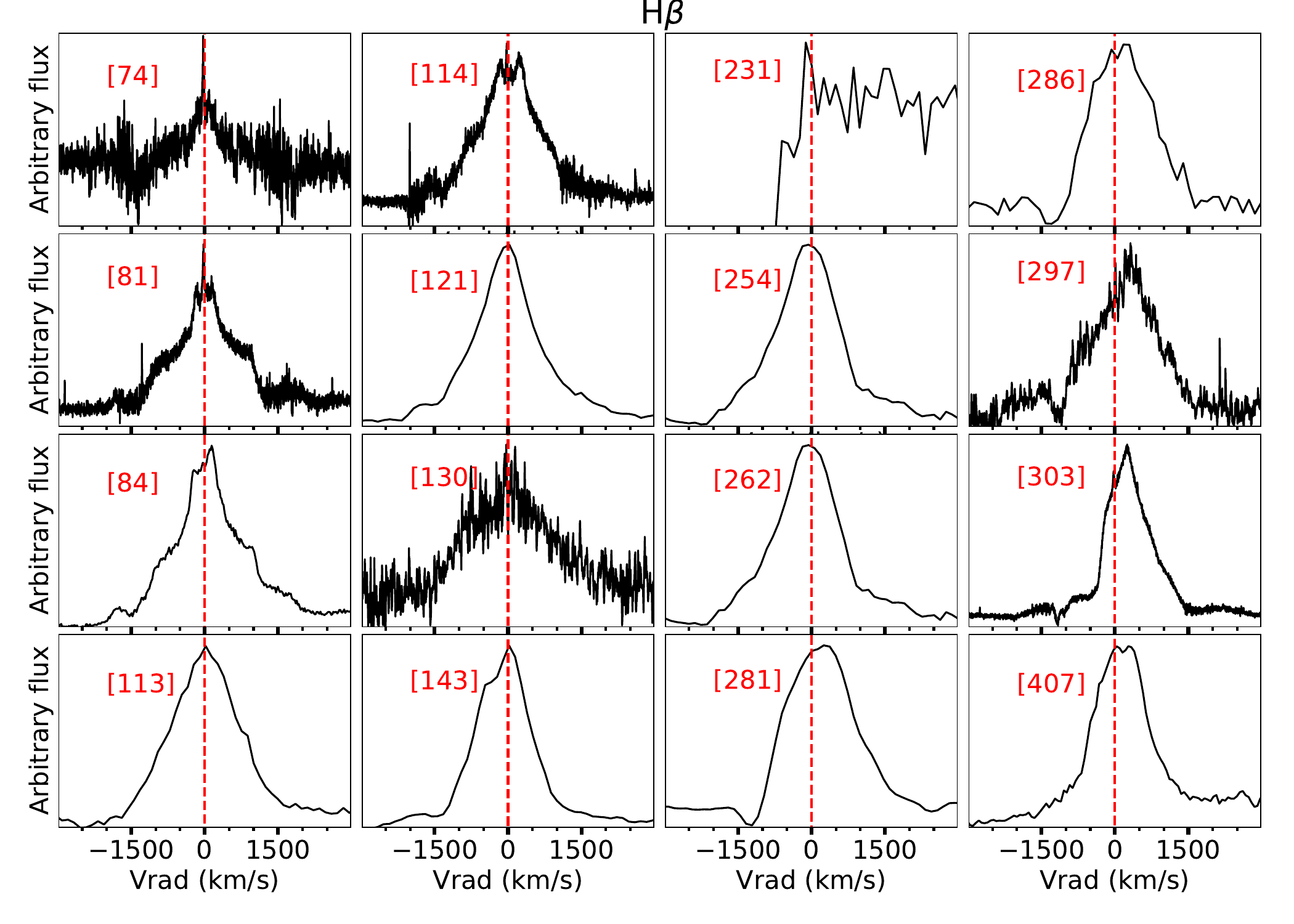}
\caption{The line profile evolution of H$\beta$ throughout the outburst of V1047~Cen. The numbers between brackets are days after outburst. The red dashed lines represents the rest velocity ($V_{\mathrm{vrad}}$ = 0\,km\,s$^{-1}$). A heliocentric correction is applied to the radial velocities.} 
\label{Fig:beta}
\end{center}
\end{figure*}



\begin{table*}
\centering
\caption{Log of the OGLE optical photometry.}
\begin{tabular}{lccc}
\hline
JD & Magnitude & Err. & Band \\
\hline
2456437.64553 & 17.190 & 0.037 & \textit{I}\\
2456438.65480 & 17.325 & 0.040 & \textit{I}\\
2456465.62998 & 17.052 & 0.028 & \textit{I}\\
2456467.65411 & 17.075 & 0.033 & \textit{I}\\
2456478.47615 & 17.156 & 0.025 & \textit{I}\\
2456494.56288 & 17.111 & 0.043 & \textit{I}\\
2456500.50861 & 17.094 & 0.032 & \textit{I}\\
2456522.47809 & 17.061 & 0.036 & \textit{I}\\
2456693.83993 & 16.963 & 0.017 & \textit{I}\\
2456699.82818 & 17.020 & 0.020 & \textit{I}\\
2456702.84838 & 17.012 & 0.021 & \textit{I}\\
2456705.81657 & 16.981 & 0.027 & \textit{I}\\
2456713.85601 & 17.076 & 0.021 & \textit{I}\\
\hline
\end{tabular}
\label{table:OGLE_log}
\end{table*}

\begin{table*}
\centering
\caption{Log of the AAVSO optical photometry.}
\begin{tabular}{lccc}
\hline
JD & Magnitude & Err. & Band \\
\hline
2458655.4927 & 15.488 & 0.032 & \textit{V}\\
2458655.4938 & 15.503 & 0.026 & \textit{V}\\
2458655.4947 & 13.807 & 0.033 & \textit{I}\\
2458655.4954 & 13.846 & 0.038 & \textit{I}\\
2458655.4962 & 14.672 & 0.029 & \textit{R}\\
2458655.4970 & 14.636 & 0.030 & \textit{R}\\
2458656.4909 & 15.472 & 0.049 & \textit{V}\\
2458656.4911 & 16.674 & 0.065 & \textit{B}\\
2458656.4920 & 15.475 & 0.036 & \textit{V}\\
2458656.4929 & 13.718 & 0.051 & \textit{I}\\
2458656.4936 & 13.709 & 0.040 & \textit{I}\\
2458656.4945 & 14.542 & 0.032 & \textit{R}\\
2458656.4952 & 14.531 & 0.030 & \textit{R}\\
2458656.4963 & 16.497 & 0.056 & \textit{B}\\
2458656.4977 & 16.468 & 0.058 & \textit{B}\\
\hline
\end{tabular}
\label{table:AAVSO_log}
\end{table*}

\begin{table*}
\centering
\caption{Log of the SMARTS NIR photometry.}
\begin{tabular}{lccc}
\hline
JD & Magnitude & Err. & Band \\
\hline
2458656.49902 & 12.265 & 0.016 & \textit{H}\\
2458656.50000 & 11.864 & 0.018 & \textit{K}\\
2458656.50000 & 13.014 & 0.025 & \textit{J}\\
2458657.51172 & 11.465 & 0.019 & \textit{K}\\
2458657.51172 & 11.967 & 0.013 & \textit{H}\\
2458657.51172 & 12.682 & 0.027 & \textit{J}\\
2458657.51562 & 12.057 & 0.012 & \textit{H}\\
2458657.51660 & 11.602 & 0.018 & \textit{K}\\
2458657.51660 & 12.735 & 0.025 & \textit{J}\\
2458660.50195 & 11.327 & 0.018 & \textit{K}\\
\hline
\end{tabular}
\label{table:SMARTS_log}
\end{table*}

\begin{table*}
\centering
\caption{Log of the NEOWISE NIR photometry.}
\begin{tabular}{lccc}
\hline
JD & Magnitude & Err. & Band \\
\hline
2458685.13 & 9.224 & 0.092 & \textit{W1}\\
2458685.13 & 8.382 & 0.083 & \textit{W2}\\
2458887.15 & 8.368 & 0.083 & \textit{W1}\\
2458887.15 & 7.741 & 0.074 & \textit{W2}\\
2458894.82 & 8.585 & 0.085 & \textit{W1}\\
2458894.82 & 7.998 & 0.079 & \textit{W2}\\
2459049.28 & 12.119 & 0.121 & \textit{W1}\\
2459049.28 & 10.737 & 0.107 & \textit{W2}\\
\hline
\end{tabular}
\label{table:NEOWISE_log}
\end{table*}

\begin{table*}
\centering
\caption{Optical spectroscopic observations log.}
\begin{tabular}{cccccc}
\hline
 Telescope & Instrument & date & $ t - t_0$ & Resolving power & $\lambda$ Range \\ 
&  & &(days) &  & ($\mathrm{\AA}$)\\
\hline
SALT & HRS & 2019-06-19 & 74 & 14000 & 4000\,--\,9000\\
SALT & HRS & 2019-06-26 & 81 & 14000 & 4000\,--\,9000\\
SOAR & Goodman & 2019-06-29 & 84 & 5000 & 4500\,--\,5170\\
SOAR & Goodman & 2019-06-29 & 84 & 5000 & 6130\,--\,6710\\
SMARTS & CHIRON & 2019-07-17 & 104 & 28000 & 4080\,--\,8900\\
SOAR & Goodman & 2019-07-28 & 113 & 1000 & 3800\,--\,7800\\
SOAR & Goodman & 2019-07-28 & 113 & 5000 & 6130\,--\,6710\\
SALT & HRS & 2019-07-29 & 114 & 14000 & 4000\,--\,9000\\
SOAR & Goodman & 2019-08-05 & 121 & 1000 & 3800\,--\,7800\\
SOAR & Goodman & 2019-08-05 & 121 & 5000 & 6130\,--\,6710\\
SMARTS & CHIRON & 2019-08-14 & 130 & 28000 & 4080\,--\,8900\\
SOAR & Goodman & 2019-08-27 & 143 & 1000 & 3800\,--\,7800\\
SOAR & Goodman & 2019-08-27 & 143 & 5000 & 6130\,--\,6710\\
SMARTS & CHIRON & 2019-09-02 & 149 & 28000 & 4080\,--\,8900\\
SOAR & Goodman & 2019-11-23 & 231 & 1000 & 3800\,--\,7800\\
SOAR & Goodman & 2019-11-23 & 231 & 5000 & 6130\,--\,6710\\
SOAR & Goodman & 2019-12-16 & 254 & 1000 & 3800\,--\,7800\\
SOAR & Goodman & 2019-12-16 & 254 & 5000 & 6130\,--\,6710\\
SOAR & Goodman & 2019-12-24 & 262 & 1000 & 3800\,--\,7800\\
SOAR & Goodman & 2019-12-24 & 262 & 5000 & 6130\,--\,6710\\
SOAR & Goodman & 2020-01-12 & 281 & 1000 & 3800\,--\,7800\\
SOAR & Goodman & 2020-01-12 & 281 & 2800 & 7700\,--\,8870\\
SOAR & Goodman & 2020-01-12 & 281 & 5000 & 6130\,--\,6710\\
SOAR & Goodman & 2020-01-17 & 286 & 1000 & 3800\,--\,7800\\
SOAR & Goodman & 2020-01-17 & 286 & 2800 & 7700\,--\,8870\\
SOAR & Goodman & 2020-01-17 & 286 & 5000 & 6130\,--\,6710\\
SMARTS & CHIRON & 2020-01-28 & 297 & 28000 & 4080\,--\,8900\\
SALT & HRS & 2020-02-03 & 303 & 14000 & 4000\,--\,9000\\
SALT & RSS & 2020-05-17 & 407 & 1500 & 4100\,--\,7200\\
SOAR & Goodman & 2020-11-12 & 615 & 1000 & 3800\,--\,7800\\
SOAR & Goodman & 2021-01-08 & 643 & 1000 & 3800\,--\,7800\\
\hline
\end{tabular}
\label{table:spec_log}
\end{table*}

\begin{table*}
\centering
\caption{Gemini South / Flamingos 2 Observations of V1047~Cen.}
\begin{tabular}{ccccccc}
\hline
Date & $t-t_0$& Wavelength & $R_{\mathrm{max}}$ & Integration  & airmass (mean) & standard star \\
 & (days) & ($\mu$m) & $\lambda$/$\Delta \lambda$ & (s) & & \\
\hline
2020-02-23 & 323 & 1.17\,--\,1.30 & 3,000 & 1200 & 1.64 & (a)\\
2020-02-23 & 323 & 1.49\,--\,1.78 & 3,000 & 1200 & 1.52 & HIP63036\\ 
2020-02-23 & 323 & 1.95\,--\,2.45 & 3,000 & 1200 & 1.42 & HIP63036\\
2020-03-01 & 330 & 0.89\,--\,1.75 & 1,200 & 120 & 1.63 & HIP63036\\
\hline
\end{tabular}
\tablenotetext{a}{Spectra of HIP63036 and HIP67360 combined to match airmass of V1047 Cen.}
\label{table:IR_spec}
\end{table*}

\begin{table*}
\centering
\caption{Log of MeerKAT radio observations at 1.28 GHz.}
\begin{tabular}{lccccc}
\hline
Date & MJD & ($t-t_0$) & Peak flux  & Int. flux\tablenotemark{a} & $\alpha$ \\
 &  &  (days) & (mJy beam$^{-1}$) & (mJy) & \\
\hline
2019-11-26 & 58813.18 & 234.07 & $0.93\pm0.06$ & $0.95\pm0.08$ & $0.1\pm0.1$ \\
2019-12-15 & 58832.40 & 253.29  & $0.88\pm0.06$ & $0.92\pm0.08$ & $0.0\pm0.1$ \\
2019-12-20 & 58837.44 & 258.33 & $0.89\pm0.07$ & $0.9\pm0.1$ & $0.5\pm0.2$ \\
2019-12-28 & 58845.28 & 266.17 & $0.93\pm0.06$ & $0.90\pm0.08$ & $1.0\pm0.2$ \\
2020-01-03 & 58851.41 & 272.30 & $0.81\pm 0.05$ & $0.88\pm0.08$ & $-$ \\
2020-01-10 & 58858.42 & 279.31 & $0.78\pm 0.05$ & $0.89\pm0.08$ & $0.8\pm0.5$ \\
2020-01-20 & 58868.21 & 289.10 & $0.87\pm 0.06$ & $0.78\pm0.07$ & $0.2\pm0.3$ \\
2020-01-25 & 58873.18 & 294.07  & $0.80\pm 0.04$  & $0.91\pm0.06$ & $0.2\pm0.2$ \\
2020-02-01 & 58880.29 & 301.18  & $0.89\pm 0.05$ & $0.86\pm0.07$ & $-$ \\
2020-02-08 & 58887.20 & 308.09 & $0.84\pm0.05$ & $0.82\pm0.07$& $0.6\pm0.3$ \\
2020-02-15 & 58894.16 & 315.05 & $0.76\pm0.05$ & $0.83\pm0.07$ & $0.1\pm0.2$ \\
2020-02-21 & 58900.25 & 321.14 & $0.81\pm0.05$ & $0.79\pm0.08$ & $1.5\pm0.2$ \\
2020-03-02 & 58910.16 & 331.05 & $0.75\pm0.04$ & $0.93\pm0.06$ & $-0.2\pm0.2$ \\
2021-03-06 & 59279.20 & 700.09 & $0.91\pm0.05$ & $0.91\pm0.07$ & $-0.09\pm0.1$\\ 
\hline
\end{tabular}
\tablenotetext{a}{Integrated flux at 1.28 GHz, calculated by fitting a Gaussian to V1047~Cen and integrating the Gaussian's flux. Spectral indices were derived by imaging in eight frequency channels, fitting a point source in each sub-band image, and fitting the resulting fluxes with a power-law. We assume a 10\% flux calibration error for the flux in each sub-band, while a 5\% flux calibration error is assumed for the full band flux measurement.}
\label{table:radio_log}
\end{table*}

\begin{table*}
\centering
\caption{Log of the \textit{Swift} UVOT photometry.}
\begin{tabular}{lccc}
\hline
JD & Magnitude & Err. & Band \\
\hline
2458659.1322 & 17.533 & 0.084 & \textit{uvw1}\\
2458688.9402 & 17.525 & 0.080 & \textit{uvw1}\\
2458703.2580 & 16.927 & 0.068 & \textit{uvw1}\\
2458659.1284 & 19.492 & 0.281 & \textit{uvm2}\\
2458703.2548 & 18.816 & 0.188 & \textit{uvm2}\\
2458709.9321 & 18.564 & 0.159 & \textit{uvm2}\\
2458718.3840 & 18.099 & 0.179 & \textit{uvw2}\\
2458722.2083 & 18.415 & 0.122 & \textit{uvw2}\\
2458731.7276 & 18.178 & 0.126 & \textit{uvw2}\\
\hline
\end{tabular}
\label{table:Swift_log}
\end{table*}

\begin{deluxetable*}{lcccccc}
\tabletypesize{\small}
\tablecolumns{10}
\tablewidth{0pt}
\tablecaption{Classical and Dwarf Novae in same System \label{table:cn_results}}
\tablehead{
\colhead{Name} & 
\colhead{RAJ2000} & 
\colhead{DECJ2000} & 
\colhead{CN Epoch} &
\colhead{CN peak} & 
\colhead{DN Epoch} & 
\colhead{DN Peak}\\
\colhead{ } & \colhead{h~m~s} & 
\colhead{$^\circ$~$'$~$"$} & 
\colhead{date} &
\colhead{mag} & 
\colhead{date} & 
\colhead{mag}}
\startdata
GK Per & 03 31 12.01 & +43 54 15.5 & 2415439/1901-02-23 & $V$ = 0.2 & 1981 & 10.2 \\
V392 Per & 04 43 21.37 & +47 21 25.9 & 2458238/2018-04-29 & $V$ =6.3 & 1999-04 & 13 \\
GI Mon & 07 26 47.11 & $-$06 40 29.7 & 1918 & $V$ = 5.2 &\\
AT Cnc & 08 28 36.93 & +25 20 03.0 & 1645 & ? & 2005 & 12.5 \\
V1213 Cen & 13 31 15.81 & $-$63 57 38.0 & 2009-05-08 & $V$ = 8.5 & 2003 & $I$ = 19.5 \\
X Ser & 16 19 17.69 & $-$02 29 29.5 & 1903-03-26 & 8.9 & 2016-09-03 & 14 \\
Nova Sco 1437 & 17 01 28.15 & $-$43 06 12.3 & 1437 & ?& 2017-05-27  & $V$ = 13.2\\
V2109 Oph & 17 24 16.04 & $-$24 36 50.0 & 1969-06-08 & $B$ = 8.9 & 2021-08-24 & $G$ = 16.5\\
V908 Oph & 17 28 04.58 & $-$27 43 04.4 & 2434925/1954-07-01 & $<$ 9 & OGLE-IV $>$2010 & $I$ =18.2 \\
V728 Sco & 17 39 13.24 & -45 28 45.7 & 1862 & $V$ = 5.0 & 2013-2015 & $V$ $\approx$ 17\\
V1017 Sgr & 18 32 04.47 & $-$29 23 12.6 & 1919& $B$ = 6.4 & 1973 & 10.5 \\
V446 Her & 18 57 21.60  & $+$13 14 29.0 &  1960-03-03 & 2.8 & 1994 & 15.8 \\
V606 Aql & 19 20 24.29 & $-$00 08 07.8 &  1899-04-21 & 5.5 & 2019-10-20 & $g$ = 19  \\
WY Sge & 19 32 43.87 &+17 44 54.5 &  1783 & 5.4 & 1982-06-17 & 17.7 \\
V476 Cyg & 19 58 24.46 & +53 37 07.5 & 1920 & 2.0 & 2019-08-24 &  $ g \approx$ 16.5\\
\enddata
\tablecomments{GK Per: \cite{Sabbadin_Bianchini_1983,Zemko_etal_2017,ATel_11995}, V392 Per: \cite{Liu_Hu_2000,Munari_etal_2020,Murphy-Glaysher_etal_2022}, AT Cnc: \cite{Shara_etal_2012ApJ...758..121S}, V1017 Sgr: \cite{Salazar_etal_2017}, V908 Oph: \cite{Tappert_2016,Mroz_etal_2015},V606 Aql: \cite{Tappert_2016,Duerbeck_1987,Bellm_2019},WY Sge: \cite{Shara_1984}, X Ser: \cite{Payne-Gaposchkin_1957}, V446 Her: \cite{Cragg_1960}, Nova Sco 1437: \cite{Kochanek_etal_2017,Shara_etal_2017Natur} V1213 Cen: \cite{mroz16}, V728 Sco: \cite{Vogt_etal_2018,Kato_2022}, V476 Cyg: \cite{Kato_2022}.
For systems with multiple dwarf nova outbursts, we list the brightest detected outburst found from a literature search. The objects in this table were found by searching the AAVSO Variable Star Index (VSX) for variability types containing a classical nova plus a dwarf nova identifier. It is possible that this list is not complete.} 
\end{deluxetable*}

\end{document}